%% file: main.tex
\documentclass[letterpaper,journal,twoside]{IEEEtran}
\pdfoutput=1

\input{localstyle}

\HideNotes

\title{Aggregate Characterization of User Behavior in Twitter and
  Analysis of the Retweet Graph}

\author{
  David R.\ Bild, Yue Liu, Robert P.\ Dick, Z.\ Morley Mao, and Dan S.\ Wallach%
  \thanks{This work is supported by the National Science Foundation under grant TC-0964545}%
  \thanks{D.\ R.\ Bild, Y.\ Liu, R.\ P.\ Dick, and Z.\ M.\ Mao are with the Electrical Engineering and Computer Science Department, University of Michigan, Ann Arbor, MI 48109. E-mail: drbild,liuyue,dickrp,zmao@umich.edu}%
  \thanks{D.\ S.\ Wallach is with the Department of Computer Science, Rice University, Houston, TX 77005. E-mail: dwallach@cs.rice.edu}}

\begin{document}

\maketitle

\thispagestyle{plain}
\pagestyle{plain}

\input{abs}
\input{intro}
\input{datasets}

\input{lifetime-tweets}
\input{tweet-rates}
\input{interevent-durations}

\input{retweet-graph}

\input{implications}
\input{spam}
\input{conc}

\input{appendices}

\nocite{tange11feb}
\bibliographystyle{IEEEtran}
\bibliography{robbib,rdgroup}

\end{document}

%% file: localstyle.tex
\usepackage[utf8]{inputenc}% read source as UTF-8
\usepackage[T1]{fontenc}% use modern font encoding
\usepackage{float}
\usepackage{amsmath,amssymb,amsthm}
\usepackage{commath} % typeset differential and other common math
\usepackage{etoolbox}
\usepackage{microtype}
\usepackage{threeparttable}
\usepackage{siunitx}
\usepackage{xfrac}
\usepackage{subfig}
\usepackage{multirow}
\usepackage{booktabs}
\usepackage[super]{nth}% superscript ordinals \nth{1}, \nth{2}, ...

\usepackage[bookmarks=false,hidelinks]{hyperref}

\usepackage{RobStd}
\usepackage{algorithm}% http://ctan.org/pkg/algorithms
\usepackage{algpseudocode}% http://ctan.org/pkg/algorithmicx

\usepackage{fixltx2e}% fix latex bugs, like out-of-order floats

\graphicspath{{fig/}}

\newcommand{\figinput}[1]{%
\providecommand{\fns}{\footnotesize}%
\providecommand{\sss}{\scriptsize}%
\providecommand{\ts}{\tiny}%
{\fontsize{10}{12}\selectfont{\textsf{\input{fig/#1}}}}%
\let\fns\undefined%
\let\ss\undefined
\let\ts\undefined
}

\newcommand{\apdxref}[1]{\hyperref[#1]{Appendix~\ref{#1}}}

\robustify\bfseries
\robustify\itshape

\DeclareMathOperator{\E}{\mathsf{E}}

\DeclareMathOperator{\erfc}{\mathsf{erfc}}

\DeclareMathOperator*{\argmax}{\arg\,\max}

\renewcommand{\texttilde}{\raise.17ex\hbox{$\scriptstyle\sim$}}

\newcommand{\setnatural}{\ensuremath{\mathbb{N}\xspace}}
\newcommand{\setcount}{\ensuremath{\setnatural^+\xspace}}

\newcommand{\est}[1]{\hat{#1}}

\newcommand{\likelihood}{\ensuremath{\mathcal{L}}\xspace}

\newcommand{\appropto}{\mathrel{\vcenter{
  \offinterlineskip\halign{\hfil$##$\cr
    \propto\cr\noalign{\kern2pt}\sim\cr\noalign{\kern-2pt}}}}}

%% file: abs.tex
\begin{abstract}
  Most previous analysis of Twitter user behavior is focused on
  individual information cascades and the social \emph{followers}
  graph. We instead study aggregate user behavior and the retweet
  graph with a focus on quantitative descriptions.  We find that the
  lifetime tweet distribution is a type-II discrete Weibull stemming
  from a power law hazard function, the tweet rate distribution,
  although asymptotically power law, exhibits a lognormal cutoff over
  finite sample intervals, and the inter-tweet interval distribution
  is power law with exponential cutoff.  The retweet graph is
  small-world and scale-free, like the social graph, but is less
  disassortative and has much stronger clustering.  These differences
  are consistent with it better capturing the real-world social
  relationships of and trust between users. Beyond just understanding
  and modeling human communication patterns and social networks,
  applications for alternative, decentralized microblogging
  systems---both predicting real-word performance and detecting
  spam---are discussed.
\end{abstract}

%% file: intro.tex
\section{Introduction}
\label{sec:intro}

\Note{Prior work has focused on cascades/social, but we need tweet
  behavior.}

Quantitative modeling of Twitter usage is important both for
understanding human communication patterns and optimizing the
performance of other microblogging-esque communication
platforms. However, prior analysis is focused on the social
graph~\cite{kwak10apr,bliss12sep,teutle10feb,gabielkov12dec,ghosh12apr}
or on individual information cascades that represent a small fraction
of all tweets~\cite{suh10aug,java07aug,wu11mar,lotan11,galuba10jun}.
Descriptions of basic behaviors are missing from the literature.  For
example, the qualitative distributions of the number of
\emph{followers} and \emph{friends} is available~\cite{kwak10apr}, but
not the distribution of tweet rates.  Common factors of tweets that
are heavily retweeted are known~\cite{suh10aug}, but propensity of
users to retweet, i.e., distribution of retweet rates, is not. We
begin to fill these gaps by considering user behavior \emph{as a
  whole}, providing quantitative descriptions of the distributions of
lifetime tweets, tweet rates, and inter-tweet times.

\Note{Motivated by designing new microblogging systems. Shout/ FETHER/
  Twister/ Etc.}

We are motivated by increasing interest in decentralized microblogging
systems designed to protect user privacy and resist censorship.
FETHER~\cite{sandler09apr}, Cuckoo~\cite{xu10jun}, and
Litter~\cite{stjuste11oct} reduce dependence on a single provider,
while Shout~\cite{shout} and Twister~\cite{twister} are explicitly
designed to avoid censorship and reprisal by government agencies.
Designing a decentralized system capable of handling the message rates
and volumes of Twitter is already a significant challenge and is
nearly impossible without a good understanding of those usage
patterns.

\Note{Need analysis/simulations with correct tweet volumes and distributions}

Given the complexity of these systems, understanding of the trade-offs
in the performance and cost metrics---throughput, latency, energy
consumption---is obtained through simulations, but such simulations
are only as accurate as the data and models driving them.  Consider
fair allocation of network resources---fairness looks very different
when the \emph{expected} distribution of tweets is 80--20 power law
and not uniform. Or, consider measuring delivery latencies, with
messages queuing at intermediate nodes, a metric dependent on the
(non-Poisson) arrival process, i.e., the inter-tweet duration
distribution.  Quantitative models of these basic behaviors are
needed.

\Note{Borrow/compare from other systems.  Found
  problems/contradictions with those analysis.}

The underlying human behaviors should extend across communication
platforms---tweet rates should mirror call rates in the telephone
network and total lifetime tweets should mirror total lifetime
contributions to Wikipedia or YouTube---suggesting that models of
those behaviors~\cite{wilkinson08jul,seshadri08aug,candia08jun} be
used in proxy for microblogging design.  However, our analysis of the
Twitter data shows differing behavior, indicating possible faults in
several of these models.  Our results for Twitter should enable future
work to identify or refine further commonalities in human
communication.

\Note{How does information spread.  Gossip instead of social?  Graph structure}

Tweets generally travel via the explicit social \emph{followers}
graph~\cite{kwak10apr}, which has been well-studied. Surprisingly, the
\emph{retweet} graph, in which a directed edge connects two users if
the source has retweeted the destination, has received almost no
attention. This \emph{implicit} graph may be actually more relevant to
information propagation in decentralized systems. A throughput-limited
system needs some way of prioritizing messages. People are usually
more selective in what they say than to whom they listen, so the
retweet graph may better encode true interest and trust relationships
among users.  For example, Shout\footnote{Shout~\cite{shout} is
  decentralized, geographic microblogging system in which messages are
  broadcast to users within radio range of the sender. Other users may
  re-shout the message, extending its reach, but the protocol does not
  directly support multi-hop delivery.} does not support
friend/follower relationships, so the retweet graph is the only
available social graph.  We conduct the first study of the retweet
graph obtained from a 4-month sample of 10\% of all tweets and compare
it to the social followers graph.

\Note{Applications}

These results have wide applicability.  The quantification of
communication behaviors and the social graph, beyond allowing direct
comparison with other already-characterized platforms, enables the
development of generative models explaining the underlying processes.
In a more direct view, knowing the number of tweets, tweet rates, and
inter-tweet times are sufficient for simulating and optimizing
microblogging platform performance and the confirmation that the
retweet graph is scale-free and small-word enables the generation of
random retweet graphs for empirical evaluation.  We focus on two such
applications, the design of distributed microblogging systems and the
detection of spammers using connectivity in the retweet graph.

\Note{Summary of specific contributions/results}

We have the following findings.  
\begin{itemize}
\item The distribution of lifetime tweets is discrete Weibull
  (type-II), generalizing a power law form shown by Wilkinson for
  other online communities~\cite{wilkinson08jul}. We conjecture that
  the Weibull shape parameter reflects the average amount of (positive
  or negative) feedback available to
  contributors. (\autoref{sec:lifetime-contributions})
\item The distribution of tweet (and retweet) rates is asymptotically
  power law, but exhibits a lognormal cutoff over finite-duration
  samples. Thus, high tweet rates are much more rare in practice than
  the asymptotic distribution would suggest.  We also discount a
  double Pareto lognormal (DPLN) explanation previously advanced in
  the context of call
  rates~\cite{seshadri08aug}. (\autoref{sec:tweet-rates})
\item The distribution of inter-tweet durations is power law with
  exponential cutoff, mirroring that of telephone
  calls~\cite{candia08jun}. (\autoref{sec:intertweet-durations})
\item The retweet graph is small-world and (roughly) scale-free, like
  the social followers graph, but less disassortative and more highly
  clustered.  It is more similar than the followers graph to
  real-world social networks, consistent with better reflection of
  real-world relationships and trust. (\autoref{sec:retweet-graph})
\end{itemize}

In \autoref{sec:implications}, we discuss the implications of these
results for decentralized microblogging architectures and in
\autoref{sec:spam} we consider using the structure of the retweet
graph for spammer detection.

%% file: datasets.tex
\section{Datasets}
\label{sec:datasets}

\Note{Why multiple data sets; corrections; etc.}

The Twitter API rate limits and terms of service prevent collection
and sharing of a single complete tweet dataset suitable for all our
queries~\cite{watters11mar}. Our largest and most recent
dataset---10\% of all tweets sent between June and September 2012---is
the focus of our analysis, but we supplement with sets from other
researchers as necessary. This section summarizes these datasets and
describes our main procedure for inferring population statistics from
the 10\% sample.

\subsection{2009 Social Graph}
\label{sec:dataset-social-graph}

Kwak \etal's 2009 crawl~\cite{kwak10apr} remains the largest and most
complete public snapshot of the Twitter social followers graph,
covering 41.7 million users and 1.47 billion relations.  The data is
dated, but still the best available. Repeating this crawl is
infeasible under current rate limits and feasible sampling strategies
(e.g., snowball-sampling~\cite{goodman61mar}) lead to results that are
difficult to interpret~\cite{lee06jan}.  We use this social graph
snapshot for all comparisons with the retweet graph.

\subsection{Lifetime Contribution Dataset}
\label{sec:dataset-lifetime}

No tweet dataset is complete enough to compute lifetime contributions,
the number of tweets sent before quitting Twitter, but the Twitter API
exposes (subject to rate limits) the necessary information. We
collected account age, date of last tweet, and total tweet count (as
of June 2013) for \num{1318683} users selected uniformly randomly from
the 2009 social graph set\footnote{The 2009 social graph dataset is
  the closest to a uniform random sample of Twitter users we could
  find. More recent sets are biased towards users that tweet more
  often.}.  \num{525779} of these users were inactive,\footnote{The
  creation dates of protected tweets are hidden, so all users with
  protected tweets were excluded.} i.e., had not tweeted in the prior
six months~\cite{wilkinson08jul}. Their ages and tweet counts form the
lifetime contribution set used in
\autoref{sec:lifetime-contributions}.

\subsection{SNAP Tweet Dataset}
\label{sec:dataset-snap}

Computing inter-tweet intervals requires consecutive tweets---a random
sample is insufficient\footnote{A random sample would be sufficient if
  the process were Poisson, but it is not.}.  For our inter-tweet
distribution analysis in \autoref{sec:intertweet-durations}, we use a
collection of 467 million tweets gathered by the SNAP team in
2009~\cite{yang11feb}. The full dataset is no longer publicly
available per request from Twitter, but the authors kindly shared the
inter-tweet metadata.

\subsection{10\% Sample (Gardenhose) Dataset}
\label{sec:dataset-gardenhose}

\begin{table}
\begin{center}
  \caption{10\% Sample (Gardenhose) Dataset}
\label{tbl:dataset-gardenhose}
  \begin{threeparttable}
    \begin{tabular}{@{}lS[table-format=11]S[table-format=11]@{}}
      \toprule
      & {\bfseries 10\% Sample}  & {\bfseries Actual Value\tnote{\textdagger}} \\
      \midrule
      \# of Tweets   & 4097787713        & 41256584408 \\
      \# of Retweets &  953457874        &  9664691519 \\
      \midrule
      \# of Tweeters   & 104083457       &   166335390 \\
      \# of Retweeters &  51319979       &    84278086 \\
      \# of Retweetees &  38975108       &    69224526 \\
      \bottomrule
    \end{tabular}
    \begin{tablenotes}
    \item [\textdagger] Estimated using the described EM procedure.
  \end{tablenotes}
\end{threeparttable}
\end{center}
\end{table}

Our primary dataset is a uniform random 10\% sample\footnote{More
  precisely, each tweet is included in this sample with 10\%
  probability.}  of all tweets (the ``gardenhose'' stream) sent in the
four month period spanning June through September
2012. \autoref{tbl:dataset-gardenhose} shows the scope of the dataset,
using the following definitions. A \emph{tweeter} is a user that sends
a \emph{tweet}, an original message. A \emph{retweeter} is a user that
sends a \emph{retweet}, forwarding a previous tweet.  A
\emph{retweetee} is a user whose tweet was \emph{retweeted}. Retweets
were identified using both Twitter-provided metadata and analysis of
the message contents for retweet syntax, e.g., ``RT@'', as described
in \apdxref{sec:retweet-identification}.

The sampled data poses a challenge for drawing quantitative
conclusions about user behavior and the structure of the retweet
graph. For many of the distributions we wish to quantify, the sample
is biased towards users that tweeted more frequently. In fact, most
users with fewer than ten tweets will not appear at all. Much prior
work in the social network and graph analysis literature has focused
on qualitatively characterizing the errors introduced by subsampling,
motivated by quicker analysis~\cite{lee06jan,son12oct}. We instead
develop an approach to accurately estimate quantitative population
statistics from the 10\% random sample.

\subsection{Estimating Population Distributions from the 10\% Sample
  Dataset}
\label{sec:em-summary}

For simplicity, we describe the method for a concrete problem:
determining the distribution of tweets per user during the four month
window.  The method is trivially adapted to a variety of such
problems, including multivariate joint distributions as in
\autoref{sec:reciprocity}. Similar approaches are used in other
fields~\cite{duffield05oct}.  We wish to determine the number of
users, $f_i$, with $i\in \setcount$ tweets given the number of users,
$g_j$, with $j\in \setcount$ tweets observed in the sample.  $g_j$
includes some users from each $f_{i\geq j}$, with the binomial
distribution $B_{0.1}(i, j)$ describing how the users in $f_i$ are
partitioned among the various $g_{j\leq i}$.  Intuitively, a good
estimate $\est{f}$ is that which maximizes the probability of the
observation $g$, i.e., standard maximum likelihood estimation.

The corresponding likelihood function is not analytically tractable,
so we employ an expectation maximization
algorithm~\cite{dempster77,borman09jan} to compute the estimate
$\est{f}$, summarized here (see \apdxref{sec:em-derivation} for
details). Let $\phi_i$ be the probability that a user sends $i$ tweets
conditional on at least one of them being observed and $c_{i,j}$ be
the probability a user with $i$ tweets has $j$ of them observed
conditional on $j\geq 1$ (i.e., the binomial probability conditional
on at least one success). The log-likelihood function to maximize is
\begin{equation}
  \label{eqn:em-likelihood}
  \likelihood(\phi|f,g) = \sum_{1\leq j\leq i}f_{i,j}\log\left(\phi_ic_{i,j}\right),
\end{equation}
where $\phi$ are the parameters to estimate and $f$ and $g$ are
the hidden and observed variables, respectively.  We compute the
parameter estimate by iteratively selecting a new estimate
$\phi^{k+1}$ that maximizes the \emph{expected} likelihood
under the previous estimate $\phi^k$, i.e.,
\begin{equation}
  \label{eqn:em-max}
  \phi^{k+1} \triangleq \argmax_{\phi} \mathcal{Q}\left(\phi, \phi^k\right),
\end{equation}
where
\begin{equation}
  \label{eqn:em-expectation}
  \mathcal{Q}\left(\phi, \phi^k\right) \triangleq \E_{f|g,\phi^k}\left[\likelihood(\phi|f,g)\right].
\end{equation}
This process is known to converge~\cite{mclachlan08}.  Letting $\gamma
\triangleq \sum_{1\leq j}g_j$ be the total number of observed users,
\autoref{eqn:em-max} can be solved using Lagrangian multipliers to
yield
\begin{equation}
  \phi^{k+1}_i = \frac{1}{\gamma}\E_{\phi^k}\left[f_i|g\right]
\end{equation}
and the hidden original frequencies recovered from the final estimate
$\est{\phi}$ as
\begin{equation}
  \est{f}_i = \gamma\est{\phi_i}\frac{1}{1 - B_{0.1}(i,0)}.
\end{equation}

\begin{figure}
  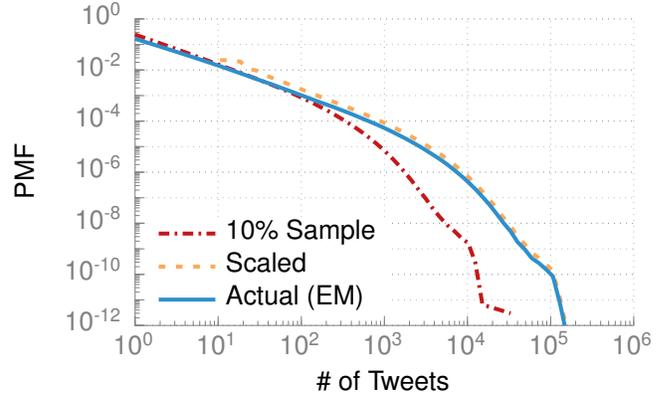
  \caption{Distribution of tweets per user for the 10\% sample, the
    scaled sample ($j$ observed tweets maps to $10j$ sent tweets), and
    the underlying population as estimated by the EM algorithm. The
    differences illustrate the importance of recovering the actual
    distribution via, for example, our EM algorithm.}
  \label{fig:mle-example}
\end{figure}

\autoref{fig:mle-example} shows the result using the distribution of
tweets sent during our four-month collection window as an example.
The correct distribution computed via the EM algorithm is
substantially different, particularly in the lower decades, from the
uncorrected or scaled (i.e., assuming that observing $j$ tweets
implies $10j$ were sent) distributions.

%% file: fig/mle-example.pdf_tex
% GNUPLOT: LaTeX picture with Postscript
\begingroup
  \makeatletter
  \providecommand\color[2][]{%
    \GenericError{(gnuplot) \space\space\space\@spaces}{%
      Package color not loaded in conjunction with
      terminal option `colourtext'%
    }{See the gnuplot documentation for explanation.%
    }{Either use 'blacktext' in gnuplot or load the package
      color.sty in LaTeX.}%
    \renewcommand\color[2][]{}%
  }%
  \providecommand\includegraphics[2][]{%
    \GenericError{(gnuplot) \space\space\space\@spaces}{%
      Package graphicx or graphics not loaded%
    }{See the gnuplot documentation for explanation.%
    }{The gnuplot epslatex terminal needs graphicx.sty or graphics.sty.}%
    \renewcommand\includegraphics[2][]{}%
  }%
  \providecommand\rotatebox[2]{#2}%
  \@ifundefined{ifGPcolor}{%
    \newif\ifGPcolor
    \GPcolortrue
  }{}%
  \@ifundefined{ifGPblacktext}{%
    \newif\ifGPblacktext
    \GPblacktextfalse
  }{}%
  % define a \g@addto@macro without @ in the name:
  \let\gplgaddtomacro\g@addto@macro
  % define empty templates for all commands taking text:
  \gdef\gplbacktext{}%
  \gdef\gplfronttext{}%
  \makeatother
  \ifGPblacktext
    % no textcolor at all
    \def\colorrgb#1{}%
    \def\colorgray#1{}%
  \else
    % gray or color?
    \ifGPcolor
      \def\colorrgb#1{\color[rgb]{#1}}%
      \def\colorgray#1{\color[gray]{#1}}%
      \expandafter\def\csname LTw\endcsname{\color{white}}%
      \expandafter\def\csname LTb\endcsname{\color{black}}%
      \expandafter\def\csname LTa\endcsname{\color{black}}%
      \expandafter\def\csname LT0\endcsname{\color[rgb]{1,0,0}}%
      \expandafter\def\csname LT1\endcsname{\color[rgb]{0,1,0}}%
      \expandafter\def\csname LT2\endcsname{\color[rgb]{0,0,1}}%
      \expandafter\def\csname LT3\endcsname{\color[rgb]{1,0,1}}%
      \expandafter\def\csname LT4\endcsname{\color[rgb]{0,1,1}}%
      \expandafter\def\csname LT5\endcsname{\color[rgb]{1,1,0}}%
      \expandafter\def\csname LT6\endcsname{\color[rgb]{0,0,0}}%
      \expandafter\def\csname LT7\endcsname{\color[rgb]{1,0.3,0}}%
      \expandafter\def\csname LT8\endcsname{\color[rgb]{0.5,0.5,0.5}}%
    \else
      % gray
      \def\colorrgb#1{\color{black}}%
      \def\colorgray#1{\color[gray]{#1}}%
      \expandafter\def\csname LTw\endcsname{\color{white}}%
      \expandafter\def\csname LTb\endcsname{\color{black}}%
      \expandafter\def\csname LTa\endcsname{\color{black}}%
      \expandafter\def\csname LT0\endcsname{\color{black}}%
      \expandafter\def\csname LT1\endcsname{\color{black}}%
      \expandafter\def\csname LT2\endcsname{\color{black}}%
      \expandafter\def\csname LT3\endcsname{\color{black}}%
      \expandafter\def\csname LT4\endcsname{\color{black}}%
      \expandafter\def\csname LT5\endcsname{\color{black}}%
      \expandafter\def\csname LT6\endcsname{\color{black}}%
      \expandafter\def\csname LT7\endcsname{\color{black}}%
      \expandafter\def\csname LT8\endcsname{\color{black}}%
    \fi
  \fi
    \setlength{\unitlength}{0.0500bp}%
    \ifx\gptboxheight\undefined%
      \newlength{\gptboxheight}%
      \newlength{\gptboxwidth}%
      \newsavebox{\gptboxtext}%
    \fi%
    \setlength{\fboxrule}{0.5pt}%
    \setlength{\fboxsep}{1pt}%
\begin{picture}(4960.00,3060.00)%
    \gplgaddtomacro\gplbacktext{%
      \colorrgb{0.50,0.50,0.50}%
      \put(845,569){\makebox(0,0)[r]{\strut{}10\textsuperscript{-12}}}%
      \colorrgb{0.50,0.50,0.50}%
      \put(845,954){\makebox(0,0)[r]{\strut{}10\textsuperscript{-10}}}%
      \colorrgb{0.50,0.50,0.50}%
      \put(845,1340){\makebox(0,0)[r]{\strut{}10\textsuperscript{-8}}}%
      \colorrgb{0.50,0.50,0.50}%
      \put(845,1725){\makebox(0,0)[r]{\strut{}10\textsuperscript{-6}}}%
      \colorrgb{0.50,0.50,0.50}%
      \put(845,2110){\makebox(0,0)[r]{\strut{}10\textsuperscript{-4}}}%
      \colorrgb{0.50,0.50,0.50}%
      \put(845,2496){\makebox(0,0)[r]{\strut{}10\textsuperscript{-2}}}%
      \colorrgb{0.50,0.50,0.50}%
      \put(845,2881){\makebox(0,0)[r]{\strut{}10\textsuperscript{0}}}%
      \colorrgb{0.50,0.50,0.50}%
      \put(845,762){\makebox(0,0)[r]{\strut{}}}%
      \colorrgb{0.50,0.50,0.50}%
      \put(845,1147){\makebox(0,0)[r]{\strut{}}}%
      \colorrgb{0.50,0.50,0.50}%
      \put(845,1532){\makebox(0,0)[r]{\strut{}}}%
      \colorrgb{0.50,0.50,0.50}%
      \put(845,1918){\makebox(0,0)[r]{\strut{}}}%
      \colorrgb{0.50,0.50,0.50}%
      \put(845,2303){\makebox(0,0)[r]{\strut{}}}%
      \colorrgb{0.50,0.50,0.50}%
      \put(845,2688){\makebox(0,0)[r]{\strut{}}}%
      \colorrgb{0.50,0.50,0.50}%
      \put(934,391){\makebox(0,0){\strut{}10\textsuperscript{0}}}%
      \colorrgb{0.50,0.50,0.50}%
      \put(1560,391){\makebox(0,0){\strut{}10\textsuperscript{1}}}%
      \colorrgb{0.50,0.50,0.50}%
      \put(2187,391){\makebox(0,0){\strut{}10\textsuperscript{2}}}%
      \colorrgb{0.50,0.50,0.50}%
      \put(2813,391){\makebox(0,0){\strut{}10\textsuperscript{3}}}%
      \colorrgb{0.50,0.50,0.50}%
      \put(3439,391){\makebox(0,0){\strut{}10\textsuperscript{4}}}%
      \colorrgb{0.50,0.50,0.50}%
      \put(4066,391){\makebox(0,0){\strut{}10\textsuperscript{5}}}%
      \colorrgb{0.50,0.50,0.50}%
      \put(4692,391){\makebox(0,0){\strut{}10\textsuperscript{6}}}%
    }%
    \gplgaddtomacro\gplfronttext{%
      \csname LTb\endcsname%
      \put(133,1725){\rotatebox{-270}{\makebox(0,0){\strut{}PMF}}}%
      \csname LTb\endcsname%
      \put(2813,124){\makebox(0,0){\strut{}\# of Tweets}}%
      \csname LTb\endcsname%
      \put(1628,1262){\makebox(0,0)[l]{\strut{}10\% Sample}}%
      \csname LTb\endcsname%
      \put(1628,1013){\makebox(0,0)[l]{\strut{}Scaled}}%
      \csname LTb\endcsname%
      \put(1628,764){\makebox(0,0)[l]{\strut{}Actual (EM)}}%
    }%
    \gplbacktext
    \put(0,0){\includegraphics{mle-example}}%
    \gplfronttext
  \end{picture}%
\endgroup

%% file: lifetime-tweets.tex
\section{Distribution of Lifetime Tweets}
\label{sec:lifetime-contributions}

\Note{Introduction to Lifetime Contributions and Prior Work}

Strong regularities in participation behavior have been observed
across many online peer production systems, suggesting a common
underlying dynamic.  Wilkinson found that for Bugzilla, Essembly,
Wikipedia, and Digg, the probability that a user makes no further
contributions is inversely proportional to the number of contributions
already made, suggesting a notion of \emph{participation
  momentum}~\cite{wilkinson08jul}.  Huberman \etal observed the same
in YouTube~\cite{huberman09dec}.  We look for a similar effect in
Twitter.

We quantify contribution as the number of tweets sent\footnote{One
  could instead consider retweets, replies, or direct messages, but
  obtaining data for these is more difficult.}, so the lifetime
contribution is the tweet count when the user becomes inactive.
Following Wilkinson~\cite{wilkinson08jul}, a user that has not tweeted
for six months (as of June 2013 when our lifetime contributions
dataset was collected) is \emph{inactive}.

\Note{Results and Analysis}

\begin{figure}
  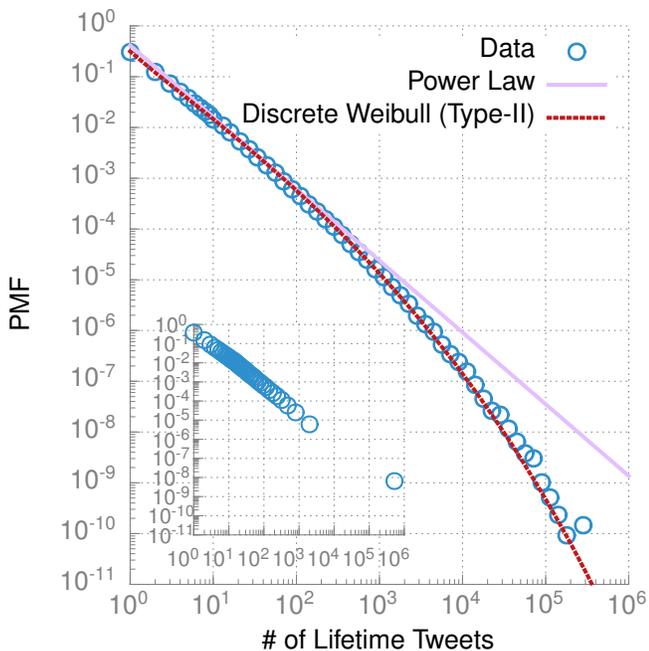
  \caption{Distribution of total lifetime tweets.  Distribution
    parameters (\autoref{tbl:lifetime-tweets-params}) were obtained by
    maximum likelihood estimation. In the inset, equal-count binning
    obscures the cutoff. The sparse upper tail causes a wide and thus
    seemingly-outlying last bin.}
  \label{fig:lifetime-tweets}
\end{figure}

\Note{Graph of distributions}

\autoref{fig:lifetime-tweets} plots the
logarithmically-binned~\cite{milojevic10dec} empirical distribution.
It is heavy-tailed, but decays more quickly in the upper tail than a
true power law.  The higher density in the last bin
(\texttilde\num{200000} tweets) is due to Twitter's rate limits of
1000 tweets per day and 100 tweets per hour, because users that would
occupy the upper tail (>\num{200000} tweets) are forced into this
bin.\footnote{The rate limit means that the lifetime contribution
  distribution can be viewed as a censored~\cite{johnson05censoring}
  version of the ``natural'' distribution.} YouTube exhibits the same
non-power law, upper tail cutoff~\cite{huberman09dec}, consistent with
a common dynamic underlying both systems.

\subsection{Critique of Previously-Reported Power Law Behavior}

\Note{Why not the power law of Wilkinson}

Surprisingly, the cutoff does not match the strong power law evidence
reported for Bugzilla, Essembly, Wikipedia, and
Digg~\cite{wilkinson08jul}.  We believe those systems do contain a
similar cutoff, but it was obscured by the analysis methods used.  We
observe three weaknesses of the prior approach.  First, the
equal-count binning\footnote{In equal-count binning, each bin is sized
  to contain the same number of samples and thus the same area under
  the density function.  For $B$ bins, the height of a bin $b_i$ is
  computed as $B/w(b_i),$ where $w(b_i)$ is the width of $b_i$.}
method used obscures the upper tail behavior; logarithmic binning is
preferred~\cite{milojevic10dec}. Second, maximum likelihood
estimation, not binned regression, should be used for
fitting~\cite{clauset09}.  Finally, the goodness-of-fit should be
computed against the empirical distribution function
(Kolmogorov--Smirnov or Anderson--Darling test)~\cite{clauset09}, not
against binned data (the G-test).

\begin{table}
  \centering
  \caption{Power-Law Exponents for Lifetime Contributions in Various
    Online Communities, Computed Incorrectly Using Equal-Count Binning}
  \label{tbl:momentum-exponents}
  \begin{threeparttable}
    \begin{tabular}{@{}lS[table-format=1.2]S[table-format=1.2]S[table-format=2]@{}}
      \toprule
      \bfseries Contribution Type & {\bfseries $\boldsymbol{\alpha}$} & {\bfseries $\mathbf{p}$-value} & {\bfseries min. $\mathbf{k}$} \\
      \midrule
      Essembly votes\tnote{\textdagger} & 1.47 & 0.59 & 3 \\
      Digg votes\tnote{\textdagger} & 1.53 & 0.64 & 15 \\
      \itshape Twitter tweets & \itshape 1.54 & \itshape 0.96 & \itshape 12 \\
      Bugzilla comments\tnote{\textdagger} & 1.98 & 0.74 & 5 \\
      Essembly submissions\tnote{\textdagger} & 2.02 & 0.25 & 7 \\
      Wikipedia edits\tnote{\textdagger} & 2.28 & 0.69 & 10 \\
      Digg submissions\tnote{\textdagger} & 2.40 & 0.04 & 15 \\
      Youtube submissions\tnote{\textdaggerdbl} & 2.46 & {---} & {---} \\
      \bottomrule
    \end{tabular}
    \begin{tablenotes}
      \item [\textdagger] from Wilkinson~\cite{wilkinson08jul}
      \item [\textdaggerdbl] from Huberman, Romero, and Wu~\cite{huberman09dec}
    \end{tablenotes}
  \end{threeparttable}
\end{table}

The original datasets are unavailable\footnote{Emails to the author
  bounced as undeliverable.}, so we tested our hypothesis by applying
the same methods to our Twitter data.  As expected, equal-count
binning, shown in the inset of \autoref{fig:lifetime-tweets}, hides
the known cutoff.  The G-test for a power law fit by regression to the
improperly binned data indicates a good match
(\autoref{tbl:momentum-exponents}), despite the obvious mismatch in
the real data. Clearly, these methods can obscure any underlying
cutoff.  Our results are consistent with Bugzilla, Essembly,
Wikipedia, and Digg contributions containing the same cutoffs as
Twitter and YouTube, but the original data would be needed to prove
this conclusion.

\begin{table}
  \centering
  \caption{Parameters for Distributions of Lifetime Tweets}
  \label{tbl:lifetime-tweets-params}
  \begin{tabular}{@{}lllS[table-format=-1.2e-1,table-align-exponent=false]@{}}
    \toprule
    \multicolumn{2}{@{}l}{\bfseries Distribution} & \multicolumn{2}{c}{\bfseries Parameters} \\
    \cmidrule{1-2}
    Name & PMF & \multicolumn{2}{c}{(fit by MLE)}\\
    \midrule
    \multirow{2}{*}{Power Law} & \multirow{2}{*}{$\displaystyle\frac{1}{\zeta(\alpha,x_\textrm{min})}\cdot\frac{1}{x^{\alpha}}$} & $\alpha$ & 1.54 \\
                               &                                                                        & $x_\textrm{min}$ & 12.00 \\
    \addlinespace
    Type-II & \multirow{3}{*}{$\displaystyle\frac{c}{x^{1-\beta}}\prod_{n=1}^{x-1}\left(1-\frac{c}{n^{1-\beta}}\right)$}  & $\beta$ & 0.17 \\
    Discrete &                                        & $c$ & 0.32 \\
    Weibull~\cite{stein84jun} \\
    \bottomrule
  \end{tabular}
\end{table}

\subsection{Lifetime Tweets Follow a Weibull Distribution}

\Note{Fit of distributions}

\begin{figure}
  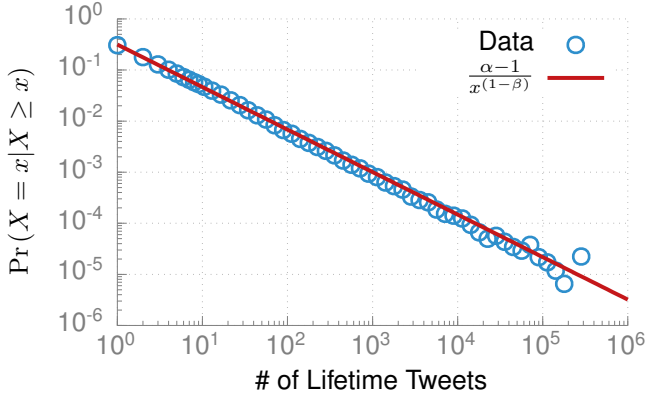
  \caption{The probability that a user who has sent $x$ tweets quits
    without sending another, i.e., the hazard rate.  The decreasing
    trend suggests a sort of momentum; the more times a user has
    tweeted, the more likely he is to tweet again. The power law
    parameters are calculated from
    \autoref{tbl:lifetime-tweets-params}, not fit to the data.}
  \label{fig:lifetime-momentum}
\end{figure}

If the distribution is not power law, what is it? Examining the
\emph{hazard function}, or probability that a user who has made $x$
contributions quits without another, provides the answer.  Shown in
\autoref{fig:lifetime-momentum}, the hazard function is an obvious
power law.  Wilkinson referred to this behavior in other online
communities as \emph{participation momentum}~\cite{wilkinson08jul}; we
will return to that interpretation later.

The power law hazard function $\frac{\alpha-1}{x^{1-\beta}}$ is that
of the Weibull distribution\footnote{The Weibull distribution is
  sometimes called the \emph{stretched exponential}.}, for continuous
support.  For discrete support, the distribution with a power law
hazard function is called a Type II Discrete Weibull\footnote{The much
  more common Type I Discrete Weibull~\cite{nakagawa75dec} instead
  preserves the exponential form of the complementary cumulative
  density function.}~\cite{stein84jun} and has mass function
\begin{equation}
  \Pr(X=x) = \frac{\alpha-1}{x^{1-\beta}}\prod_{n=1}^{x-1}\left(1-\frac{\alpha-1}{n^{1-\beta}}\right).
\end{equation}
A maximum likelihood fit to the lifetime contribution data yields
$\beta=0.17$ and $\alpha=1.32$, as shown in
\autoref{fig:lifetime-tweets}. The upper tail deviates slightly, which
we attribute to Twitter's rate limit policy.  Some users that would
have tweeted more than \texttilde\num{200000} times were artificially
limited to fewer tweets, increasing the weight in that portion of the
upper tail.

\subsection{Interpreting the Hazard Function as Participation
  Momentum}

\Note{Momentum interpretation}

Wilkinson~\cite{wilkinson08jul} used a notion of participation
momentum to explain the power law hazard function. For his assumed
power law distribution, $C\frac{1}{x^{\alpha}}$, the hazard function
is $\frac{\alpha - 1}{x}$ and $\alpha$ can be seen as a metric for the
effort needed to contribute.  Higher required effort leads to a higher
probability of quitting. \autoref{tbl:momentum-exponents} shows the
$\alpha$'s for several systems.  Intuitively, tweeting seems more
taxing than voting on Digg stories but less so than commenting on
Bugzilla reports. And indeed, we find that $\alpha_\text{Digg} <
\alpha_\text{Twitter} < \alpha_\text{Bugzilla}$.

Alternatively, the hazard function might be more directly related to
account age than total contributions. To reject this possibility, we
compared the Kendall tau rank correlations~\cite{kendall38jun} between
lifetime contributions, age, and average tweet rate (lifetime
contributions/age).  Unsurprisingly, age (i.e., longer life)
correlates with increased lifetime contributions ($\tau =
\num{0.4708}$, $p = \num{0.00}$, 95\% CI [$\num{0.4690}$,
$\num{0.4726}$]).  In contrast, the tweet rate is essentially
uncorrelated with lifetime contributions ($\tau = \num{-0.0067}$, $p =
0.00$, 95\% CI [$\num{-0.0085}$, $\num{-0.0049}$]), indicating that
the momentum function is not driven by age. If it were, the
correlation would be strongly positive because faster tweeters would
generate more tweets in their (independently determined) lifetimes.
The strong negative relationship between tweet rate and age ($\tau =
\num{-0.5687}$, $p = 0.00$, 95\% CI [$\num{-0.5705}$,
$\num{-0.5669}$]) further supports this conclusion. The hazard rate is
determined by the current total contributions, so users with higher
tweet rates must have shorter lifetime ages.

The hazard function we observe ($\frac{\alpha-1}{x^{1 - \beta}}$
instead of Wilkinson's $\frac{\alpha - 1}{x}$) invites additional
thought.  The new parameter $\beta$ ($\beta=0$ in Wilkinson's model)
models momentum gain---a higher $\beta$ translates to more momentum
gain per contribution.  For example, one could imagine that $\beta$
reflects the effect of feedback. Positive (negative) viewer-generated
feedback like retweets and replies in Twitter or comments and view
counts in YouTube might accelerate (decelerate) momentum gains
relative to systems without such visible feedback, like Digg votes or
Wikipedia edits.\footnote{Wilkinson's reported results are consistent
  with this hypothesis.  The contribution types with the most visible
  feedback---Essembly and Digg submissions---show little support for a
  power law, with p-values of 0.25 and 0.04. $\beta > 0$ would explain
  the non-power law behavior. The distribution for YouTube by Huberman
  \etal also shows a cutoff~\cite{huberman09dec} consistent with a
  hazard function with $\beta > 0$.} Refinement of this interpretation
is a promising area for future work.

In summary, lifetime contributions in Twitter are driven by a power
law hazard function $\left(\frac{\alpha-1}{x^{1-\beta}}\right)$ viewed
as participation momentum. $\alpha$ reflects the effort needed to
contribute and $\beta$ the amount of feedback provided by system.  The
power law momentum leads to a Type II Discrete Weibull distribute for
lifetime contributions.  This dynamic holds across a variety of online
communities~\cite{wilkinson08jul,huberman09dec}.

%% file: fig/lifetime-tweets.pdf_tex
% GNUPLOT: LaTeX picture with Postscript
\begingroup
  \makeatletter
  \providecommand\color[2][]{%
    \GenericError{(gnuplot) \space\space\space\@spaces}{%
      Package color not loaded in conjunction with
      terminal option `colourtext'%
    }{See the gnuplot documentation for explanation.%
    }{Either use 'blacktext' in gnuplot or load the package
      color.sty in LaTeX.}%
    \renewcommand\color[2][]{}%
  }%
  \providecommand\includegraphics[2][]{%
    \GenericError{(gnuplot) \space\space\space\@spaces}{%
      Package graphicx or graphics not loaded%
    }{See the gnuplot documentation for explanation.%
    }{The gnuplot epslatex terminal needs graphicx.sty or graphics.sty.}%
    \renewcommand\includegraphics[2][]{}%
  }%
  \providecommand\rotatebox[2]{#2}%
  \@ifundefined{ifGPcolor}{%
    \newif\ifGPcolor
    \GPcolortrue
  }{}%
  \@ifundefined{ifGPblacktext}{%
    \newif\ifGPblacktext
    \GPblacktextfalse
  }{}%
  % define a \g@addto@macro without @ in the name:
  \let\gplgaddtomacro\g@addto@macro
  % define empty templates for all commands taking text:
  \gdef\gplbacktext{}%
  \gdef\gplfronttext{}%
  \makeatother
  \ifGPblacktext
    % no textcolor at all
    \def\colorrgb#1{}%
    \def\colorgray#1{}%
  \else
    % gray or color?
    \ifGPcolor
      \def\colorrgb#1{\color[rgb]{#1}}%
      \def\colorgray#1{\color[gray]{#1}}%
      \expandafter\def\csname LTw\endcsname{\color{white}}%
      \expandafter\def\csname LTb\endcsname{\color{black}}%
      \expandafter\def\csname LTa\endcsname{\color{black}}%
      \expandafter\def\csname LT0\endcsname{\color[rgb]{1,0,0}}%
      \expandafter\def\csname LT1\endcsname{\color[rgb]{0,1,0}}%
      \expandafter\def\csname LT2\endcsname{\color[rgb]{0,0,1}}%
      \expandafter\def\csname LT3\endcsname{\color[rgb]{1,0,1}}%
      \expandafter\def\csname LT4\endcsname{\color[rgb]{0,1,1}}%
      \expandafter\def\csname LT5\endcsname{\color[rgb]{1,1,0}}%
      \expandafter\def\csname LT6\endcsname{\color[rgb]{0,0,0}}%
      \expandafter\def\csname LT7\endcsname{\color[rgb]{1,0.3,0}}%
      \expandafter\def\csname LT8\endcsname{\color[rgb]{0.5,0.5,0.5}}%
    \else
      % gray
      \def\colorrgb#1{\color{black}}%
      \def\colorgray#1{\color[gray]{#1}}%
      \expandafter\def\csname LTw\endcsname{\color{white}}%
      \expandafter\def\csname LTb\endcsname{\color{black}}%
      \expandafter\def\csname LTa\endcsname{\color{black}}%
      \expandafter\def\csname LT0\endcsname{\color{black}}%
      \expandafter\def\csname LT1\endcsname{\color{black}}%
      \expandafter\def\csname LT2\endcsname{\color{black}}%
      \expandafter\def\csname LT3\endcsname{\color{black}}%
      \expandafter\def\csname LT4\endcsname{\color{black}}%
      \expandafter\def\csname LT5\endcsname{\color{black}}%
      \expandafter\def\csname LT6\endcsname{\color{black}}%
      \expandafter\def\csname LT7\endcsname{\color{black}}%
      \expandafter\def\csname LT8\endcsname{\color{black}}%
    \fi
  \fi
  \setlength{\unitlength}{0.0500bp}%
  \begin{picture}(4960.00,4960.00)%
    \gplgaddtomacro\gplbacktext{%
      \colorrgb{0.50,0.50,0.50}%
      \put(845,569){\makebox(0,0)[r]{\strut{}10\textsuperscript{-11}}}%
      \colorrgb{0.50,0.50,0.50}%
      \put(845,952){\makebox(0,0)[r]{\strut{}10\textsuperscript{-10}}}%
      \colorrgb{0.50,0.50,0.50}%
      \put(845,1335){\makebox(0,0)[r]{\strut{}10\textsuperscript{-9}}}%
      \colorrgb{0.50,0.50,0.50}%
      \put(845,1718){\makebox(0,0)[r]{\strut{}10\textsuperscript{-8}}}%
      \colorrgb{0.50,0.50,0.50}%
      \put(845,2101){\makebox(0,0)[r]{\strut{}10\textsuperscript{-7}}}%
      \colorrgb{0.50,0.50,0.50}%
      \put(845,2484){\makebox(0,0)[r]{\strut{}10\textsuperscript{-6}}}%
      \colorrgb{0.50,0.50,0.50}%
      \put(845,2866){\makebox(0,0)[r]{\strut{}10\textsuperscript{-5}}}%
      \colorrgb{0.50,0.50,0.50}%
      \put(845,3249){\makebox(0,0)[r]{\strut{}10\textsuperscript{-4}}}%
      \colorrgb{0.50,0.50,0.50}%
      \put(845,3632){\makebox(0,0)[r]{\strut{}10\textsuperscript{-3}}}%
      \colorrgb{0.50,0.50,0.50}%
      \put(845,4015){\makebox(0,0)[r]{\strut{}10\textsuperscript{-2}}}%
      \colorrgb{0.50,0.50,0.50}%
      \put(845,4398){\makebox(0,0)[r]{\strut{}10\textsuperscript{-1}}}%
      \colorrgb{0.50,0.50,0.50}%
      \put(845,4781){\makebox(0,0)[r]{\strut{}10\textsuperscript{0}}}%
      \colorrgb{0.50,0.50,0.50}%
      \put(934,391){\makebox(0,0){\strut{}10\textsuperscript{0}}}%
      \colorrgb{0.50,0.50,0.50}%
      \put(1560,391){\makebox(0,0){\strut{}10\textsuperscript{1}}}%
      \colorrgb{0.50,0.50,0.50}%
      \put(2187,391){\makebox(0,0){\strut{}10\textsuperscript{2}}}%
      \colorrgb{0.50,0.50,0.50}%
      \put(2813,391){\makebox(0,0){\strut{}10\textsuperscript{3}}}%
      \colorrgb{0.50,0.50,0.50}%
      \put(3439,391){\makebox(0,0){\strut{}10\textsuperscript{4}}}%
      \colorrgb{0.50,0.50,0.50}%
      \put(4066,391){\makebox(0,0){\strut{}10\textsuperscript{5}}}%
      \colorrgb{0.50,0.50,0.50}%
      \put(4692,391){\makebox(0,0){\strut{}10\textsuperscript{6}}}%
      \csname LTb\endcsname%
      \put(133,2675){\rotatebox{-270}{\makebox(0,0){\strut{}PMF}}}%
      \csname LTb\endcsname%
      \put(2813,124){\makebox(0,0){\strut{}\# of Lifetime Tweets}}%
    }%
    \gplgaddtomacro\gplfronttext{%
      \csname LTb\endcsname%
      \put(3998,4586){\makebox(0,0)[r]{\strut{}Data}}%
      \csname LTb\endcsname%
      \put(3998,4337){\makebox(0,0)[r]{\strut{}Power Law}}%
      \csname LTb\endcsname%
      \put(3998,4088){\makebox(0,0)[r]{\strut{}Discrete Weibull (Type-II)}}%
    }%
    \gplgaddtomacro\gplbacktext{%
      \colorrgb{0.50,0.50,0.50}%
      \put(1324,942){\makebox(0,0){\fns 10\textsuperscript{-11}\ \ \ }}%
      \colorrgb{0.50,0.50,0.50}%
      \put(1324,1086){\makebox(0,0){\fns 10\textsuperscript{-10}\ \ \ }}%
      \colorrgb{0.50,0.50,0.50}%
      \put(1324,1231){\makebox(0,0){\fns 10\textsuperscript{-9}\ \ \ }}%
      \colorrgb{0.50,0.50,0.50}%
      \put(1324,1375){\makebox(0,0){\fns 10\textsuperscript{-8}\ \ \ }}%
      \colorrgb{0.50,0.50,0.50}%
      \put(1324,1519){\makebox(0,0){\fns 10\textsuperscript{-7}\ \ \ }}%
      \colorrgb{0.50,0.50,0.50}%
      \put(1324,1663){\makebox(0,0){\fns 10\textsuperscript{-6}\ \ \ }}%
      \colorrgb{0.50,0.50,0.50}%
      \put(1324,1808){\makebox(0,0){\fns 10\textsuperscript{-5}\ \ \ }}%
      \colorrgb{0.50,0.50,0.50}%
      \put(1324,1952){\makebox(0,0){\fns 10\textsuperscript{-4}\ \ \ }}%
      \colorrgb{0.50,0.50,0.50}%
      \put(1324,2096){\makebox(0,0){\fns 10\textsuperscript{-3}\ \ \ }}%
      \colorrgb{0.50,0.50,0.50}%
      \put(1324,2240){\makebox(0,0){\fns 10\textsuperscript{-2}\ \ \ }}%
      \colorrgb{0.50,0.50,0.50}%
      \put(1324,2385){\makebox(0,0){\fns 10\textsuperscript{-1}\ \ \ }}%
      \colorrgb{0.50,0.50,0.50}%
      \put(1324,2529){\makebox(0,0){\fns 10\textsuperscript{0}\ \ \ }}%
      \colorrgb{0.50,0.50,0.50}%
      \put(1413,764){\makebox(0,0){\fns 10\textsuperscript{0}\ \ \ }}%
      \colorrgb{0.50,0.50,0.50}%
      \put(1678,764){\makebox(0,0){\fns 10\textsuperscript{1}\ \ \ }}%
      \colorrgb{0.50,0.50,0.50}%
      \put(1942,764){\makebox(0,0){\fns 10\textsuperscript{2}\ \ \ }}%
      \colorrgb{0.50,0.50,0.50}%
      \put(2207,764){\makebox(0,0){\fns 10\textsuperscript{3}\ \ \ }}%
      \colorrgb{0.50,0.50,0.50}%
      \put(2471,764){\makebox(0,0){\fns 10\textsuperscript{4}\ \ \ }}%
      \colorrgb{0.50,0.50,0.50}%
      \put(2736,764){\makebox(0,0){\fns 10\textsuperscript{5}\ \ \ }}%
      \colorrgb{0.50,0.50,0.50}%
      \put(3000,764){\makebox(0,0){\fns 10\textsuperscript{6}\ \ \ }}%
    }%
    \gplgaddtomacro\gplfronttext{%
    }%
    \gplbacktext
    \put(0,0){\includegraphics{lifetime-tweets}}%
    \gplfronttext
  \end{picture}%
\endgroup

%% file: fig/lifetime-momentum.pdf_tex
% GNUPLOT: LaTeX picture with Postscript
\begingroup
  \makeatletter
  \providecommand\color[2][]{%
    \GenericError{(gnuplot) \space\space\space\@spaces}{%
      Package color not loaded in conjunction with
      terminal option `colourtext'%
    }{See the gnuplot documentation for explanation.%
    }{Either use 'blacktext' in gnuplot or load the package
      color.sty in LaTeX.}%
    \renewcommand\color[2][]{}%
  }%
  \providecommand\includegraphics[2][]{%
    \GenericError{(gnuplot) \space\space\space\@spaces}{%
      Package graphicx or graphics not loaded%
    }{See the gnuplot documentation for explanation.%
    }{The gnuplot epslatex terminal needs graphicx.sty or graphics.sty.}%
    \renewcommand\includegraphics[2][]{}%
  }%
  \providecommand\rotatebox[2]{#2}%
  \@ifundefined{ifGPcolor}{%
    \newif\ifGPcolor
    \GPcolortrue
  }{}%
  \@ifundefined{ifGPblacktext}{%
    \newif\ifGPblacktext
    \GPblacktextfalse
  }{}%
  % define a \g@addto@macro without @ in the name:
  \let\gplgaddtomacro\g@addto@macro
  % define empty templates for all commands taking text:
  \gdef\gplbacktext{}%
  \gdef\gplfronttext{}%
  \makeatother
  \ifGPblacktext
    % no textcolor at all
    \def\colorrgb#1{}%
    \def\colorgray#1{}%
  \else
    % gray or color?
    \ifGPcolor
      \def\colorrgb#1{\color[rgb]{#1}}%
      \def\colorgray#1{\color[gray]{#1}}%
      \expandafter\def\csname LTw\endcsname{\color{white}}%
      \expandafter\def\csname LTb\endcsname{\color{black}}%
      \expandafter\def\csname LTa\endcsname{\color{black}}%
      \expandafter\def\csname LT0\endcsname{\color[rgb]{1,0,0}}%
      \expandafter\def\csname LT1\endcsname{\color[rgb]{0,1,0}}%
      \expandafter\def\csname LT2\endcsname{\color[rgb]{0,0,1}}%
      \expandafter\def\csname LT3\endcsname{\color[rgb]{1,0,1}}%
      \expandafter\def\csname LT4\endcsname{\color[rgb]{0,1,1}}%
      \expandafter\def\csname LT5\endcsname{\color[rgb]{1,1,0}}%
      \expandafter\def\csname LT6\endcsname{\color[rgb]{0,0,0}}%
      \expandafter\def\csname LT7\endcsname{\color[rgb]{1,0.3,0}}%
      \expandafter\def\csname LT8\endcsname{\color[rgb]{0.5,0.5,0.5}}%
    \else
      % gray
      \def\colorrgb#1{\color{black}}%
      \def\colorgray#1{\color[gray]{#1}}%
      \expandafter\def\csname LTw\endcsname{\color{white}}%
      \expandafter\def\csname LTb\endcsname{\color{black}}%
      \expandafter\def\csname LTa\endcsname{\color{black}}%
      \expandafter\def\csname LT0\endcsname{\color{black}}%
      \expandafter\def\csname LT1\endcsname{\color{black}}%
      \expandafter\def\csname LT2\endcsname{\color{black}}%
      \expandafter\def\csname LT3\endcsname{\color{black}}%
      \expandafter\def\csname LT4\endcsname{\color{black}}%
      \expandafter\def\csname LT5\endcsname{\color{black}}%
      \expandafter\def\csname LT6\endcsname{\color{black}}%
      \expandafter\def\csname LT7\endcsname{\color{black}}%
      \expandafter\def\csname LT8\endcsname{\color{black}}%
    \fi
  \fi
    \setlength{\unitlength}{0.0500bp}%
    \ifx\gptboxheight\undefined%
      \newlength{\gptboxheight}%
      \newlength{\gptboxwidth}%
      \newsavebox{\gptboxtext}%
    \fi%
    \setlength{\fboxrule}{0.5pt}%
    \setlength{\fboxsep}{1pt}%
\begin{picture}(4960.00,3060.00)%
    \gplgaddtomacro\gplbacktext{%
      \colorrgb{0.50,0.50,0.50}%
      \put(756,569){\makebox(0,0)[r]{\strut{}10\textsuperscript{-6}}}%
      \colorrgb{0.50,0.50,0.50}%
      \put(756,954){\makebox(0,0)[r]{\strut{}10\textsuperscript{-5}}}%
      \colorrgb{0.50,0.50,0.50}%
      \put(756,1340){\makebox(0,0)[r]{\strut{}10\textsuperscript{-4}}}%
      \colorrgb{0.50,0.50,0.50}%
      \put(756,1725){\makebox(0,0)[r]{\strut{}10\textsuperscript{-3}}}%
      \colorrgb{0.50,0.50,0.50}%
      \put(756,2110){\makebox(0,0)[r]{\strut{}10\textsuperscript{-2}}}%
      \colorrgb{0.50,0.50,0.50}%
      \put(756,2496){\makebox(0,0)[r]{\strut{}10\textsuperscript{-1}}}%
      \colorrgb{0.50,0.50,0.50}%
      \put(756,2881){\makebox(0,0)[r]{\strut{}10\textsuperscript{0}}}%
      \colorrgb{0.50,0.50,0.50}%
      \put(845,391){\makebox(0,0){\strut{}10\textsuperscript{0}}}%
      \colorrgb{0.50,0.50,0.50}%
      \put(1486,391){\makebox(0,0){\strut{}10\textsuperscript{1}}}%
      \colorrgb{0.50,0.50,0.50}%
      \put(2127,391){\makebox(0,0){\strut{}10\textsuperscript{2}}}%
      \colorrgb{0.50,0.50,0.50}%
      \put(2769,391){\makebox(0,0){\strut{}10\textsuperscript{3}}}%
      \colorrgb{0.50,0.50,0.50}%
      \put(3410,391){\makebox(0,0){\strut{}10\textsuperscript{4}}}%
      \colorrgb{0.50,0.50,0.50}%
      \put(4051,391){\makebox(0,0){\strut{}10\textsuperscript{5}}}%
      \colorrgb{0.50,0.50,0.50}%
      \put(4692,391){\makebox(0,0){\strut{}10\textsuperscript{6}}}%
    }%
    \gplgaddtomacro\gplfronttext{%
      \csname LTb\endcsname%
      \put(133,1725){\rotatebox{-270}{\makebox(0,0){\strut{}$\Pr\left(X=x \middle\vert X\geq x\right)$}}}%
      \csname LTb\endcsname%
      \put(2768,124){\makebox(0,0){\strut{}\# of Lifetime Tweets}}%
      \csname LTb\endcsname%
      \put(3998,2685){\makebox(0,0)[r]{\strut{}Data}}%
      \csname LTb\endcsname%
      \put(3998,2436){\makebox(0,0)[r]{\strut{}$\frac{\alpha-1}{x^{(1-\beta)}}$}}%
    }%
    \gplbacktext
    \put(0,0){\includegraphics{lifetime-momentum}}%
    \gplfronttext
  \end{picture}%
\endgroup

%% file: tweet-rates.tex
\section{Distribution of Tweet Rates}
\label{sec:tweet-rates}

The distribution of tweet rates is arguably the most important
statistic for microblogging system design.  An architecture designed
for uniform messaging rates across the network will struggle with a
heavy-tailed rate distribution. In this section, we describe an
analytical model and generative mechanism for the rate distribution
and reject a model previously proposed for telephone call rates.
Although we are most interested in the tweet rate distribution, we
model the easier-to-consider tweet count distribution. The former is
easily recovered by dividing out the 4-month sampling duration.

\subsection{An Analytical Approximation of the Tweet Rate
  Distribution}

\begin{figure}
  \centering
  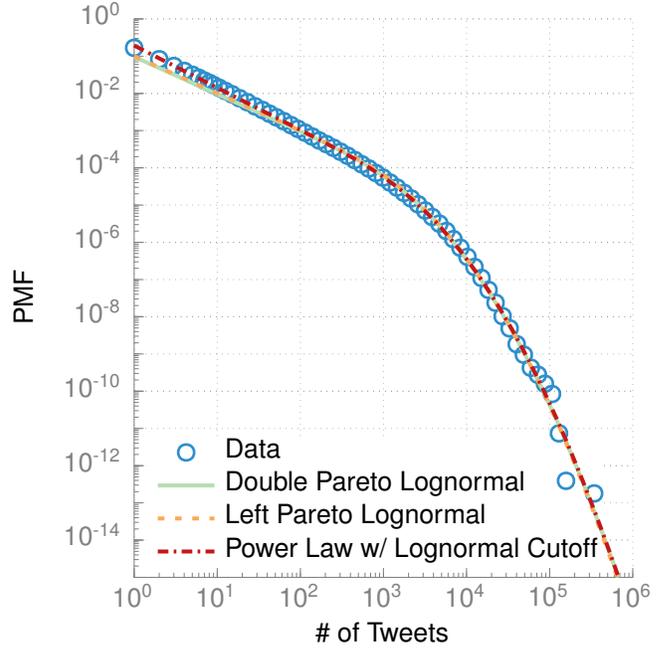
  \caption{Distribution of tweets per user for the four month period
    from June through September 2012.  }
  \label{fig:tweet-distribution-4month}
\end{figure}

\Note{Introduction of the data and graph.}

\autoref{fig:tweet-distribution-4month} plots the
logarithmically-binned empirical tweet distribution.  It is
heavy-tailed, consistent with other forms of
authorship~\cite{lotka26}. The tails form two different regimes
meeting at $X=\texttilde2000$, each heavy-tailed but with different
exponents.  We show in \autoref{sec:tweet-rates-generative} that this
phase change is a dynamic effect related to the sample period length
(i.e., four months)---the crossing point increases with the square of
the sample period length.

\Note{Discussion of best fit form.  ERFC is lognormal for x>>1.}

Simulating microblogging performance and comparing rates across
communication systems benefits from a closed-form of the
distribution. The forthcoming generative model in
\autoref{sec:tweet-rates-generative} is not analytically tractable, so
we describe an analytical approximation first.
\autoref{fig:tweet-distribution-4month} suggests a cutoff power law,
but the upper tail is heavier than the common exponential
cutoff~\cite{clauset09}.  Instead, the cutoff appears lognormal,
suggesting the following density function\footnote{We use a continuous
  model for simplicity. The integral data can be viewed as a rounded
  version of the product of the true tweet rate and sampling period.},
\begin{equation}
  \label{eqn:tweet-rates-distribution}
  p(x) = cx^{-\beta}\Phi^c\left(\frac{\ln x - \mu}{\sigma}\right),
\end{equation}
where $\Phi^c$ is the complementary CDF of the standard normal
distribution and $c$ is a normalizing constant. The maximum likelihood
fit is shown in \autoref{fig:tweet-distribution-4month}, with
$\beta=1.13,$ $\mu=7.6,$ $\sigma=1.06,$ and $c=0.19.$ The lognormal
cutoff shape is seen by noting that
\begin{equation*}
  \Phi^c(z) \propto \erfc\left(\frac{z}{\sqrt{2}}\right) \text{ and } \erfc(z) \approx \frac{1}{\sqrt{\pi}}\frac{e^{-z^2}}{z} \text{ for } z \gg 1,
\end{equation*}
leading to the approximately lognormal form
\begin{equation}
  \Phi^c\left(\frac{\ln x - \mu}{\sigma}\right) \appropto \frac{\sigma}{\ln x - \mu}e^{-\frac{(\ln x - \mu)^2}{2\sigma^2}} \text{ for } \frac{\ln x - \mu}{\sigma} \gg 1.
\end{equation}
The power law exponent in the lower tail is $\beta$, the phase change
to the cutoff regime occurs at $\exp(\mu)$, and the upper tail
steepness is controlled by $\sigma$.

\subsection{The Distribution is Not Double Pareto--Lognormal}
\label{sec:tweet-rates-dpln}

At first glance, \autoref{fig:tweet-distribution-4month} appears to be
Double Pareto-Lognormal (DPLN), a recently-discovered distribution
that has found wide-spread popularity across many fields, perhaps due
to its clear generative interpretation~\cite{reed04apr}.  Seshadri
\etal suggested its use for communication rates, specifically call
rates in a cellular network, interpreting the generative process as
evolving \emph{social wealth}~\cite{seshadri08aug}.  However, in this
section we show that the DPLN does not correctly capture the lower
tail behavior of tweet rates (or call rates). In the next section, we
describe a different mechanism to explain the shape.

We first summarize the origin of the DPLN distribution~\cite{reed04apr}.
Given a stochastic process $X$ evolving via Geometric Brownian motion
(GBM)
\begin{equation}
\dif X = \mu X + \sigma X \dif W,
\end{equation}
where $W$ is the Wiener process, with lognormally distributed initial
state, $\log X_0 \sim \mathcal{N}(\nu, \tau^2),$ then $X_t$ is also
lognormally distributed, $\log X_t \sim \mathcal{N}(\nu + \frac{\mu -
  \sigma^2}{2}t, \tau^2+\sigma^2t).$ If the observation (or killing)
time $t \triangleq T$ is exponentially distributed, $T \sim
\mathrm{Exp}(\lambda),$ then the observed (or final) state has DPLN
distribution, $X_T \sim \mathrm{DPLN}(\alpha, \beta, \nu, \tau),$
where $\alpha > 0$ and $-\beta < 0$ are the roots of the
characteristic equation
\begin{equation}
  \frac{\sigma^2}{2}z^2 + \left(\mu - \frac{\sigma^2}{2}\right)z - \lambda = 0.
\end{equation}
Seshadri \etal~\cite{seshadri08aug} proposed that the number of calls
made by an individual reflects an underlying \emph{social wealth} that
evolves via such a GBM.  For an exponentially growing population, the
age distribution of the sampled users is exponential and the resulting
distribution of calls (or social wealth) will be DPLN.  This model
seems qualitatively reasonable for Twitter as well, but cannot capture
the correct power law exponent in the lower tail (see
\autoref{fig:tweet-distribution-4month}). The call distribution data
exhibits a similar mismatch, challenging the model's suitability there
as well.

The density function of $\mathrm{DPLN}(\alpha, \beta, \nu, \tau)$ is
\begin{equation}
  f(x) = \frac{\beta}{\alpha + \beta}f_1(x) + \frac{\alpha}{\alpha + \beta}f_2(x),
\end{equation}
where
\begin{equation}
  f_1(x) = \alpha x^{-\alpha - 1}A(\alpha,\nu,\tau)\Phi\left(\frac{\ln x - \nu - \alpha\tau^2}{\tau}\right),
\end{equation}
\begin{equation}
  f_2(x) = \beta x^{\beta - 1}A(-\beta,\nu,\tau)\Phi^c\left(\frac{\ln x - \nu + \beta\tau^2}{\tau}\right),
\end{equation}
\begin{equation}
  A(\theta,\nu,\tau) = \exp\left(\frac{\theta \nu + \theta^2\tau^2}{2}\right),
\end{equation}
and $\Phi$ and $\Phi^c$ are the CDF and complementary CDF of the
standard normal distribution.  $f_1$ and $f_2$ are the limiting
densities as $\alpha \to \infty$ and $\beta \to \infty$, respectively,
and are called the \emph{right Pareto lognormal} and \emph{left Pareto
  lognormal} distributions.

Two observations stand out.  First, the distribution is Pareto in both
tails, with minimum slope of $-1$ in the lower. Second, the left
Pareto lognormal form is nearly equivalent to our expression
\autoref{eqn:tweet-rates-distribution}, which differs only by
accommodating lower tail exponents below
$-1$. \autoref{fig:tweet-distribution-4month} shows maximum likelihood
fits of both the DPLN and left Pareto lognormal distributions.  The
lower tail is steeper than allowed by the DPLN ($-1.13 < -1$) and fits
poorly. The call distribution data shows a similar mismatch. Although
the DPLN has widely applicable, it does not model these communication
patterns.  Our model from the following section should better fit the
call data~\cite{seshadri08aug} as well.

In the upper tail, both distributions fit equally well (i.e., a
likelihood ratio test does not favor either fit). The data are
insufficient to distinguish a lognormal from a power law upper tail, a
common issue~\cite{clauset09}. We favor the lognormal form for
\autoref{eqn:tweet-rates-distribution} because it is simpler (i.e.,
has fewer parameters) and most real world ``power laws'' exhibit some
cutoff~\cite{clauset09}.

\subsection{An Urn Process Generating the Tweet Rate Distribution}
\label{sec:tweet-rates-generative}

\begin{figure}
  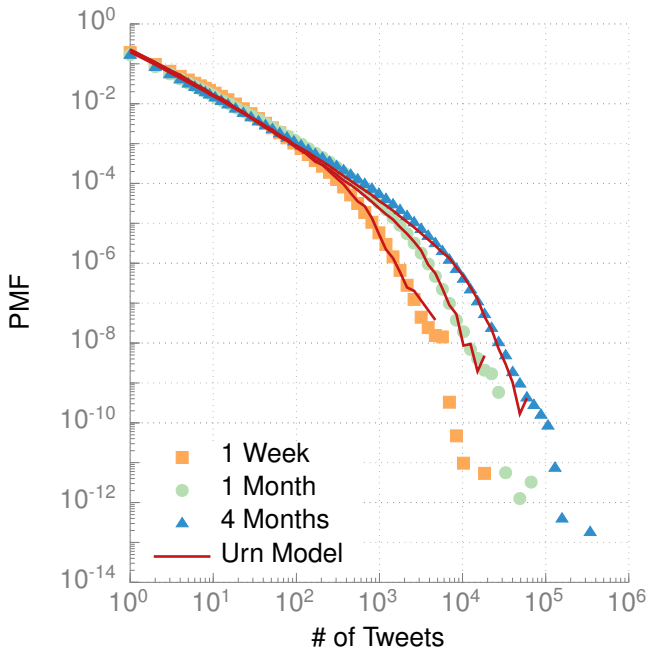
  \caption{Distribution of tweet counts over various sample periods,
    showing the time-dependent cutoff. The asymptotic distribution is
    Pareto.  Traces for the urn model describing this effect were
    obtained by simulation.}
  \label{fig:tweet-distribution-scales}
\end{figure}

In this section we develop an urn process to describe tweet
distribution in \autoref{fig:tweet-distribution-4month}.  The phase
change is a dynamic effect governed by the sampling period. As the
period increases, the distribution approaches that of the lower
tail---approximately Pareto with exponent $-1.13$.  In practical
terms, high-rate tweeters are much rarer in a finite sample than the
asymptotic distribution would predict.

\autoref{fig:tweet-distribution-scales} shows the distribution for two
sample periods, illustrating the dynamic phase change.  The lower tail
extends further with the longer period.  Degree distributions in
growing networks evolve similarly.  Although simple preferential
attachment of new nodes leads to a straight power
law~\cite{barabasi99oct}, when existing nodes also generate new edges
via preferential attachment, the distribution is double Pareto (with
exponents -2 and -3) with a time-dependent crossing point ($k_c =
[b^2t(2 + \alpha
t)]^{\sfrac{1}{2}}$)~\cite{barabasi02aug}.\footnote{In a network that
  allows self-edges, the exponents are -$\sfrac{3}{2}$ and -3 with
  crossing time $k_c \approx
  \sqrt{ct}(2+ct)^{\sfrac{3}{2}}$~\cite{dorogovtsev00oct}.} A similar
model describes the tweet distribution.

Consider the evolution of the sample of tweets. Users join the sample
upon their first tweet (during the sample period) and then continue to
produce additional tweets at some rate.  Discretize time relative to
new users joining the sample, i.e., one user joins at each time step
so there are $t$ users at time $t$. Let $k(s,t)$ be the (expected)
tweet count at time $t$ for the user first observed at time
$s$. Assume new tweets are generated at a constant average rate $c$,
i.e., $ct$ new tweets appear at each time step, distributed among
existing users with frequency proportional to $A + k(s,t)^\alpha$. $A$
is some initial attractiveness and $\alpha$ is the non-linearity of
the preference~\cite{dorogovtsev02jun}.  The resulting continuum
equation\footnote{We use the notation and continuous approximation of
  Dorogovtsev and Mendes~\cite{dorogovtsev01dec}.} is
\begin{equation}
  \pd{k(s,t)}{t} = (1 +ct)\frac{A + k(s,t)^\alpha}{\int_0^t \! A + k(u,t)^\alpha \, \dif u }
\end{equation}

An analytical solution exists when $A=0$ and
$\alpha=1$~\cite{dorogovtsev01dec}, but for the general case we resort
to Monte Carlo simulations. \autoref{fig:tweet-distribution-scales}
shows the close match to the empirical density when $A = 1$ and
$\alpha = 0.88$.\footnote{Parameters were chosen by a coarse, manual
  exploration of the space.  Fine-tuning might further improve the
  fit.} Assuming the power law form of the asymptotic density, $p(k)
\propto k^{-\beta},$ the power law form of the rate distribution can
be recovered. Taking $\lambda$ as the tweet rate and noting that
$\lambda \propto k^{-\alpha}$ when $k \gg A$, then
\begin{equation}
  p(\lambda) = p(k^{-\alpha}) \propto \frac{1}{\alpha}\lambda^{-\frac{-1+\alpha+\beta}{\alpha}}.
\end{equation}
Thus, for $\alpha$ close to 1, the power law exponent recovered from
\autoref{fig:tweet-distribution-4month} slightly overestimates that of
the tweet rate.

Relating back to the analytical approximation of
\autoref{eqn:tweet-rates-distribution}, $\mu$ is related to $ct$
by $\mu \approx 1.32\ln(ct) + 0.56$.  $\beta = 1.13$ and $\sigma =
1.06$ are constants best determined by fitting.

\subsection{Distributions of Retweeter and Retweetee Rates}

\Note{Discuss the graph for retweet and retweeted as well.}

\begin{figure}
  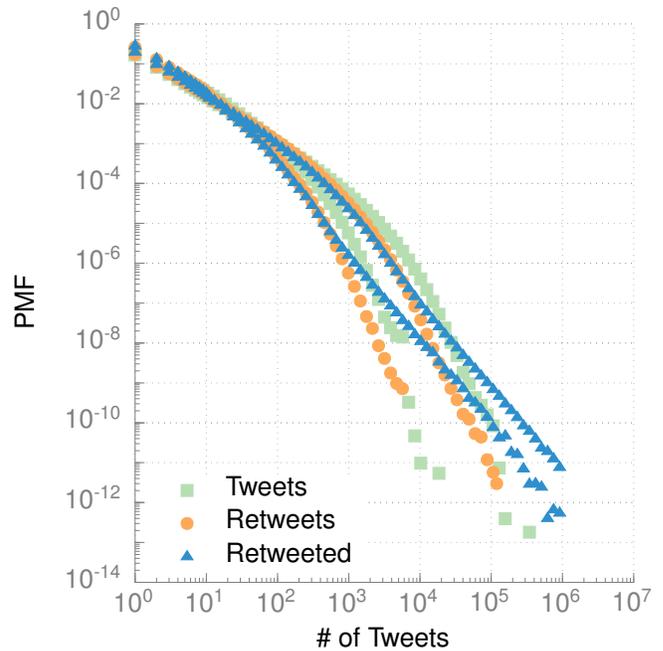
  \caption{Distributions for tweets sent, retweets sent, and times
    retweeted for the 1 week and 4 month samples. All categories show
    similar time-dependent phase changes, suggesting the same
    underlying mechanism.  Retweets differ from tweets only in a lower
    average rate (parameter $c$ in the urn model).}
  \label{fig:all-distribution-scales}
\end{figure}

The retweet and retweetee rates show a similar dynamic behavior in
\autoref{fig:all-distribution-scales}. The retweet behavior differs
only in the average rate $c$, which is about $2\times$ lower.  The
retweetee distribution exhibits two interesting differences.  First,
it extends further to the right, indicating that retweets of popular
users outnumber tweets from extensive users. Second, the slopes of the
power law regimes are more consistent with a pure preferential
attachment process (i.e., $\alpha = 1$). The retweetee rate comes
directly from a preferential attachment process---initial retweets
increase exposure, begetting additional retweets---and thus should
match the linear form seen in other systems. The power law form of the
tweet and retweet rates describes the underlying propensity to tweet,
but without the same generative interpretation.

%% file: fig/tweet-distribution-4month.pdf_tex
% GNUPLOT: LaTeX picture with Postscript
\begingroup
  \makeatletter
  \providecommand\color[2][]{%
    \GenericError{(gnuplot) \space\space\space\@spaces}{%
      Package color not loaded in conjunction with
      terminal option `colourtext'%
    }{See the gnuplot documentation for explanation.%
    }{Either use 'blacktext' in gnuplot or load the package
      color.sty in LaTeX.}%
    \renewcommand\color[2][]{}%
  }%
  \providecommand\includegraphics[2][]{%
    \GenericError{(gnuplot) \space\space\space\@spaces}{%
      Package graphicx or graphics not loaded%
    }{See the gnuplot documentation for explanation.%
    }{The gnuplot epslatex terminal needs graphicx.sty or graphics.sty.}%
    \renewcommand\includegraphics[2][]{}%
  }%
  \providecommand\rotatebox[2]{#2}%
  \@ifundefined{ifGPcolor}{%
    \newif\ifGPcolor
    \GPcolortrue
  }{}%
  \@ifundefined{ifGPblacktext}{%
    \newif\ifGPblacktext
    \GPblacktextfalse
  }{}%
  % define a \g@addto@macro without @ in the name:
  \let\gplgaddtomacro\g@addto@macro
  % define empty templates for all commands taking text:
  \gdef\gplbacktext{}%
  \gdef\gplfronttext{}%
  \makeatother
  \ifGPblacktext
    % no textcolor at all
    \def\colorrgb#1{}%
    \def\colorgray#1{}%
  \else
    % gray or color?
    \ifGPcolor
      \def\colorrgb#1{\color[rgb]{#1}}%
      \def\colorgray#1{\color[gray]{#1}}%
      \expandafter\def\csname LTw\endcsname{\color{white}}%
      \expandafter\def\csname LTb\endcsname{\color{black}}%
      \expandafter\def\csname LTa\endcsname{\color{black}}%
      \expandafter\def\csname LT0\endcsname{\color[rgb]{1,0,0}}%
      \expandafter\def\csname LT1\endcsname{\color[rgb]{0,1,0}}%
      \expandafter\def\csname LT2\endcsname{\color[rgb]{0,0,1}}%
      \expandafter\def\csname LT3\endcsname{\color[rgb]{1,0,1}}%
      \expandafter\def\csname LT4\endcsname{\color[rgb]{0,1,1}}%
      \expandafter\def\csname LT5\endcsname{\color[rgb]{1,1,0}}%
      \expandafter\def\csname LT6\endcsname{\color[rgb]{0,0,0}}%
      \expandafter\def\csname LT7\endcsname{\color[rgb]{1,0.3,0}}%
      \expandafter\def\csname LT8\endcsname{\color[rgb]{0.5,0.5,0.5}}%
    \else
      % gray
      \def\colorrgb#1{\color{black}}%
      \def\colorgray#1{\color[gray]{#1}}%
      \expandafter\def\csname LTw\endcsname{\color{white}}%
      \expandafter\def\csname LTb\endcsname{\color{black}}%
      \expandafter\def\csname LTa\endcsname{\color{black}}%
      \expandafter\def\csname LT0\endcsname{\color{black}}%
      \expandafter\def\csname LT1\endcsname{\color{black}}%
      \expandafter\def\csname LT2\endcsname{\color{black}}%
      \expandafter\def\csname LT3\endcsname{\color{black}}%
      \expandafter\def\csname LT4\endcsname{\color{black}}%
      \expandafter\def\csname LT5\endcsname{\color{black}}%
      \expandafter\def\csname LT6\endcsname{\color{black}}%
      \expandafter\def\csname LT7\endcsname{\color{black}}%
      \expandafter\def\csname LT8\endcsname{\color{black}}%
    \fi
  \fi
    \setlength{\unitlength}{0.0500bp}%
    \ifx\gptboxheight\undefined%
      \newlength{\gptboxheight}%
      \newlength{\gptboxwidth}%
      \newsavebox{\gptboxtext}%
    \fi%
    \setlength{\fboxrule}{0.5pt}%
    \setlength{\fboxsep}{1pt}%
\begin{picture}(4960.00,4960.00)%
    \gplgaddtomacro\gplbacktext{%
      \colorrgb{0.50,0.50,0.50}%
      \put(845,850){\makebox(0,0)[r]{\strut{}10\textsuperscript{-14}}}%
      \colorrgb{0.50,0.50,0.50}%
      \put(845,1411){\makebox(0,0)[r]{\strut{}10\textsuperscript{-12}}}%
      \colorrgb{0.50,0.50,0.50}%
      \put(845,1973){\makebox(0,0)[r]{\strut{}10\textsuperscript{-10}}}%
      \colorrgb{0.50,0.50,0.50}%
      \put(845,2535){\makebox(0,0)[r]{\strut{}10\textsuperscript{-8}}}%
      \colorrgb{0.50,0.50,0.50}%
      \put(845,3096){\makebox(0,0)[r]{\strut{}10\textsuperscript{-6}}}%
      \colorrgb{0.50,0.50,0.50}%
      \put(845,3658){\makebox(0,0)[r]{\strut{}10\textsuperscript{-4}}}%
      \colorrgb{0.50,0.50,0.50}%
      \put(845,4219){\makebox(0,0)[r]{\strut{}10\textsuperscript{-2}}}%
      \colorrgb{0.50,0.50,0.50}%
      \put(845,4781){\makebox(0,0)[r]{\strut{}10\textsuperscript{0}}}%
      \colorrgb{0.50,0.50,0.50}%
      \put(845,569){\makebox(0,0)[r]{\strut{}}}%
      \colorrgb{0.50,0.50,0.50}%
      \put(845,1131){\makebox(0,0)[r]{\strut{}}}%
      \colorrgb{0.50,0.50,0.50}%
      \put(845,1692){\makebox(0,0)[r]{\strut{}}}%
      \colorrgb{0.50,0.50,0.50}%
      \put(845,2254){\makebox(0,0)[r]{\strut{}}}%
      \colorrgb{0.50,0.50,0.50}%
      \put(845,2815){\makebox(0,0)[r]{\strut{}}}%
      \colorrgb{0.50,0.50,0.50}%
      \put(845,3377){\makebox(0,0)[r]{\strut{}}}%
      \colorrgb{0.50,0.50,0.50}%
      \put(845,3939){\makebox(0,0)[r]{\strut{}}}%
      \colorrgb{0.50,0.50,0.50}%
      \put(845,4500){\makebox(0,0)[r]{\strut{}}}%
      \colorrgb{0.50,0.50,0.50}%
      \put(934,391){\makebox(0,0){\strut{}10\textsuperscript{0}}}%
      \colorrgb{0.50,0.50,0.50}%
      \put(1560,391){\makebox(0,0){\strut{}10\textsuperscript{1}}}%
      \colorrgb{0.50,0.50,0.50}%
      \put(2187,391){\makebox(0,0){\strut{}10\textsuperscript{2}}}%
      \colorrgb{0.50,0.50,0.50}%
      \put(2813,391){\makebox(0,0){\strut{}10\textsuperscript{3}}}%
      \colorrgb{0.50,0.50,0.50}%
      \put(3439,391){\makebox(0,0){\strut{}10\textsuperscript{4}}}%
      \colorrgb{0.50,0.50,0.50}%
      \put(4066,391){\makebox(0,0){\strut{}10\textsuperscript{5}}}%
      \colorrgb{0.50,0.50,0.50}%
      \put(4692,391){\makebox(0,0){\strut{}10\textsuperscript{6}}}%
    }%
    \gplgaddtomacro\gplfronttext{%
      \csname LTb\endcsname%
      \put(133,2675){\rotatebox{-270}{\makebox(0,0){\strut{}PMF}}}%
      \csname LTb\endcsname%
      \put(2813,124){\makebox(0,0){\strut{}\# of Tweets}}%
      \csname LTb\endcsname%
      \put(1628,1511){\makebox(0,0)[l]{\strut{}Data}}%
      \csname LTb\endcsname%
      \put(1628,1262){\makebox(0,0)[l]{\strut{}Double Pareto Lognormal}}%
      \csname LTb\endcsname%
      \put(1628,1013){\makebox(0,0)[l]{\strut{}Left Pareto Lognormal}}%
      \csname LTb\endcsname%
      \put(1628,764){\makebox(0,0)[l]{\strut{}Power Law w/ Lognormal Cutoff}}%
    }%
    \gplbacktext
    \put(0,0){\includegraphics{tweet-distribution-4month}}%
    \gplfronttext
  \end{picture}%
\endgroup

%% file: fig/tweet-distribution-scales.pdf_tex
% GNUPLOT: LaTeX picture with Postscript
\begingroup
  \makeatletter
  \providecommand\color[2][]{%
    \GenericError{(gnuplot) \space\space\space\@spaces}{%
      Package color not loaded in conjunction with
      terminal option `colourtext'%
    }{See the gnuplot documentation for explanation.%
    }{Either use 'blacktext' in gnuplot or load the package
      color.sty in LaTeX.}%
    \renewcommand\color[2][]{}%
  }%
  \providecommand\includegraphics[2][]{%
    \GenericError{(gnuplot) \space\space\space\@spaces}{%
      Package graphicx or graphics not loaded%
    }{See the gnuplot documentation for explanation.%
    }{The gnuplot epslatex terminal needs graphicx.sty or graphics.sty.}%
    \renewcommand\includegraphics[2][]{}%
  }%
  \providecommand\rotatebox[2]{#2}%
  \@ifundefined{ifGPcolor}{%
    \newif\ifGPcolor
    \GPcolortrue
  }{}%
  \@ifundefined{ifGPblacktext}{%
    \newif\ifGPblacktext
    \GPblacktextfalse
  }{}%
  % define a \g@addto@macro without @ in the name:
  \let\gplgaddtomacro\g@addto@macro
  % define empty templates for all commands taking text:
  \gdef\gplbacktext{}%
  \gdef\gplfronttext{}%
  \makeatother
  \ifGPblacktext
    % no textcolor at all
    \def\colorrgb#1{}%
    \def\colorgray#1{}%
  \else
    % gray or color?
    \ifGPcolor
      \def\colorrgb#1{\color[rgb]{#1}}%
      \def\colorgray#1{\color[gray]{#1}}%
      \expandafter\def\csname LTw\endcsname{\color{white}}%
      \expandafter\def\csname LTb\endcsname{\color{black}}%
      \expandafter\def\csname LTa\endcsname{\color{black}}%
      \expandafter\def\csname LT0\endcsname{\color[rgb]{1,0,0}}%
      \expandafter\def\csname LT1\endcsname{\color[rgb]{0,1,0}}%
      \expandafter\def\csname LT2\endcsname{\color[rgb]{0,0,1}}%
      \expandafter\def\csname LT3\endcsname{\color[rgb]{1,0,1}}%
      \expandafter\def\csname LT4\endcsname{\color[rgb]{0,1,1}}%
      \expandafter\def\csname LT5\endcsname{\color[rgb]{1,1,0}}%
      \expandafter\def\csname LT6\endcsname{\color[rgb]{0,0,0}}%
      \expandafter\def\csname LT7\endcsname{\color[rgb]{1,0.3,0}}%
      \expandafter\def\csname LT8\endcsname{\color[rgb]{0.5,0.5,0.5}}%
    \else
      % gray
      \def\colorrgb#1{\color{black}}%
      \def\colorgray#1{\color[gray]{#1}}%
      \expandafter\def\csname LTw\endcsname{\color{white}}%
      \expandafter\def\csname LTb\endcsname{\color{black}}%
      \expandafter\def\csname LTa\endcsname{\color{black}}%
      \expandafter\def\csname LT0\endcsname{\color{black}}%
      \expandafter\def\csname LT1\endcsname{\color{black}}%
      \expandafter\def\csname LT2\endcsname{\color{black}}%
      \expandafter\def\csname LT3\endcsname{\color{black}}%
      \expandafter\def\csname LT4\endcsname{\color{black}}%
      \expandafter\def\csname LT5\endcsname{\color{black}}%
      \expandafter\def\csname LT6\endcsname{\color{black}}%
      \expandafter\def\csname LT7\endcsname{\color{black}}%
      \expandafter\def\csname LT8\endcsname{\color{black}}%
    \fi
  \fi
    \setlength{\unitlength}{0.0500bp}%
    \ifx\gptboxheight\undefined%
      \newlength{\gptboxheight}%
      \newlength{\gptboxwidth}%
      \newsavebox{\gptboxtext}%
    \fi%
    \setlength{\fboxrule}{0.5pt}%
    \setlength{\fboxsep}{1pt}%
\begin{picture}(4960.00,4960.00)%
    \gplgaddtomacro\gplbacktext{%
      \colorrgb{0.50,0.50,0.50}%
      \put(845,569){\makebox(0,0)[r]{\strut{}10\textsuperscript{-14}}}%
      \colorrgb{0.50,0.50,0.50}%
      \put(845,1171){\makebox(0,0)[r]{\strut{}10\textsuperscript{-12}}}%
      \colorrgb{0.50,0.50,0.50}%
      \put(845,1772){\makebox(0,0)[r]{\strut{}10\textsuperscript{-10}}}%
      \colorrgb{0.50,0.50,0.50}%
      \put(845,2374){\makebox(0,0)[r]{\strut{}10\textsuperscript{-8}}}%
      \colorrgb{0.50,0.50,0.50}%
      \put(845,2976){\makebox(0,0)[r]{\strut{}10\textsuperscript{-6}}}%
      \colorrgb{0.50,0.50,0.50}%
      \put(845,3578){\makebox(0,0)[r]{\strut{}10\textsuperscript{-4}}}%
      \colorrgb{0.50,0.50,0.50}%
      \put(845,4179){\makebox(0,0)[r]{\strut{}10\textsuperscript{-2}}}%
      \colorrgb{0.50,0.50,0.50}%
      \put(845,4781){\makebox(0,0)[r]{\strut{}10\textsuperscript{0}}}%
      \colorrgb{0.50,0.50,0.50}%
      \put(845,870){\makebox(0,0)[r]{\strut{}}}%
      \colorrgb{0.50,0.50,0.50}%
      \put(845,1472){\makebox(0,0)[r]{\strut{}}}%
      \colorrgb{0.50,0.50,0.50}%
      \put(845,2073){\makebox(0,0)[r]{\strut{}}}%
      \colorrgb{0.50,0.50,0.50}%
      \put(845,2675){\makebox(0,0)[r]{\strut{}}}%
      \colorrgb{0.50,0.50,0.50}%
      \put(845,3277){\makebox(0,0)[r]{\strut{}}}%
      \colorrgb{0.50,0.50,0.50}%
      \put(845,3878){\makebox(0,0)[r]{\strut{}}}%
      \colorrgb{0.50,0.50,0.50}%
      \put(845,4480){\makebox(0,0)[r]{\strut{}}}%
      \colorrgb{0.50,0.50,0.50}%
      \put(934,391){\makebox(0,0){\strut{}10\textsuperscript{0}}}%
      \colorrgb{0.50,0.50,0.50}%
      \put(1560,391){\makebox(0,0){\strut{}10\textsuperscript{1}}}%
      \colorrgb{0.50,0.50,0.50}%
      \put(2187,391){\makebox(0,0){\strut{}10\textsuperscript{2}}}%
      \colorrgb{0.50,0.50,0.50}%
      \put(2813,391){\makebox(0,0){\strut{}10\textsuperscript{3}}}%
      \colorrgb{0.50,0.50,0.50}%
      \put(3439,391){\makebox(0,0){\strut{}10\textsuperscript{4}}}%
      \colorrgb{0.50,0.50,0.50}%
      \put(4066,391){\makebox(0,0){\strut{}10\textsuperscript{5}}}%
      \colorrgb{0.50,0.50,0.50}%
      \put(4692,391){\makebox(0,0){\strut{}10\textsuperscript{6}}}%
    }%
    \gplgaddtomacro\gplfronttext{%
      \csname LTb\endcsname%
      \put(133,2675){\rotatebox{-270}{\makebox(0,0){\strut{}PMF}}}%
      \csname LTb\endcsname%
      \put(2813,124){\makebox(0,0){\strut{}\# of Tweets}}%
      \csname LTb\endcsname%
      \put(1628,1511){\makebox(0,0)[l]{\strut{}1 Week}}%
      \csname LTb\endcsname%
      \put(1628,1262){\makebox(0,0)[l]{\strut{}1 Month}}%
      \csname LTb\endcsname%
      \put(1628,1013){\makebox(0,0)[l]{\strut{}4 Months}}%
      \csname LTb\endcsname%
      \put(1628,764){\makebox(0,0)[l]{\strut{}Urn Model}}%
    }%
    \gplbacktext
    \put(0,0){\includegraphics{tweet-distribution-scales}}%
    \gplfronttext
  \end{picture}%
\endgroup

%% file: fig/all-distribution-scales.pdf_tex
% GNUPLOT: LaTeX picture with Postscript
\begingroup
  \makeatletter
  \providecommand\color[2][]{%
    \GenericError{(gnuplot) \space\space\space\@spaces}{%
      Package color not loaded in conjunction with
      terminal option `colourtext'%
    }{See the gnuplot documentation for explanation.%
    }{Either use 'blacktext' in gnuplot or load the package
      color.sty in LaTeX.}%
    \renewcommand\color[2][]{}%
  }%
  \providecommand\includegraphics[2][]{%
    \GenericError{(gnuplot) \space\space\space\@spaces}{%
      Package graphicx or graphics not loaded%
    }{See the gnuplot documentation for explanation.%
    }{The gnuplot epslatex terminal needs graphicx.sty or graphics.sty.}%
    \renewcommand\includegraphics[2][]{}%
  }%
  \providecommand\rotatebox[2]{#2}%
  \@ifundefined{ifGPcolor}{%
    \newif\ifGPcolor
    \GPcolortrue
  }{}%
  \@ifundefined{ifGPblacktext}{%
    \newif\ifGPblacktext
    \GPblacktextfalse
  }{}%
  % define a \g@addto@macro without @ in the name:
  \let\gplgaddtomacro\g@addto@macro
  % define empty templates for all commands taking text:
  \gdef\gplbacktext{}%
  \gdef\gplfronttext{}%
  \makeatother
  \ifGPblacktext
    % no textcolor at all
    \def\colorrgb#1{}%
    \def\colorgray#1{}%
  \else
    % gray or color?
    \ifGPcolor
      \def\colorrgb#1{\color[rgb]{#1}}%
      \def\colorgray#1{\color[gray]{#1}}%
      \expandafter\def\csname LTw\endcsname{\color{white}}%
      \expandafter\def\csname LTb\endcsname{\color{black}}%
      \expandafter\def\csname LTa\endcsname{\color{black}}%
      \expandafter\def\csname LT0\endcsname{\color[rgb]{1,0,0}}%
      \expandafter\def\csname LT1\endcsname{\color[rgb]{0,1,0}}%
      \expandafter\def\csname LT2\endcsname{\color[rgb]{0,0,1}}%
      \expandafter\def\csname LT3\endcsname{\color[rgb]{1,0,1}}%
      \expandafter\def\csname LT4\endcsname{\color[rgb]{0,1,1}}%
      \expandafter\def\csname LT5\endcsname{\color[rgb]{1,1,0}}%
      \expandafter\def\csname LT6\endcsname{\color[rgb]{0,0,0}}%
      \expandafter\def\csname LT7\endcsname{\color[rgb]{1,0.3,0}}%
      \expandafter\def\csname LT8\endcsname{\color[rgb]{0.5,0.5,0.5}}%
    \else
      % gray
      \def\colorrgb#1{\color{black}}%
      \def\colorgray#1{\color[gray]{#1}}%
      \expandafter\def\csname LTw\endcsname{\color{white}}%
      \expandafter\def\csname LTb\endcsname{\color{black}}%
      \expandafter\def\csname LTa\endcsname{\color{black}}%
      \expandafter\def\csname LT0\endcsname{\color{black}}%
      \expandafter\def\csname LT1\endcsname{\color{black}}%
      \expandafter\def\csname LT2\endcsname{\color{black}}%
      \expandafter\def\csname LT3\endcsname{\color{black}}%
      \expandafter\def\csname LT4\endcsname{\color{black}}%
      \expandafter\def\csname LT5\endcsname{\color{black}}%
      \expandafter\def\csname LT6\endcsname{\color{black}}%
      \expandafter\def\csname LT7\endcsname{\color{black}}%
      \expandafter\def\csname LT8\endcsname{\color{black}}%
    \fi
  \fi
    \setlength{\unitlength}{0.0500bp}%
    \ifx\gptboxheight\undefined%
      \newlength{\gptboxheight}%
      \newlength{\gptboxwidth}%
      \newsavebox{\gptboxtext}%
    \fi%
    \setlength{\fboxrule}{0.5pt}%
    \setlength{\fboxsep}{1pt}%
\begin{picture}(4960.00,4960.00)%
    \gplgaddtomacro\gplbacktext{%
      \colorrgb{0.50,0.50,0.50}%
      \put(845,569){\makebox(0,0)[r]{\strut{}10\textsuperscript{-14}}}%
      \colorrgb{0.50,0.50,0.50}%
      \put(845,1171){\makebox(0,0)[r]{\strut{}10\textsuperscript{-12}}}%
      \colorrgb{0.50,0.50,0.50}%
      \put(845,1772){\makebox(0,0)[r]{\strut{}10\textsuperscript{-10}}}%
      \colorrgb{0.50,0.50,0.50}%
      \put(845,2374){\makebox(0,0)[r]{\strut{}10\textsuperscript{-8}}}%
      \colorrgb{0.50,0.50,0.50}%
      \put(845,2976){\makebox(0,0)[r]{\strut{}10\textsuperscript{-6}}}%
      \colorrgb{0.50,0.50,0.50}%
      \put(845,3578){\makebox(0,0)[r]{\strut{}10\textsuperscript{-4}}}%
      \colorrgb{0.50,0.50,0.50}%
      \put(845,4179){\makebox(0,0)[r]{\strut{}10\textsuperscript{-2}}}%
      \colorrgb{0.50,0.50,0.50}%
      \put(845,4781){\makebox(0,0)[r]{\strut{}10\textsuperscript{0}}}%
      \colorrgb{0.50,0.50,0.50}%
      \put(845,870){\makebox(0,0)[r]{\strut{}}}%
      \colorrgb{0.50,0.50,0.50}%
      \put(845,1472){\makebox(0,0)[r]{\strut{}}}%
      \colorrgb{0.50,0.50,0.50}%
      \put(845,2073){\makebox(0,0)[r]{\strut{}}}%
      \colorrgb{0.50,0.50,0.50}%
      \put(845,2675){\makebox(0,0)[r]{\strut{}}}%
      \colorrgb{0.50,0.50,0.50}%
      \put(845,3277){\makebox(0,0)[r]{\strut{}}}%
      \colorrgb{0.50,0.50,0.50}%
      \put(845,3878){\makebox(0,0)[r]{\strut{}}}%
      \colorrgb{0.50,0.50,0.50}%
      \put(845,4480){\makebox(0,0)[r]{\strut{}}}%
      \colorrgb{0.50,0.50,0.50}%
      \put(934,391){\makebox(0,0){\strut{}10\textsuperscript{0}}}%
      \colorrgb{0.50,0.50,0.50}%
      \put(1471,391){\makebox(0,0){\strut{}10\textsuperscript{1}}}%
      \colorrgb{0.50,0.50,0.50}%
      \put(2008,391){\makebox(0,0){\strut{}10\textsuperscript{2}}}%
      \colorrgb{0.50,0.50,0.50}%
      \put(2545,391){\makebox(0,0){\strut{}10\textsuperscript{3}}}%
      \colorrgb{0.50,0.50,0.50}%
      \put(3081,391){\makebox(0,0){\strut{}10\textsuperscript{4}}}%
      \colorrgb{0.50,0.50,0.50}%
      \put(3618,391){\makebox(0,0){\strut{}10\textsuperscript{5}}}%
      \colorrgb{0.50,0.50,0.50}%
      \put(4155,391){\makebox(0,0){\strut{}10\textsuperscript{6}}}%
      \colorrgb{0.50,0.50,0.50}%
      \put(4692,391){\makebox(0,0){\strut{}10\textsuperscript{7}}}%
    }%
    \gplgaddtomacro\gplfronttext{%
      \csname LTb\endcsname%
      \put(133,2675){\rotatebox{-270}{\makebox(0,0){\strut{}PMF}}}%
      \csname LTb\endcsname%
      \put(2813,124){\makebox(0,0){\strut{}\# of Tweets}}%
      \csname LTb\endcsname%
      \put(1628,1262){\makebox(0,0)[l]{\strut{}Tweets}}%
      \csname LTb\endcsname%
      \put(1628,1013){\makebox(0,0)[l]{\strut{}Retweets}}%
      \csname LTb\endcsname%
      \put(1628,764){\makebox(0,0)[l]{\strut{}Retweeted}}%
    }%
    \gplbacktext
    \put(0,0){\includegraphics{all-distribution-scales}}%
    \gplfronttext
  \end{picture}%
\endgroup

%% file: interevent-durations.tex
\section{Distribution of Intertweet Durations}
\label{sec:intertweet-durations}

\begin{figure}[t]
  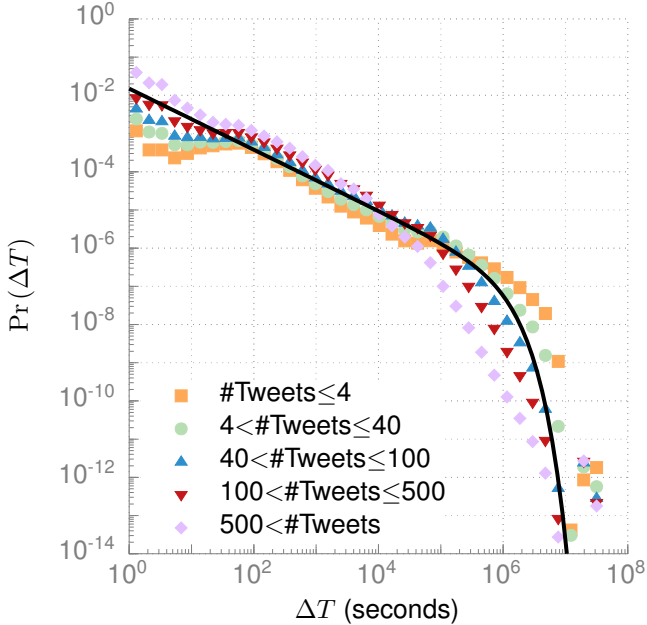
  \caption{The interevent distributions with users grouped by number
    of tweets for the three month period covering June through August
    2009. The line is a best-fit power law with exponential cutoff.}
  \label{fig:intertweet-duration}
\end{figure}

\begin{figure}[t]
  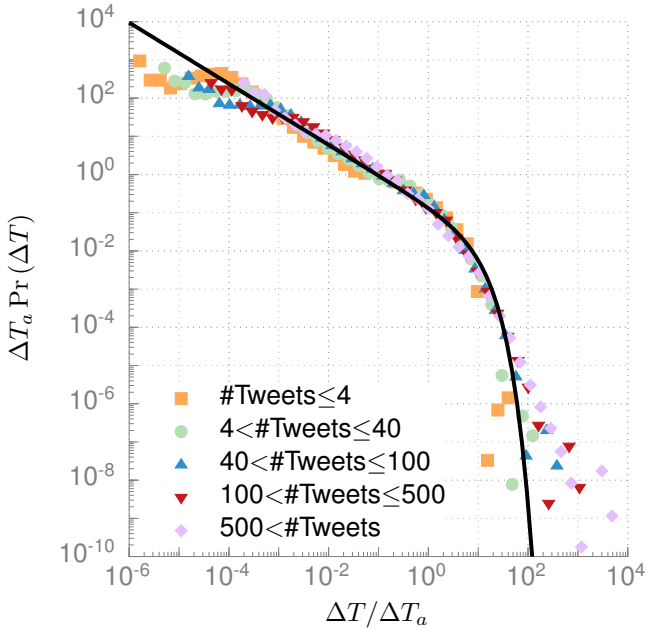
  \caption{The interevent distributions of
    \autoref{fig:intertweet-duration} collapse when scaled by the
    group's average interevent duration, $\Delta T_a$. The line is a
    best-fit power law with exponential cutoff.}
  \label{fig:intertweet-duration-scaled}
\end{figure}

Arrival processes in communication systems are traditionally assumed
to be Poisson~\cite{brown05}, but \emph{per-individual} interval
distributions for various activities including email, printing, and
telephone calls are
heavy-tailed~\cite{oliveira05oct,harder06feb,candia08jun}.  We show
that this same behavior holds in Twitter, with our analysis mirroring
that of Candia \etal for telephone calls~\cite{candia08jun} to enable
easy comparison. The SNAP tweet dataset is used for this analysis.

We group the users by their total tweets to isolate the effects of
differing tweet rates.  \autoref{fig:intertweet-duration} plots the
empirical distributions. Scaling by the group's average interevent
time ($\Delta t_a$) collapses the distributions to a single curve,
shown in \autoref{fig:intertweet-duration-scaled}.  This universal
trait is also found in email and telephone
systems~\cite{goh08feb,candia08jun}: the distribution is described by
$\Pr(\Delta T) = \frac{1}{\Delta T_a} F\left(\frac{\Delta T}{\Delta
    T_a}\right)$, where $F(\cdot)$ is independent of the average rate.
The best-fit cutoff power law is
\begin{equation}
  \Pr(\Delta T) \propto (\Delta T)^{-\alpha} \exp\left(-\frac{\Delta T}{\tau_c}\right),
\end{equation}
with exponent $\alpha \approx 0.8$ and cutoff $\tau_c \approx
\SI{8.1}{\day}$, shown as the black line in
\autoref{fig:intertweet-duration-scaled}. $\Delta T_a$ is taken as the
whole population average here.

%% file: fig/intertweet-duration.pdf_tex
% GNUPLOT: LaTeX picture with Postscript
\begingroup
  \makeatletter
  \providecommand\color[2][]{%
    \GenericError{(gnuplot) \space\space\space\@spaces}{%
      Package color not loaded in conjunction with
      terminal option `colourtext'%
    }{See the gnuplot documentation for explanation.%
    }{Either use 'blacktext' in gnuplot or load the package
      color.sty in LaTeX.}%
    \renewcommand\color[2][]{}%
  }%
  \providecommand\includegraphics[2][]{%
    \GenericError{(gnuplot) \space\space\space\@spaces}{%
      Package graphicx or graphics not loaded%
    }{See the gnuplot documentation for explanation.%
    }{The gnuplot epslatex terminal needs graphicx.sty or graphics.sty.}%
    \renewcommand\includegraphics[2][]{}%
  }%
  \providecommand\rotatebox[2]{#2}%
  \@ifundefined{ifGPcolor}{%
    \newif\ifGPcolor
    \GPcolortrue
  }{}%
  \@ifundefined{ifGPblacktext}{%
    \newif\ifGPblacktext
    \GPblacktextfalse
  }{}%
  % define a \g@addto@macro without @ in the name:
  \let\gplgaddtomacro\g@addto@macro
  % define empty templates for all commands taking text:
  \gdef\gplbacktext{}%
  \gdef\gplfronttext{}%
  \makeatother
  \ifGPblacktext
    % no textcolor at all
    \def\colorrgb#1{}%
    \def\colorgray#1{}%
  \else
    % gray or color?
    \ifGPcolor
      \def\colorrgb#1{\color[rgb]{#1}}%
      \def\colorgray#1{\color[gray]{#1}}%
      \expandafter\def\csname LTw\endcsname{\color{white}}%
      \expandafter\def\csname LTb\endcsname{\color{black}}%
      \expandafter\def\csname LTa\endcsname{\color{black}}%
      \expandafter\def\csname LT0\endcsname{\color[rgb]{1,0,0}}%
      \expandafter\def\csname LT1\endcsname{\color[rgb]{0,1,0}}%
      \expandafter\def\csname LT2\endcsname{\color[rgb]{0,0,1}}%
      \expandafter\def\csname LT3\endcsname{\color[rgb]{1,0,1}}%
      \expandafter\def\csname LT4\endcsname{\color[rgb]{0,1,1}}%
      \expandafter\def\csname LT5\endcsname{\color[rgb]{1,1,0}}%
      \expandafter\def\csname LT6\endcsname{\color[rgb]{0,0,0}}%
      \expandafter\def\csname LT7\endcsname{\color[rgb]{1,0.3,0}}%
      \expandafter\def\csname LT8\endcsname{\color[rgb]{0.5,0.5,0.5}}%
    \else
      % gray
      \def\colorrgb#1{\color{black}}%
      \def\colorgray#1{\color[gray]{#1}}%
      \expandafter\def\csname LTw\endcsname{\color{white}}%
      \expandafter\def\csname LTb\endcsname{\color{black}}%
      \expandafter\def\csname LTa\endcsname{\color{black}}%
      \expandafter\def\csname LT0\endcsname{\color{black}}%
      \expandafter\def\csname LT1\endcsname{\color{black}}%
      \expandafter\def\csname LT2\endcsname{\color{black}}%
      \expandafter\def\csname LT3\endcsname{\color{black}}%
      \expandafter\def\csname LT4\endcsname{\color{black}}%
      \expandafter\def\csname LT5\endcsname{\color{black}}%
      \expandafter\def\csname LT6\endcsname{\color{black}}%
      \expandafter\def\csname LT7\endcsname{\color{black}}%
      \expandafter\def\csname LT8\endcsname{\color{black}}%
    \fi
  \fi
    \setlength{\unitlength}{0.0500bp}%
    \ifx\gptboxheight\undefined%
      \newlength{\gptboxheight}%
      \newlength{\gptboxwidth}%
      \newsavebox{\gptboxtext}%
    \fi%
    \setlength{\fboxrule}{0.5pt}%
    \setlength{\fboxsep}{1pt}%
\begin{picture}(4960.00,4960.00)%
    \gplgaddtomacro\gplbacktext{%
      \colorrgb{0.50,0.50,0.50}%
      \put(845,747){\makebox(0,0)[r]{\strut{}10\textsuperscript{-14}}}%
      \colorrgb{0.50,0.50,0.50}%
      \put(845,1323){\makebox(0,0)[r]{\strut{}10\textsuperscript{-12}}}%
      \colorrgb{0.50,0.50,0.50}%
      \put(845,1900){\makebox(0,0)[r]{\strut{}10\textsuperscript{-10}}}%
      \colorrgb{0.50,0.50,0.50}%
      \put(845,2476){\makebox(0,0)[r]{\strut{}10\textsuperscript{-8}}}%
      \colorrgb{0.50,0.50,0.50}%
      \put(845,3052){\makebox(0,0)[r]{\strut{}10\textsuperscript{-6}}}%
      \colorrgb{0.50,0.50,0.50}%
      \put(845,3628){\makebox(0,0)[r]{\strut{}10\textsuperscript{-4}}}%
      \colorrgb{0.50,0.50,0.50}%
      \put(845,4205){\makebox(0,0)[r]{\strut{}10\textsuperscript{-2}}}%
      \colorrgb{0.50,0.50,0.50}%
      \put(845,4781){\makebox(0,0)[r]{\strut{}10\textsuperscript{0}}}%
      \colorrgb{0.50,0.50,0.50}%
      \put(845,1035){\makebox(0,0)[r]{\strut{}}}%
      \colorrgb{0.50,0.50,0.50}%
      \put(845,1611){\makebox(0,0)[r]{\strut{}}}%
      \colorrgb{0.50,0.50,0.50}%
      \put(845,2188){\makebox(0,0)[r]{\strut{}}}%
      \colorrgb{0.50,0.50,0.50}%
      \put(845,2764){\makebox(0,0)[r]{\strut{}}}%
      \colorrgb{0.50,0.50,0.50}%
      \put(845,3340){\makebox(0,0)[r]{\strut{}}}%
      \colorrgb{0.50,0.50,0.50}%
      \put(845,3917){\makebox(0,0)[r]{\strut{}}}%
      \colorrgb{0.50,0.50,0.50}%
      \put(845,4493){\makebox(0,0)[r]{\strut{}}}%
      \colorrgb{0.50,0.50,0.50}%
      \put(934,569){\makebox(0,0){\strut{}10\textsuperscript{0}}}%
      \colorrgb{0.50,0.50,0.50}%
      \put(1874,569){\makebox(0,0){\strut{}10\textsuperscript{2}}}%
      \colorrgb{0.50,0.50,0.50}%
      \put(2813,569){\makebox(0,0){\strut{}10\textsuperscript{4}}}%
      \colorrgb{0.50,0.50,0.50}%
      \put(3752,569){\makebox(0,0){\strut{}10\textsuperscript{6}}}%
      \colorrgb{0.50,0.50,0.50}%
      \put(4692,569){\makebox(0,0){\strut{}10\textsuperscript{8}}}%
      \colorrgb{0.50,0.50,0.50}%
      \put(1404,569){\makebox(0,0){\strut{}}}%
      \colorrgb{0.50,0.50,0.50}%
      \put(2343,569){\makebox(0,0){\strut{}}}%
      \colorrgb{0.50,0.50,0.50}%
      \put(3283,569){\makebox(0,0){\strut{}}}%
      \colorrgb{0.50,0.50,0.50}%
      \put(4222,569){\makebox(0,0){\strut{}}}%
    }%
    \gplgaddtomacro\gplfronttext{%
      \csname LTb\endcsname%
      \put(133,2764){\rotatebox{-270}{\makebox(0,0){\strut{}$\Pr\left(\Delta T\right)$}}}%
      \csname LTb\endcsname%
      \put(2813,302){\makebox(0,0){\strut{}$\Delta T\vphantom{T_a}$ (seconds)}}%
      \csname LTb\endcsname%
      \put(1628,1938){\makebox(0,0)[l]{\strut{}\#Tweets$\leq$4}}%
      \csname LTb\endcsname%
      \put(1628,1689){\makebox(0,0)[l]{\strut{}4$<$\#Tweets$\leq$40}}%
      \csname LTb\endcsname%
      \put(1628,1440){\makebox(0,0)[l]{\strut{}40$<$\#Tweets$\leq$100}}%
      \csname LTb\endcsname%
      \put(1628,1191){\makebox(0,0)[l]{\strut{}100$<$\#Tweets$\leq$500}}%
      \csname LTb\endcsname%
      \put(1628,942){\makebox(0,0)[l]{\strut{}500$<$\#Tweets}}%
    }%
    \gplbacktext
    \put(0,0){\includegraphics{intertweet-duration}}%
    \gplfronttext
  \end{picture}%
\endgroup

%% file: fig/intertweet-duration-scaled.pdf_tex
% GNUPLOT: LaTeX picture with Postscript
\begingroup
  \makeatletter
  \providecommand\color[2][]{%
    \GenericError{(gnuplot) \space\space\space\@spaces}{%
      Package color not loaded in conjunction with
      terminal option `colourtext'%
    }{See the gnuplot documentation for explanation.%
    }{Either use 'blacktext' in gnuplot or load the package
      color.sty in LaTeX.}%
    \renewcommand\color[2][]{}%
  }%
  \providecommand\includegraphics[2][]{%
    \GenericError{(gnuplot) \space\space\space\@spaces}{%
      Package graphicx or graphics not loaded%
    }{See the gnuplot documentation for explanation.%
    }{The gnuplot epslatex terminal needs graphicx.sty or graphics.sty.}%
    \renewcommand\includegraphics[2][]{}%
  }%
  \providecommand\rotatebox[2]{#2}%
  \@ifundefined{ifGPcolor}{%
    \newif\ifGPcolor
    \GPcolortrue
  }{}%
  \@ifundefined{ifGPblacktext}{%
    \newif\ifGPblacktext
    \GPblacktextfalse
  }{}%
  % define a \g@addto@macro without @ in the name:
  \let\gplgaddtomacro\g@addto@macro
  % define empty templates for all commands taking text:
  \gdef\gplbacktext{}%
  \gdef\gplfronttext{}%
  \makeatother
  \ifGPblacktext
    % no textcolor at all
    \def\colorrgb#1{}%
    \def\colorgray#1{}%
  \else
    % gray or color?
    \ifGPcolor
      \def\colorrgb#1{\color[rgb]{#1}}%
      \def\colorgray#1{\color[gray]{#1}}%
      \expandafter\def\csname LTw\endcsname{\color{white}}%
      \expandafter\def\csname LTb\endcsname{\color{black}}%
      \expandafter\def\csname LTa\endcsname{\color{black}}%
      \expandafter\def\csname LT0\endcsname{\color[rgb]{1,0,0}}%
      \expandafter\def\csname LT1\endcsname{\color[rgb]{0,1,0}}%
      \expandafter\def\csname LT2\endcsname{\color[rgb]{0,0,1}}%
      \expandafter\def\csname LT3\endcsname{\color[rgb]{1,0,1}}%
      \expandafter\def\csname LT4\endcsname{\color[rgb]{0,1,1}}%
      \expandafter\def\csname LT5\endcsname{\color[rgb]{1,1,0}}%
      \expandafter\def\csname LT6\endcsname{\color[rgb]{0,0,0}}%
      \expandafter\def\csname LT7\endcsname{\color[rgb]{1,0.3,0}}%
      \expandafter\def\csname LT8\endcsname{\color[rgb]{0.5,0.5,0.5}}%
    \else
      % gray
      \def\colorrgb#1{\color{black}}%
      \def\colorgray#1{\color[gray]{#1}}%
      \expandafter\def\csname LTw\endcsname{\color{white}}%
      \expandafter\def\csname LTb\endcsname{\color{black}}%
      \expandafter\def\csname LTa\endcsname{\color{black}}%
      \expandafter\def\csname LT0\endcsname{\color{black}}%
      \expandafter\def\csname LT1\endcsname{\color{black}}%
      \expandafter\def\csname LT2\endcsname{\color{black}}%
      \expandafter\def\csname LT3\endcsname{\color{black}}%
      \expandafter\def\csname LT4\endcsname{\color{black}}%
      \expandafter\def\csname LT5\endcsname{\color{black}}%
      \expandafter\def\csname LT6\endcsname{\color{black}}%
      \expandafter\def\csname LT7\endcsname{\color{black}}%
      \expandafter\def\csname LT8\endcsname{\color{black}}%
    \fi
  \fi
    \setlength{\unitlength}{0.0500bp}%
    \ifx\gptboxheight\undefined%
      \newlength{\gptboxheight}%
      \newlength{\gptboxwidth}%
      \newsavebox{\gptboxtext}%
    \fi%
    \setlength{\fboxrule}{0.5pt}%
    \setlength{\fboxsep}{1pt}%
\begin{picture}(4960.00,4960.00)%
    \gplgaddtomacro\gplbacktext{%
      \colorrgb{0.50,0.50,0.50}%
      \put(845,747){\makebox(0,0)[r]{\strut{}10\textsuperscript{-10}}}%
      \colorrgb{0.50,0.50,0.50}%
      \put(845,1323){\makebox(0,0)[r]{\strut{}10\textsuperscript{-8}}}%
      \colorrgb{0.50,0.50,0.50}%
      \put(845,1900){\makebox(0,0)[r]{\strut{}10\textsuperscript{-6}}}%
      \colorrgb{0.50,0.50,0.50}%
      \put(845,2476){\makebox(0,0)[r]{\strut{}10\textsuperscript{-4}}}%
      \colorrgb{0.50,0.50,0.50}%
      \put(845,3052){\makebox(0,0)[r]{\strut{}10\textsuperscript{-2}}}%
      \colorrgb{0.50,0.50,0.50}%
      \put(845,3628){\makebox(0,0)[r]{\strut{}10\textsuperscript{0}}}%
      \colorrgb{0.50,0.50,0.50}%
      \put(845,4205){\makebox(0,0)[r]{\strut{}10\textsuperscript{2}}}%
      \colorrgb{0.50,0.50,0.50}%
      \put(845,4781){\makebox(0,0)[r]{\strut{}10\textsuperscript{4}}}%
      \colorrgb{0.50,0.50,0.50}%
      \put(845,1035){\makebox(0,0)[r]{\strut{}}}%
      \colorrgb{0.50,0.50,0.50}%
      \put(845,1611){\makebox(0,0)[r]{\strut{}}}%
      \colorrgb{0.50,0.50,0.50}%
      \put(845,2188){\makebox(0,0)[r]{\strut{}}}%
      \colorrgb{0.50,0.50,0.50}%
      \put(845,2764){\makebox(0,0)[r]{\strut{}}}%
      \colorrgb{0.50,0.50,0.50}%
      \put(845,3340){\makebox(0,0)[r]{\strut{}}}%
      \colorrgb{0.50,0.50,0.50}%
      \put(845,3917){\makebox(0,0)[r]{\strut{}}}%
      \colorrgb{0.50,0.50,0.50}%
      \put(845,4493){\makebox(0,0)[r]{\strut{}}}%
      \colorrgb{0.50,0.50,0.50}%
      \put(934,569){\makebox(0,0){\strut{}10\textsuperscript{-6}}}%
      \colorrgb{0.50,0.50,0.50}%
      \put(1686,569){\makebox(0,0){\strut{}10\textsuperscript{-4}}}%
      \colorrgb{0.50,0.50,0.50}%
      \put(2437,569){\makebox(0,0){\strut{}10\textsuperscript{-2}}}%
      \colorrgb{0.50,0.50,0.50}%
      \put(3189,569){\makebox(0,0){\strut{}10\textsuperscript{0}}}%
      \colorrgb{0.50,0.50,0.50}%
      \put(3940,569){\makebox(0,0){\strut{}10\textsuperscript{2}}}%
      \colorrgb{0.50,0.50,0.50}%
      \put(4692,569){\makebox(0,0){\strut{}10\textsuperscript{4}}}%
      \colorrgb{0.50,0.50,0.50}%
      \put(1310,569){\makebox(0,0){\strut{}}}%
      \colorrgb{0.50,0.50,0.50}%
      \put(2061,569){\makebox(0,0){\strut{}}}%
      \colorrgb{0.50,0.50,0.50}%
      \put(2813,569){\makebox(0,0){\strut{}}}%
      \colorrgb{0.50,0.50,0.50}%
      \put(3565,569){\makebox(0,0){\strut{}}}%
      \colorrgb{0.50,0.50,0.50}%
      \put(4316,569){\makebox(0,0){\strut{}}}%
    }%
    \gplgaddtomacro\gplfronttext{%
      \csname LTb\endcsname%
      \put(133,2764){\rotatebox{-270}{\makebox(0,0){\strut{}$\Delta T_a\Pr\left(\Delta T\right)$}}}%
      \csname LTb\endcsname%
      \put(2813,302){\makebox(0,0){\strut{}$\Delta T/\Delta T_a$}}%
      \csname LTb\endcsname%
      \put(1628,1938){\makebox(0,0)[l]{\strut{}\#Tweets$\leq$4}}%
      \csname LTb\endcsname%
      \put(1628,1689){\makebox(0,0)[l]{\strut{}4$<$\#Tweets$\leq$40}}%
      \csname LTb\endcsname%
      \put(1628,1440){\makebox(0,0)[l]{\strut{}40$<$\#Tweets$\leq$100}}%
      \csname LTb\endcsname%
      \put(1628,1191){\makebox(0,0)[l]{\strut{}100$<$\#Tweets$\leq$500}}%
      \csname LTb\endcsname%
      \put(1628,942){\makebox(0,0)[l]{\strut{}500$<$\#Tweets}}%
    }%
    \gplbacktext
    \put(0,0){\includegraphics{intertweet-duration-scaled}}%
    \gplfronttext
  \end{picture}%
\endgroup

%% file: retweet-graph.tex
\section{Characteristics of the Retweet Graph}
\label{sec:retweet-graph}

The natural and explicit network in Twitter---the social graph in
which a directed edge represents the \emph{follower}
relationship---has been well-studied.  Kwak \etal first reported on
basic network properties like degree distribution, reciprocity, and
average path length~\cite{kwak10apr}, and later works have studied
these and other characteristics in more
detail~\cite{bliss12sep,teutle10feb,gabielkov12dec,ghosh12apr}. However,
an alternative, implicit network---the retweet graph in which a
directed edge indicates that the source retweeted the
destination---has been neglected.  We conduct the first
characterization of the retweet graph and confirm that it, like many
real-world networks, is small-world and scale-free. The reported
metrics are useful for generating random retweet graphs using general
parametric models like R-MAT~\cite{chakrabarti04apr} ($a=0.52$,
$b=0.18$, $c=0.17$, $d=0.13$) or other specific generative
models~\cite{bollobas03jan}.

We pay particular attention to contrasting the social
following\footnote{The social following graph is simply the social
  follower graph with the edge direction reversed to match that of the
  retweet graph.} and retweet graphs. Intuitively, they should be
similar because retweets are usually sent by followers.  However, we
conjecture that the retweet graph more closely models the real-world
social and trust relationships among users, because it derives from a
more forceful action---not just listening to others' ideas, but
forwarding them to one's own friends.  Using the follower graph as a
trust proxy has been proposed for applications ranging from spam
filtering~\cite{benevenuto10jul,song11sep,yang12apr} to Sybil
detection~\cite{yu08jun,yu08may}. We conjecture that the retweet graph
is a better choice and provide some supporting evidence. Full
treatment of this conjecture is beyond our scope.

\subsection{Analyzing a Random Subsample of the Retweet Graph}

\begin{figure}
  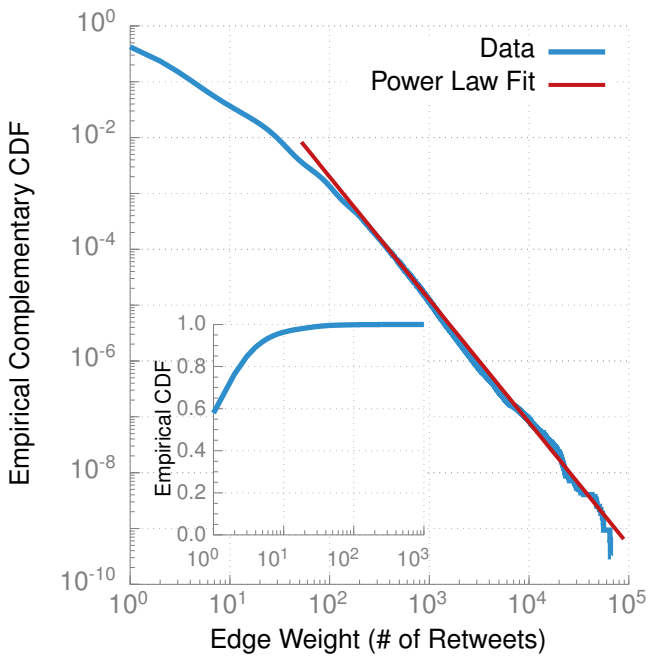
  \caption{Distribution of number of edge weights in the retweet
    graph, corrected using the EM method.  A directed edge indicates
    that one user retweeted another and the weight is the number of
    such retweets.}
  \label{fig:retweet-edge-weights}
\end{figure}

The retweet graph is constructed from our largest dataset, the 10\%
sample, and thus does not contain all edges. An edge is included with
probability proportional to the number of retweets sent along
it. However, 60\% of edges have a single retweet and 98\% have fewer
than 10 (see \autoref{fig:retweet-edge-weights}), so for simplicity we
assume each edge is included with 10\% probability.  Many measured
properties in an edge-sampled graph differ from the original graph.
When possible, we use the EM-based method from
\autoref{sec:em-summary} to correct our results. When not, we estimate
the errors using the literature on sampled
graphs~\cite{lee06jan,son12oct,stumpf05mar}.

\subsection{Degree Distributions}
\label{sec:degree-distributions}

We begin with the in- and out-degree distributions.  The in-degree
$k_\mathrm{in}^i$ of a node $i$ is the number of unique users who
retweeted $i$ and the out-degree $k_\mathrm{out}^i$ is the number of
unique users retweeted by $i$.  The average in-degree $\langle
k_\mathrm{in}\rangle \triangleq N^{-1}\sum_{i\in V}k_\mathrm{in}^i =
88.4$ and the similarly-defined average out-degree $\langle
k_\mathrm{out}\rangle = 74.3$. $V$ is the set of nodes and $N$ their
cardinality. In reality $\langle k_\mathrm{in} \rangle = \langle
k_\mathrm{out} \rangle;$ the observed difference is an artifact of the
EM-based population estimation.  The degree standard deviations are
$\sigma_\mathrm{in} = 4187.3$ and $\sigma_\mathrm{out} = 228.4.$
Higher in-degree variance is expected because, as with real-world
networks~\cite{son12oct}, \emph{popularity} (the number of users who
retweeted an individual) is more variable than \emph{extensivity} (the
number of users an individual retweeted).

\begin{figure}
  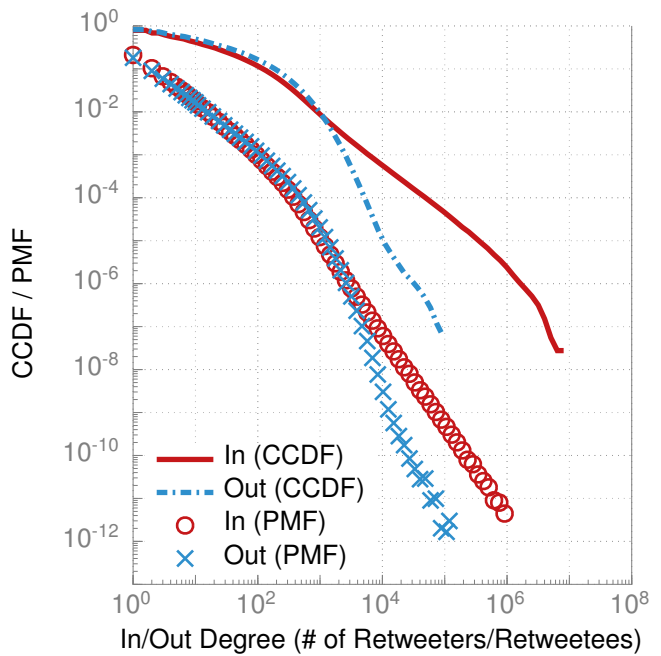
  \caption{In and out degree distributions for the retweet graph. Both
    exhibit the double-Pareto behavior common to evolving
    networks~\cite{barabasi02aug,dorogovtsev01dec}. In the upper tail,
    the in-degree power-law exponent is 2.2 and 3.75 for the
    out-degree.}
  \label{fig:retweet-degrees}
\end{figure}

The distributions, shown in \autoref{fig:retweet-degrees}, are similar
to the social following graph~\cite{kwak10apr}. Both are heavy-tailed
and exhibit the same two-phase power law common to such
networks. Similarly to the tweet rate distribution
(\autoref{sec:tweet-rates-generative}), the two phases are a dynamic
effect arising from two forms of evolution in the
graph~\cite{barabasi02aug,dorogovtsev01dec}---the addition of new
nodes and preferentially-attached new edges between existing nodes.
The outgoing (incoming) node $i$ for a new edge is selected with
relative probability $d_\mathrm{out}(i) + \delta_\mathrm{out}$
($d_\mathrm{in}(i) + \delta_\mathrm{in}$), where $\delta_\mathrm{out}$
and $\delta_\mathrm{in}$ are the initial attractiveness constants and
$d(\cdot)$ returns the node degree.  Bollob{\'a}s \etal elucidate this
process for a general context~\cite{bollobas03jan}.

The power law exponents are determined by $\delta_\mathrm{out}$
($\delta_\mathrm{in}$).  The lower tails are similar with $\alpha
\approx 1.3$.  In the upper tail, $\alpha_\mathrm{out} = 3.75$ and
$\alpha_\mathrm{in} = 2.2.$ $a_\mathrm{in}$ matches the followers
graph (2.3)~\cite{kwak10apr} and is in the range of most real-world
networks (2--3). $\alpha_\mathrm{out}$ exceeds that range because
extensivity is not inherently preferential (like popularity).

\subsection{Reciprocity}
\label{sec:reciprocity}

Reciprocity is the fraction of links that are bidirectional.  Many
social networks have high reciprocity---most relationships are
bidirectional (68\% in Flickr~\cite{cha09apr} and 84\% on Yahoo!
360~\cite{kumar06aug}). In the Twitter follow graph, reciprocity is
lower at just 22.1\%~\cite{kwak10apr}. If retweeting is more
discriminating than following, the retweet reciprocity should be
lower. Indeed, it is just 11.1\%.\footnote{We estimated the
  distribution of all non-zero pairwise edge weight tuples (the number
  of retweets in both directions) from the 10\% sample using the EM
  algorithm. The fraction that are non-zero in both directions is the
  reciprocity.}  Higher reciprocity in the follower graph may stem
from the popularity of follow-back schemes in which a user, in an
attempt to gain followers, promises to follow back anyone who follows
him.  The low reciprocity suggests that using the retweet graph as a
proxy for trust is promising. Although a malicious node can establish
many outgoing links, it has little control over the incoming
structure.

\subsection{Average Shortest Path Length (Degree of Separation)}
\label{sec:apl}

\begin{figure}
  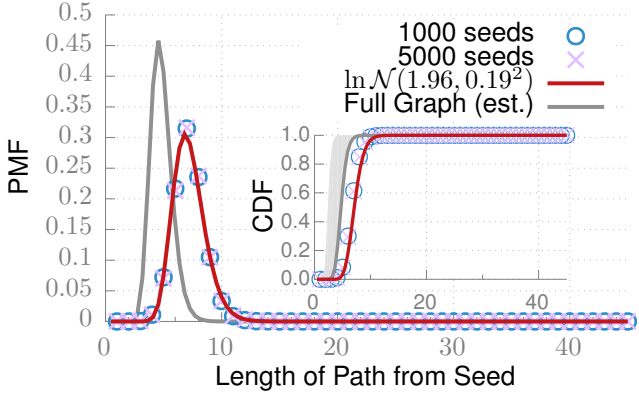
  \caption{Distribution of average path length (degree of separation)
    in edge-sampled retweet graph. The gray line is the estimated
    distribution for the full graph.}
  \label{fig:retweet-average-path-length}
\end{figure}

The real-world human social network has a small average shortest path
length (APL) of about six, shown most famously by Stanley
Milgram~\cite{milgram67may,travers69dec}. Many online networks are
similar~\cite{watts98jun,leskovec08apr}, but the social followers
graph is denser with an APL of 4.12~\cite{kwak10apr}.  Kwak \etal
attribute this difference to Twitter's additional role as an
information source. Edges are more dense because users follow
\emph{both} social acquaintances and sources of interesting content.

We determined the path length distribution of the 10\% edge-sampled
graph by computing all shortest paths for both 1000 and 5000 random
starting nodes. The obtained distributions (shown in
\autoref{fig:retweet-average-path-length}) overlap, indicating a
sufficient sample size.  Lee \etal showed that edge sampling increases
the APL by $1.5$--$3\times$ (the gray range in the inset plot)
depending on the graph structure~\cite{lee06jan}. We use $1.5\times$,
determined by sampling the followers graph\footnote{The 2009
  crawl~\cite{kwak10apr} is complete, so we compared the true
  statistic against that of a 10\% subsample.}, to estimate the full
distribution (grey line in plot).  The estimated APL is 4.8 and the
90th-percentile or \emph{effective diameter}~\cite{palmer01may} is
8.5.  The difference from the followers graph is within estimation
error.

The best-fit distribution (solid line in plot) is
log-normal\footnote{We compared with the Weibull, Gumbel, Fr\'{e}chet,
  and encompassing generalized extreme value distributions.} with
$\mu=1.5$ and $\sigma=0.27$. This differs from undirected
Erd\"{o}s-R\'{e}nyi (ER) graphs, for which the limiting distribution
is Weibull~\cite{bauckhage13aug}, but we do not know of similar
theoretical results for directed graphs.

\subsection{Assortativity (Node Degree Correlation)}
\label{sec:assortativity}

Degree assortativity---the tendency of nodes to connect with others of
similar degree---summarizes the structural characteristics that in
part determine how content (e.g., retweets or disease) spreads and
resilience to node removal~\cite{newman02nov}.  In an assortative
network, content easily propagates through a connected component of
tightly clustered, high degree nodes that is resistant to node
removal, but may not reach the low degree boundary of the network.
Conversely, a disassortative network has a larger connected component
so content propagates further, but can be partitioned by the removal
of a high degree node.

For undirected networks, assortativity is simply the Pearson
correlation between the degrees of adjacent nodes. The concept
generalizes to directed networks by considering all possible
directional degree pairs as separate assortativity
metrics~\cite{foster10jun}, $r(in,in)$, $r(in,out)$, $r(out,in)$,
$r(out,out)$, again using the Pearson correlation
\begin{equation}
  r(\alpha, \beta) \triangleq \frac{\langle k_\alpha^i k_\beta^j \rangle - \langle k_\alpha^i\rangle \langle k_\beta^j \rangle}{\sigma_{k_\alpha}\sigma_{k_\beta}}
\end{equation}
where $\alpha, \beta \in \{in,out\},$ $k_\alpha^i$ ($k_\beta^j$) is
the $\alpha$-degree ($\beta$-degree) of source (destination) node $i$
($j$), the averages $\langle\cdot\rangle$ are taken over all
directional edges $(i\rightarrow j),$ and $\sigma_{k_\alpha}$
($\sigma_{k_\beta}$) is the variance of the $\alpha$-degree
($\beta$-degree).

\begin{figure}
  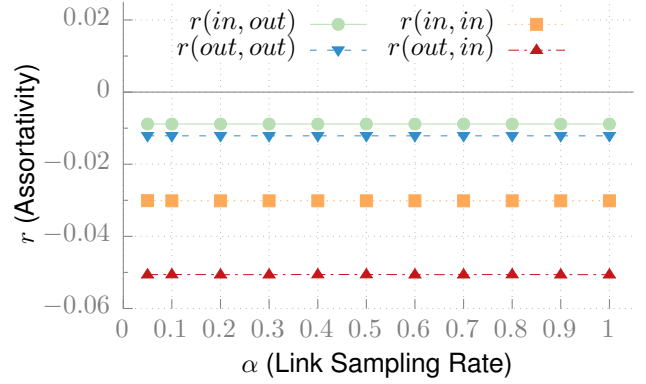
  \caption{Directed assortativities $r$ as a function of edge sampling
    rate. Edge sampling does not affect assortativity because all node
    degrees are sampled independently and identically.}
  \label{fig:follower-graph-sampled-assortativity}
\end{figure}

We characterize and contrast these metrics for both the Twitter social
following graph~\cite{kwak10apr} and retweet graph. Edge sampling
impacts the degrees of all nodes identically and thus does not effect
assortativity (see
\autoref{fig:follower-graph-sampled-assortativity})~\cite{lee06jan}.

\begin{figure}
  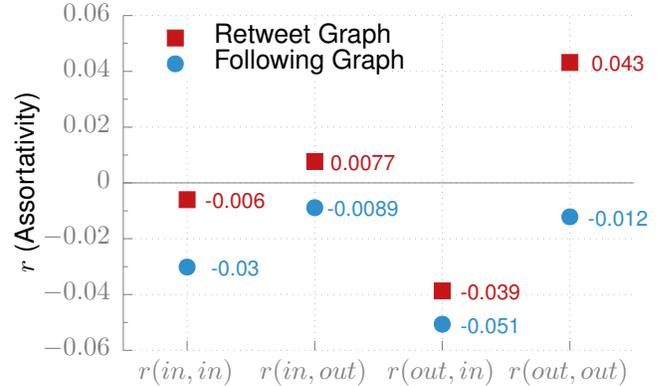
  \caption{Directed assortativity $r$ of the retweet graph and the
    social following graph. The retweet graph has higher assortative,
    more consistent with real world social networks than most online
    social networks.}
  \label{fig:retweet-assortativity}
\end{figure}

\autoref{fig:retweet-assortativity} plots the assortativities for both
networks.  Although most real-world social networks are
assortative~\cite{newman02nov}, online social networks are instead
disassortative~\cite{hu09apr}. The social followers graph is no
exception, showing weak disassortativity across all measures.  In
contrast, the retweet graph is more assortative across all
measures. It is near-neutral for both $r(in,\cdot)$ metrics,
indicating independence between one's own retweet behavior and the
number of retweets received. This is consistent with the graph
containing useful trust information, because a user cannot influence
the quantity of retweets received by selectively retweeting popular
($r(in,in)$) or extensive ($r(in,out)$) users.  The high $(out,out)$
assortativity is more consistent with real-world social networks and
indicates that extensive retweeters retweet each other. Interestingly,
they are not tightly clustered (or the $(in,out)$-assortativity would
be higher).

In Twitter, tweets propagate to followers, so the social graph
disassortativity is helpful. The connected component is larger and
tweets disseminate further more quickly. Increased susceptibility to
node failure is acceptable in a centralized system. In a decentralized
system that relies more heavily on the retweet graph for propagation,
e.g., Shout~\cite{shout}, the resilience to node failure implied by
its neutral and positive assortativities would instead be helpful.

\subsection{Clustering Coefficient}
\label{sec:clustering}

A clustering coefficient quantifies the tendency of neighboring nodes
to form highly connected clusters. Many real-world networks exhibit
tighter clustering than would be expected in similar random
graphs~\cite{watts98jun}. We consider the \emph{global clustering
  coefficient}\footnote{Sometimes called the \emph{transitivity} or
  \emph{transitivity ratio}.}, defined for undirected graphs as
\begin{equation}
  C \triangleq \frac{3N_\bigtriangleup}{N_3},
\end{equation}
where $N_3$ is the number of open \emph{or} closed triplets (three
vertices connected by two or three edges) and $N_\bigtriangleup$ is
the number of closed triplets (3-vertex cliques). Unlike the
alternative \emph{local clustering coefficient}, this definition is
suitable for networks with isolated nodes~\cite{kaiser08aug}.  In
essence, $C$ gives the probability that any two neighbors of a node
are themselves connected.

\begin{figure}
  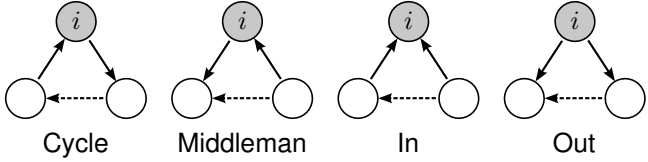
  \caption{The four types of open (solid edges) and closed (solid and
    dashed edges) directed triplets used for cluster analysis.  A
    vertex can form up to eight such triplets with each pair of
    neighbors, two of each type. The clustering coefficient
    $C_{\beta\in\{\text{cycle, middleman, in, out}\}}$ is the
    fraction of $\beta$-triplets (open and closed) that are closed.}
  \label{fig:cluster-triangle-types}
\end{figure}

Following the approach introduced by Fagiolo for the local clustering
coefficient~\cite{fagiolo07aug}, we extend the metric to directed
graphs by separately considering the four types of directed triplets,
shown in \autoref{fig:cluster-triangle-types}. The four clustering
coefficients $C_{\beta \in \{\text{cycle, middleman, in, out}\}}$ are
the fraction of $\beta$-triplets that are closed.

\begin{figure}
  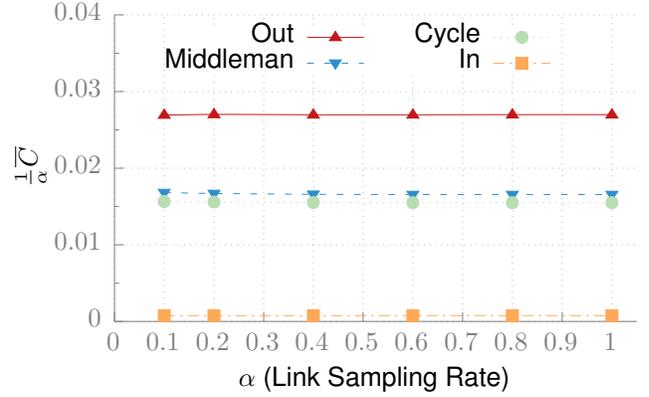
  \caption{The clustering coefficient estimator $\widehat{C}
    \triangleq \frac{1}{\alpha}\overline{C}$ as a function of edge
    sampling rate on the social ``following'' graph. Although
    potentially biased, the estimator is quite accurate for such
    graphs.}
  \label{fig:follower-graph-sampled-clustering-coefficient}
\end{figure}

An estimator from the sample clustering coefficient of an
$\alpha$-edge sampled graph ($\alpha = 0.1$ for us) is
\begin{equation}
  \widehat{C} \triangleq \frac{1}{\alpha}\overline{C},
\end{equation}
seen by noting that a triplet is included in the sample with
$\alpha^2$ probability and as a closed triplet with $\alpha^3$
probability. This estimate is biased, because the triplets are not
independent and edges can be concentrated towards open (or closed)
triplets. In practice however, it performs well on large samples, as
shown in \autoref{fig:follower-graph-sampled-clustering-coefficient}
for the social following graph.

\begin{figure}
  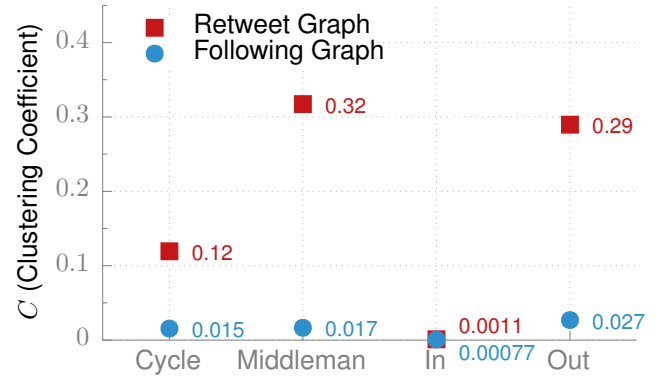
  \caption{Clustering coefficients for the social ``following'' graph
    and the retweet graph.  Clustering is significantly more prominent
    in the retweet graph and more consistent with real-world social
    networks.}
  \label{fig:clustering-coefficient}
\end{figure}

\autoref{fig:clustering-coefficient} plots the results for both the
social and retweet graphs.  The former has low clustering, but
clustering in the retweet graph is significant in metrics except
\emph{in}.  \emph{Cycle} is the only fully (transitively) connected
triplet type, and thus \emph{cycle}-clustering should best reflect
true clustering in the underlying social groups and interest
topics. The higher clustering in the retweet graph indicates that
retweet relationships are more concentrated than \emph{following}
relationships, consistent with our hypothesis of higher trust.

Although the \emph{middleman}, \emph{in}, and \emph{out} cycle are all
rotations of the same basic non-transitive triplet, their coefficients
differ due to the non-uniform degree distribution. $C_\text{in}$ is
low because the majority of $(in,in)$ edge pairs are from a few
popular users who are retweeted by many otherwise-unrelated users.
The high $C_\text{middleman}$ and $C_\text{out}$ coefficients are
reflections of the same phenomenon---transitive retweeting. User $f$
retweeting user $i$'s retweet of user $a$ is recorded by Twitter as
$f$ retweeting $a$ (hence the name middleman). Often $f$ will also
retweet some of $i$'s original content, closing the triplet. In the
out case, node $i$ plays the role of $f$ instead of the middleman.
Surprisingly, such transitive retweeting happens frequently
($C_\text{middleman} = 0.32$ and $C_\text{out} = 0.29$). In other
words, $~30\%$ of these possible two-degree retweet relationships
exist.

\subsection{Summary}
\label{sec:retweet-graph-summary}

We have confirmed that the retweet graph is scale-free and
small-world, like many social networks.  Interestingly, its clustering
and assortativity are closer to real-world networks than typical
online networks, indicating that it may better capture real-world
relationships and have application as a proxy for trust.  Full
treatment of this conjecture is beyond our scope.  The scale-free,
small-world confirmation enables the generation of random instances,
e.g., using R-MAT~\cite{chakrabarti04apr}, for empirical study. We use
this approach in \autoref{sec:spam} to evaluate the use of
connectivity in the retweet graph to detect spammers.

%% file: fig/retweet-edge-weights.pdf_tex
% GNUPLOT: LaTeX picture with Postscript
\begingroup
  \makeatletter
  \providecommand\color[2][]{%
    \GenericError{(gnuplot) \space\space\space\@spaces}{%
      Package color not loaded in conjunction with
      terminal option `colourtext'%
    }{See the gnuplot documentation for explanation.%
    }{Either use 'blacktext' in gnuplot or load the package
      color.sty in LaTeX.}%
    \renewcommand\color[2][]{}%
  }%
  \providecommand\includegraphics[2][]{%
    \GenericError{(gnuplot) \space\space\space\@spaces}{%
      Package graphicx or graphics not loaded%
    }{See the gnuplot documentation for explanation.%
    }{The gnuplot epslatex terminal needs graphicx.sty or graphics.sty.}%
    \renewcommand\includegraphics[2][]{}%
  }%
  \providecommand\rotatebox[2]{#2}%
  \@ifundefined{ifGPcolor}{%
    \newif\ifGPcolor
    \GPcolortrue
  }{}%
  \@ifundefined{ifGPblacktext}{%
    \newif\ifGPblacktext
    \GPblacktextfalse
  }{}%
  % define a \g@addto@macro without @ in the name:
  \let\gplgaddtomacro\g@addto@macro
  % define empty templates for all commands taking text:
  \gdef\gplbacktext{}%
  \gdef\gplfronttext{}%
  \makeatother
  \ifGPblacktext
    % no textcolor at all
    \def\colorrgb#1{}%
    \def\colorgray#1{}%
  \else
    % gray or color?
    \ifGPcolor
      \def\colorrgb#1{\color[rgb]{#1}}%
      \def\colorgray#1{\color[gray]{#1}}%
      \expandafter\def\csname LTw\endcsname{\color{white}}%
      \expandafter\def\csname LTb\endcsname{\color{black}}%
      \expandafter\def\csname LTa\endcsname{\color{black}}%
      \expandafter\def\csname LT0\endcsname{\color[rgb]{1,0,0}}%
      \expandafter\def\csname LT1\endcsname{\color[rgb]{0,1,0}}%
      \expandafter\def\csname LT2\endcsname{\color[rgb]{0,0,1}}%
      \expandafter\def\csname LT3\endcsname{\color[rgb]{1,0,1}}%
      \expandafter\def\csname LT4\endcsname{\color[rgb]{0,1,1}}%
      \expandafter\def\csname LT5\endcsname{\color[rgb]{1,1,0}}%
      \expandafter\def\csname LT6\endcsname{\color[rgb]{0,0,0}}%
      \expandafter\def\csname LT7\endcsname{\color[rgb]{1,0.3,0}}%
      \expandafter\def\csname LT8\endcsname{\color[rgb]{0.5,0.5,0.5}}%
    \else
      % gray
      \def\colorrgb#1{\color{black}}%
      \def\colorgray#1{\color[gray]{#1}}%
      \expandafter\def\csname LTw\endcsname{\color{white}}%
      \expandafter\def\csname LTb\endcsname{\color{black}}%
      \expandafter\def\csname LTa\endcsname{\color{black}}%
      \expandafter\def\csname LT0\endcsname{\color{black}}%
      \expandafter\def\csname LT1\endcsname{\color{black}}%
      \expandafter\def\csname LT2\endcsname{\color{black}}%
      \expandafter\def\csname LT3\endcsname{\color{black}}%
      \expandafter\def\csname LT4\endcsname{\color{black}}%
      \expandafter\def\csname LT5\endcsname{\color{black}}%
      \expandafter\def\csname LT6\endcsname{\color{black}}%
      \expandafter\def\csname LT7\endcsname{\color{black}}%
      \expandafter\def\csname LT8\endcsname{\color{black}}%
    \fi
  \fi
    \setlength{\unitlength}{0.0500bp}%
    \ifx\gptboxheight\undefined%
      \newlength{\gptboxheight}%
      \newlength{\gptboxwidth}%
      \newsavebox{\gptboxtext}%
    \fi%
    \setlength{\fboxrule}{0.5pt}%
    \setlength{\fboxsep}{1pt}%
\begin{picture}(4960.00,4960.00)%
    \gplgaddtomacro\gplbacktext{%
      \colorrgb{0.50,0.50,0.50}%
      \put(845,569){\makebox(0,0)[r]{\strut{}10\textsuperscript{-10}}}%
      \colorrgb{0.50,0.50,0.50}%
      \put(845,1411){\makebox(0,0)[r]{\strut{}10\textsuperscript{-8}}}%
      \colorrgb{0.50,0.50,0.50}%
      \put(845,2254){\makebox(0,0)[r]{\strut{}10\textsuperscript{-6}}}%
      \colorrgb{0.50,0.50,0.50}%
      \put(845,3096){\makebox(0,0)[r]{\strut{}10\textsuperscript{-4}}}%
      \colorrgb{0.50,0.50,0.50}%
      \put(845,3939){\makebox(0,0)[r]{\strut{}10\textsuperscript{-2}}}%
      \colorrgb{0.50,0.50,0.50}%
      \put(845,4781){\makebox(0,0)[r]{\strut{}10\textsuperscript{0}}}%
      \colorrgb{0.50,0.50,0.50}%
      \put(845,990){\makebox(0,0)[r]{\strut{}}}%
      \colorrgb{0.50,0.50,0.50}%
      \put(845,1833){\makebox(0,0)[r]{\strut{}}}%
      \colorrgb{0.50,0.50,0.50}%
      \put(845,2675){\makebox(0,0)[r]{\strut{}}}%
      \colorrgb{0.50,0.50,0.50}%
      \put(845,3517){\makebox(0,0)[r]{\strut{}}}%
      \colorrgb{0.50,0.50,0.50}%
      \put(845,4360){\makebox(0,0)[r]{\strut{}}}%
      \colorrgb{0.50,0.50,0.50}%
      \put(934,391){\makebox(0,0){\strut{}10\textsuperscript{0}}}%
      \colorrgb{0.50,0.50,0.50}%
      \put(1686,391){\makebox(0,0){\strut{}10\textsuperscript{1}}}%
      \colorrgb{0.50,0.50,0.50}%
      \put(2437,391){\makebox(0,0){\strut{}10\textsuperscript{2}}}%
      \colorrgb{0.50,0.50,0.50}%
      \put(3189,391){\makebox(0,0){\strut{}10\textsuperscript{3}}}%
      \colorrgb{0.50,0.50,0.50}%
      \put(3940,391){\makebox(0,0){\strut{}10\textsuperscript{4}}}%
      \colorrgb{0.50,0.50,0.50}%
      \put(4692,391){\makebox(0,0){\strut{}10\textsuperscript{5}}}%
    }%
    \gplgaddtomacro\gplfronttext{%
      \csname LTb\endcsname%
      \put(133,2675){\rotatebox{-270}{\makebox(0,0){\strut{}Empirical Complementary CDF}}}%
      \csname LTb\endcsname%
      \put(2813,124){\makebox(0,0){\strut{}Edge Weight (\# of Retweets)}}%
      \csname LTb\endcsname%
      \put(3998,4585){\makebox(0,0)[r]{\strut{}Data}}%
      \csname LTb\endcsname%
      \put(3998,4336){\makebox(0,0)[r]{\strut{}Power Law Fit}}%
    }%
    \gplgaddtomacro\gplbacktext{%
      \colorrgb{0.50,0.50,0.50}%
      \put(1473,942){\makebox(0,0){\fns 0.0\ \ \ }}%
      \colorrgb{0.50,0.50,0.50}%
      \put(1473,1259){\makebox(0,0){\fns 0.2\ \ \ }}%
      \colorrgb{0.50,0.50,0.50}%
      \put(1473,1577){\makebox(0,0){\fns 0.4\ \ \ }}%
      \colorrgb{0.50,0.50,0.50}%
      \put(1473,1894){\makebox(0,0){\fns 0.6\ \ \ }}%
      \colorrgb{0.50,0.50,0.50}%
      \put(1473,2212){\makebox(0,0){\fns 0.8\ \ \ }}%
      \colorrgb{0.50,0.50,0.50}%
      \put(1473,2529){\makebox(0,0){\fns 1.0\ \ \ }}%
      \colorrgb{0.50,0.50,0.50}%
      \put(1562,764){\makebox(0,0){\fns 10\textsuperscript{0}\ \ \ }}%
      \colorrgb{0.50,0.50,0.50}%
      \put(2091,764){\makebox(0,0){\fns 10\textsuperscript{1}\ \ \ }}%
      \colorrgb{0.50,0.50,0.50}%
      \put(2619,764){\makebox(0,0){\fns 10\textsuperscript{2}\ \ \ }}%
      \colorrgb{0.50,0.50,0.50}%
      \put(3148,764){\makebox(0,0){\fns 10\textsuperscript{3}\ \ \ }}%
    }%
    \gplgaddtomacro\gplfronttext{%
      \csname LTb\endcsname%
      \put(1206,1735){\rotatebox{-270}{\makebox(0,0){\fns Empirical CDF}}}%
    }%
    \gplbacktext
    \put(0,0){\includegraphics{retweet-edge-weights}}%
    \gplfronttext
  \end{picture}%
\endgroup

%% file: fig/retweet-degrees.pdf_tex
% GNUPLOT: LaTeX picture with Postscript
\begingroup
  \makeatletter
  \providecommand\color[2][]{%
    \GenericError{(gnuplot) \space\space\space\@spaces}{%
      Package color not loaded in conjunction with
      terminal option `colourtext'%
    }{See the gnuplot documentation for explanation.%
    }{Either use 'blacktext' in gnuplot or load the package
      color.sty in LaTeX.}%
    \renewcommand\color[2][]{}%
  }%
  \providecommand\includegraphics[2][]{%
    \GenericError{(gnuplot) \space\space\space\@spaces}{%
      Package graphicx or graphics not loaded%
    }{See the gnuplot documentation for explanation.%
    }{The gnuplot epslatex terminal needs graphicx.sty or graphics.sty.}%
    \renewcommand\includegraphics[2][]{}%
  }%
  \providecommand\rotatebox[2]{#2}%
  \@ifundefined{ifGPcolor}{%
    \newif\ifGPcolor
    \GPcolortrue
  }{}%
  \@ifundefined{ifGPblacktext}{%
    \newif\ifGPblacktext
    \GPblacktextfalse
  }{}%
  % define a \g@addto@macro without @ in the name:
  \let\gplgaddtomacro\g@addto@macro
  % define empty templates for all commands taking text:
  \gdef\gplbacktext{}%
  \gdef\gplfronttext{}%
  \makeatother
  \ifGPblacktext
    % no textcolor at all
    \def\colorrgb#1{}%
    \def\colorgray#1{}%
  \else
    % gray or color?
    \ifGPcolor
      \def\colorrgb#1{\color[rgb]{#1}}%
      \def\colorgray#1{\color[gray]{#1}}%
      \expandafter\def\csname LTw\endcsname{\color{white}}%
      \expandafter\def\csname LTb\endcsname{\color{black}}%
      \expandafter\def\csname LTa\endcsname{\color{black}}%
      \expandafter\def\csname LT0\endcsname{\color[rgb]{1,0,0}}%
      \expandafter\def\csname LT1\endcsname{\color[rgb]{0,1,0}}%
      \expandafter\def\csname LT2\endcsname{\color[rgb]{0,0,1}}%
      \expandafter\def\csname LT3\endcsname{\color[rgb]{1,0,1}}%
      \expandafter\def\csname LT4\endcsname{\color[rgb]{0,1,1}}%
      \expandafter\def\csname LT5\endcsname{\color[rgb]{1,1,0}}%
      \expandafter\def\csname LT6\endcsname{\color[rgb]{0,0,0}}%
      \expandafter\def\csname LT7\endcsname{\color[rgb]{1,0.3,0}}%
      \expandafter\def\csname LT8\endcsname{\color[rgb]{0.5,0.5,0.5}}%
    \else
      % gray
      \def\colorrgb#1{\color{black}}%
      \def\colorgray#1{\color[gray]{#1}}%
      \expandafter\def\csname LTw\endcsname{\color{white}}%
      \expandafter\def\csname LTb\endcsname{\color{black}}%
      \expandafter\def\csname LTa\endcsname{\color{black}}%
      \expandafter\def\csname LT0\endcsname{\color{black}}%
      \expandafter\def\csname LT1\endcsname{\color{black}}%
      \expandafter\def\csname LT2\endcsname{\color{black}}%
      \expandafter\def\csname LT3\endcsname{\color{black}}%
      \expandafter\def\csname LT4\endcsname{\color{black}}%
      \expandafter\def\csname LT5\endcsname{\color{black}}%
      \expandafter\def\csname LT6\endcsname{\color{black}}%
      \expandafter\def\csname LT7\endcsname{\color{black}}%
      \expandafter\def\csname LT8\endcsname{\color{black}}%
    \fi
  \fi
    \setlength{\unitlength}{0.0500bp}%
    \ifx\gptboxheight\undefined%
      \newlength{\gptboxheight}%
      \newlength{\gptboxwidth}%
      \newsavebox{\gptboxtext}%
    \fi%
    \setlength{\fboxrule}{0.5pt}%
    \setlength{\fboxsep}{1pt}%
\begin{picture}(4960.00,4960.00)%
    \gplgaddtomacro\gplbacktext{%
      \colorrgb{0.50,0.50,0.50}%
      \put(845,893){\makebox(0,0)[r]{\strut{}10\textsuperscript{-12}}}%
      \colorrgb{0.50,0.50,0.50}%
      \put(845,1541){\makebox(0,0)[r]{\strut{}10\textsuperscript{-10}}}%
      \colorrgb{0.50,0.50,0.50}%
      \put(845,2189){\makebox(0,0)[r]{\strut{}10\textsuperscript{-8}}}%
      \colorrgb{0.50,0.50,0.50}%
      \put(845,2837){\makebox(0,0)[r]{\strut{}10\textsuperscript{-6}}}%
      \colorrgb{0.50,0.50,0.50}%
      \put(845,3485){\makebox(0,0)[r]{\strut{}10\textsuperscript{-4}}}%
      \colorrgb{0.50,0.50,0.50}%
      \put(845,4133){\makebox(0,0)[r]{\strut{}10\textsuperscript{-2}}}%
      \colorrgb{0.50,0.50,0.50}%
      \put(845,4781){\makebox(0,0)[r]{\strut{}10\textsuperscript{0}}}%
      \colorrgb{0.50,0.50,0.50}%
      \put(845,569){\makebox(0,0)[r]{\strut{}}}%
      \colorrgb{0.50,0.50,0.50}%
      \put(845,1217){\makebox(0,0)[r]{\strut{}}}%
      \colorrgb{0.50,0.50,0.50}%
      \put(845,1865){\makebox(0,0)[r]{\strut{}}}%
      \colorrgb{0.50,0.50,0.50}%
      \put(845,2513){\makebox(0,0)[r]{\strut{}}}%
      \colorrgb{0.50,0.50,0.50}%
      \put(845,3161){\makebox(0,0)[r]{\strut{}}}%
      \colorrgb{0.50,0.50,0.50}%
      \put(845,3809){\makebox(0,0)[r]{\strut{}}}%
      \colorrgb{0.50,0.50,0.50}%
      \put(845,4457){\makebox(0,0)[r]{\strut{}}}%
      \colorrgb{0.50,0.50,0.50}%
      \put(934,391){\makebox(0,0){\strut{}10\textsuperscript{0}}}%
      \colorrgb{0.50,0.50,0.50}%
      \put(1874,391){\makebox(0,0){\strut{}10\textsuperscript{2}}}%
      \colorrgb{0.50,0.50,0.50}%
      \put(2813,391){\makebox(0,0){\strut{}10\textsuperscript{4}}}%
      \colorrgb{0.50,0.50,0.50}%
      \put(3752,391){\makebox(0,0){\strut{}10\textsuperscript{6}}}%
      \colorrgb{0.50,0.50,0.50}%
      \put(4692,391){\makebox(0,0){\strut{}10\textsuperscript{8}}}%
      \colorrgb{0.50,0.50,0.50}%
      \put(1404,391){\makebox(0,0){\strut{}}}%
      \colorrgb{0.50,0.50,0.50}%
      \put(2343,391){\makebox(0,0){\strut{}}}%
      \colorrgb{0.50,0.50,0.50}%
      \put(3283,391){\makebox(0,0){\strut{}}}%
      \colorrgb{0.50,0.50,0.50}%
      \put(4222,391){\makebox(0,0){\strut{}}}%
    }%
    \gplgaddtomacro\gplfronttext{%
      \csname LTb\endcsname%
      \put(133,2675){\rotatebox{-270}{\makebox(0,0){\strut{}CCDF / PMF}}}%
      \csname LTb\endcsname%
      \put(2813,124){\makebox(0,0){\strut{}In/Out Degree (\# of Retweeters/Retweetees)}}%
      \csname LTb\endcsname%
      \put(1628,1511){\makebox(0,0)[l]{\strut{}In (CCDF)}}%
      \csname LTb\endcsname%
      \put(1628,1262){\makebox(0,0)[l]{\strut{}Out (CCDF)}}%
      \csname LTb\endcsname%
      \put(1628,1013){\makebox(0,0)[l]{\strut{}In (PMF)}}%
      \csname LTb\endcsname%
      \put(1628,764){\makebox(0,0)[l]{\strut{}Out (PMF)}}%
    }%
    \gplbacktext
    \put(0,0){\includegraphics{retweet-degrees}}%
    \gplfronttext
  \end{picture}%
\endgroup

%% file: fig/retweet-average-path-length.pdf_tex
% GNUPLOT: LaTeX picture with Postscript
\begingroup
  \makeatletter
  \providecommand\color[2][]{%
    \GenericError{(gnuplot) \space\space\space\@spaces}{%
      Package color not loaded in conjunction with
      terminal option `colourtext'%
    }{See the gnuplot documentation for explanation.%
    }{Either use 'blacktext' in gnuplot or load the package
      color.sty in LaTeX.}%
    \renewcommand\color[2][]{}%
  }%
  \providecommand\includegraphics[2][]{%
    \GenericError{(gnuplot) \space\space\space\@spaces}{%
      Package graphicx or graphics not loaded%
    }{See the gnuplot documentation for explanation.%
    }{The gnuplot epslatex terminal needs graphicx.sty or graphics.sty.}%
    \renewcommand\includegraphics[2][]{}%
  }%
  \providecommand\rotatebox[2]{#2}%
  \@ifundefined{ifGPcolor}{%
    \newif\ifGPcolor
    \GPcolortrue
  }{}%
  \@ifundefined{ifGPblacktext}{%
    \newif\ifGPblacktext
    \GPblacktextfalse
  }{}%
  % define a \g@addto@macro without @ in the name:
  \let\gplgaddtomacro\g@addto@macro
  % define empty templates for all commands taking text:
  \gdef\gplbacktext{}%
  \gdef\gplfronttext{}%
  \makeatother
  \ifGPblacktext
    % no textcolor at all
    \def\colorrgb#1{}%
    \def\colorgray#1{}%
  \else
    % gray or color?
    \ifGPcolor
      \def\colorrgb#1{\color[rgb]{#1}}%
      \def\colorgray#1{\color[gray]{#1}}%
      \expandafter\def\csname LTw\endcsname{\color{white}}%
      \expandafter\def\csname LTb\endcsname{\color{black}}%
      \expandafter\def\csname LTa\endcsname{\color{black}}%
      \expandafter\def\csname LT0\endcsname{\color[rgb]{1,0,0}}%
      \expandafter\def\csname LT1\endcsname{\color[rgb]{0,1,0}}%
      \expandafter\def\csname LT2\endcsname{\color[rgb]{0,0,1}}%
      \expandafter\def\csname LT3\endcsname{\color[rgb]{1,0,1}}%
      \expandafter\def\csname LT4\endcsname{\color[rgb]{0,1,1}}%
      \expandafter\def\csname LT5\endcsname{\color[rgb]{1,1,0}}%
      \expandafter\def\csname LT6\endcsname{\color[rgb]{0,0,0}}%
      \expandafter\def\csname LT7\endcsname{\color[rgb]{1,0.3,0}}%
      \expandafter\def\csname LT8\endcsname{\color[rgb]{0.5,0.5,0.5}}%
    \else
      % gray
      \def\colorrgb#1{\color{black}}%
      \def\colorgray#1{\color[gray]{#1}}%
      \expandafter\def\csname LTw\endcsname{\color{white}}%
      \expandafter\def\csname LTb\endcsname{\color{black}}%
      \expandafter\def\csname LTa\endcsname{\color{black}}%
      \expandafter\def\csname LT0\endcsname{\color{black}}%
      \expandafter\def\csname LT1\endcsname{\color{black}}%
      \expandafter\def\csname LT2\endcsname{\color{black}}%
      \expandafter\def\csname LT3\endcsname{\color{black}}%
      \expandafter\def\csname LT4\endcsname{\color{black}}%
      \expandafter\def\csname LT5\endcsname{\color{black}}%
      \expandafter\def\csname LT6\endcsname{\color{black}}%
      \expandafter\def\csname LT7\endcsname{\color{black}}%
      \expandafter\def\csname LT8\endcsname{\color{black}}%
    \fi
  \fi
    \setlength{\unitlength}{0.0500bp}%
    \ifx\gptboxheight\undefined%
      \newlength{\gptboxheight}%
      \newlength{\gptboxwidth}%
      \newsavebox{\gptboxtext}%
    \fi%
    \setlength{\fboxrule}{0.5pt}%
    \setlength{\fboxsep}{1pt}%
\begin{picture}(4960.00,3060.00)%
    \gplgaddtomacro\gplbacktext{%
      \colorrgb{0.50,0.50,0.50}%
      \put(667,569){\makebox(0,0)[r]{\strut{}$0$}}%
      \colorrgb{0.50,0.50,0.50}%
      \put(667,800){\makebox(0,0)[r]{\strut{}$0.05$}}%
      \colorrgb{0.50,0.50,0.50}%
      \put(667,1031){\makebox(0,0)[r]{\strut{}$0.1$}}%
      \colorrgb{0.50,0.50,0.50}%
      \put(667,1263){\makebox(0,0)[r]{\strut{}$0.15$}}%
      \colorrgb{0.50,0.50,0.50}%
      \put(667,1494){\makebox(0,0)[r]{\strut{}$0.2$}}%
      \colorrgb{0.50,0.50,0.50}%
      \put(667,1725){\makebox(0,0)[r]{\strut{}$0.25$}}%
      \colorrgb{0.50,0.50,0.50}%
      \put(667,1956){\makebox(0,0)[r]{\strut{}$0.3$}}%
      \colorrgb{0.50,0.50,0.50}%
      \put(667,2187){\makebox(0,0)[r]{\strut{}$0.35$}}%
      \colorrgb{0.50,0.50,0.50}%
      \put(667,2419){\makebox(0,0)[r]{\strut{}$0.4$}}%
      \colorrgb{0.50,0.50,0.50}%
      \put(667,2650){\makebox(0,0)[r]{\strut{}$0.45$}}%
      \colorrgb{0.50,0.50,0.50}%
      \put(667,2881){\makebox(0,0)[r]{\strut{}$0.5$}}%
      \colorrgb{0.50,0.50,0.50}%
      \put(756,356){\makebox(0,0){\strut{}$0$}}%
      \colorrgb{0.50,0.50,0.50}%
      \put(1631,356){\makebox(0,0){\strut{}$10$}}%
      \colorrgb{0.50,0.50,0.50}%
      \put(2505,356){\makebox(0,0){\strut{}$20$}}%
      \colorrgb{0.50,0.50,0.50}%
      \put(3380,356){\makebox(0,0){\strut{}$30$}}%
      \colorrgb{0.50,0.50,0.50}%
      \put(4255,356){\makebox(0,0){\strut{}$40$}}%
    }%
    \gplgaddtomacro\gplfronttext{%
      \csname LTb\endcsname%
      \put(133,1725){\rotatebox{-270}{\makebox(0,0){\strut{}PMF}}}%
      \csname LTb\endcsname%
      \put(2724,124){\makebox(0,0){\strut{}Length of Path from Seed}}%
      \csname LTb\endcsname%
      \put(3998,2721){\makebox(0,0)[r]{\strut{}1000 seeds}}%
      \csname LTb\endcsname%
      \put(3998,2543){\makebox(0,0)[r]{\strut{}5000 seeds}}%
      \csname LTb\endcsname%
      \put(3998,2365){\makebox(0,0)[r]{\strut{}$\ln\mathcal{N}(1.96,0.19^2)$}}%
      \csname LTb\endcsname%
      \put(3998,2187){\makebox(0,0)[r]{\strut{}Full Graph (est.)}}%
    }%
    \gplgaddtomacro\gplbacktext{%
      \colorrgb{0.50,0.50,0.50}%
      \put(2242,887){\makebox(0,0){\fns 0.0\ \ }}%
      \colorrgb{0.50,0.50,0.50}%
      \put(2242,1104){\makebox(0,0){\fns 0.2\ \ }}%
      \colorrgb{0.50,0.50,0.50}%
      \put(2242,1321){\makebox(0,0){\fns 0.4\ \ }}%
      \colorrgb{0.50,0.50,0.50}%
      \put(2242,1539){\makebox(0,0){\fns 0.6\ \ }}%
      \colorrgb{0.50,0.50,0.50}%
      \put(2242,1756){\makebox(0,0){\fns 0.8\ \ }}%
      \colorrgb{0.50,0.50,0.50}%
      \put(2242,1973){\makebox(0,0){\fns 1.0\ \ }}%
      \colorrgb{0.50,0.50,0.50}%
      \put(2331,709){\makebox(0,0){\fns 0}}%
      \colorrgb{0.50,0.50,0.50}%
      \put(3175,709){\makebox(0,0){\fns 20}}%
      \colorrgb{0.50,0.50,0.50}%
      \put(4020,709){\makebox(0,0){\fns 40}}%
    }%
    \gplgaddtomacro\gplfronttext{%
      \csname LTb\endcsname%
      \put(1975,1430){\rotatebox{-270}{\makebox(0,0){\strut{}CDF}}}%
    }%
    \gplbacktext
    \put(0,0){\includegraphics{retweet-average-path-length}}%
    \gplfronttext
  \end{picture}%
\endgroup

%% file: fig/follower-graph-sampled-assortativity.pdf_tex
% GNUPLOT: LaTeX picture with Postscript
\begingroup
  \makeatletter
  \providecommand\color[2][]{%
    \GenericError{(gnuplot) \space\space\space\@spaces}{%
      Package color not loaded in conjunction with
      terminal option `colourtext'%
    }{See the gnuplot documentation for explanation.%
    }{Either use 'blacktext' in gnuplot or load the package
      color.sty in LaTeX.}%
    \renewcommand\color[2][]{}%
  }%
  \providecommand\includegraphics[2][]{%
    \GenericError{(gnuplot) \space\space\space\@spaces}{%
      Package graphicx or graphics not loaded%
    }{See the gnuplot documentation for explanation.%
    }{The gnuplot epslatex terminal needs graphicx.sty or graphics.sty.}%
    \renewcommand\includegraphics[2][]{}%
  }%
  \providecommand\rotatebox[2]{#2}%
  \@ifundefined{ifGPcolor}{%
    \newif\ifGPcolor
    \GPcolortrue
  }{}%
  \@ifundefined{ifGPblacktext}{%
    \newif\ifGPblacktext
    \GPblacktextfalse
  }{}%
  % define a \g@addto@macro without @ in the name:
  \let\gplgaddtomacro\g@addto@macro
  % define empty templates for all commands taking text:
  \gdef\gplbacktext{}%
  \gdef\gplfronttext{}%
  \makeatother
  \ifGPblacktext
    % no textcolor at all
    \def\colorrgb#1{}%
    \def\colorgray#1{}%
  \else
    % gray or color?
    \ifGPcolor
      \def\colorrgb#1{\color[rgb]{#1}}%
      \def\colorgray#1{\color[gray]{#1}}%
      \expandafter\def\csname LTw\endcsname{\color{white}}%
      \expandafter\def\csname LTb\endcsname{\color{black}}%
      \expandafter\def\csname LTa\endcsname{\color{black}}%
      \expandafter\def\csname LT0\endcsname{\color[rgb]{1,0,0}}%
      \expandafter\def\csname LT1\endcsname{\color[rgb]{0,1,0}}%
      \expandafter\def\csname LT2\endcsname{\color[rgb]{0,0,1}}%
      \expandafter\def\csname LT3\endcsname{\color[rgb]{1,0,1}}%
      \expandafter\def\csname LT4\endcsname{\color[rgb]{0,1,1}}%
      \expandafter\def\csname LT5\endcsname{\color[rgb]{1,1,0}}%
      \expandafter\def\csname LT6\endcsname{\color[rgb]{0,0,0}}%
      \expandafter\def\csname LT7\endcsname{\color[rgb]{1,0.3,0}}%
      \expandafter\def\csname LT8\endcsname{\color[rgb]{0.5,0.5,0.5}}%
    \else
      % gray
      \def\colorrgb#1{\color{black}}%
      \def\colorgray#1{\color[gray]{#1}}%
      \expandafter\def\csname LTw\endcsname{\color{white}}%
      \expandafter\def\csname LTb\endcsname{\color{black}}%
      \expandafter\def\csname LTa\endcsname{\color{black}}%
      \expandafter\def\csname LT0\endcsname{\color{black}}%
      \expandafter\def\csname LT1\endcsname{\color{black}}%
      \expandafter\def\csname LT2\endcsname{\color{black}}%
      \expandafter\def\csname LT3\endcsname{\color{black}}%
      \expandafter\def\csname LT4\endcsname{\color{black}}%
      \expandafter\def\csname LT5\endcsname{\color{black}}%
      \expandafter\def\csname LT6\endcsname{\color{black}}%
      \expandafter\def\csname LT7\endcsname{\color{black}}%
      \expandafter\def\csname LT8\endcsname{\color{black}}%
    \fi
  \fi
    \setlength{\unitlength}{0.0500bp}%
    \ifx\gptboxheight\undefined%
      \newlength{\gptboxheight}%
      \newlength{\gptboxwidth}%
      \newsavebox{\gptboxtext}%
    \fi%
    \setlength{\fboxrule}{0.5pt}%
    \setlength{\fboxsep}{1pt}%
\begin{picture}(4960.00,3060.00)%
    \gplgaddtomacro\gplbacktext{%
      \colorrgb{0.50,0.50,0.50}%
      \put(756,569){\makebox(0,0)[r]{\strut{}$-0.06$}}%
      \colorrgb{0.50,0.50,0.50}%
      \put(756,1113){\makebox(0,0)[r]{\strut{}$-0.04$}}%
      \colorrgb{0.50,0.50,0.50}%
      \put(756,1657){\makebox(0,0)[r]{\strut{}$-0.02$}}%
      \colorrgb{0.50,0.50,0.50}%
      \put(756,2201){\makebox(0,0)[r]{\strut{}$0$}}%
      \colorrgb{0.50,0.50,0.50}%
      \put(756,2745){\makebox(0,0)[r]{\strut{}$0.02$}}%
      \colorrgb{0.50,0.50,0.50}%
      \put(845,391){\makebox(0,0){\strut{}$0$}}%
      \colorrgb{0.50,0.50,0.50}%
      \put(1211,391){\makebox(0,0){\strut{}$0.1$}}%
      \colorrgb{0.50,0.50,0.50}%
      \put(1578,391){\makebox(0,0){\strut{}$0.2$}}%
      \colorrgb{0.50,0.50,0.50}%
      \put(1944,391){\makebox(0,0){\strut{}$0.3$}}%
      \colorrgb{0.50,0.50,0.50}%
      \put(2311,391){\makebox(0,0){\strut{}$0.4$}}%
      \colorrgb{0.50,0.50,0.50}%
      \put(2677,391){\makebox(0,0){\strut{}$0.5$}}%
      \colorrgb{0.50,0.50,0.50}%
      \put(3043,391){\makebox(0,0){\strut{}$0.6$}}%
      \colorrgb{0.50,0.50,0.50}%
      \put(3410,391){\makebox(0,0){\strut{}$0.7$}}%
      \colorrgb{0.50,0.50,0.50}%
      \put(3776,391){\makebox(0,0){\strut{}$0.8$}}%
      \colorrgb{0.50,0.50,0.50}%
      \put(4142,391){\makebox(0,0){\strut{}$0.9$}}%
      \colorrgb{0.50,0.50,0.50}%
      \put(4509,391){\makebox(0,0){\strut{}$1$}}%
    }%
    \gplgaddtomacro\gplfronttext{%
      \csname LTb\endcsname%
      \put(133,1725){\rotatebox{-270}{\makebox(0,0){\strut{}$r$ (Assortativity)}}}%
      \csname LTb\endcsname%
      \put(2768,124){\makebox(0,0){\strut{}$\alpha$ (Link Sampling Rate)}}%
      \csname LTb\endcsname%
      \put(2163,2712){\makebox(0,0)[r]{\strut{}$r(in,out)$}}%
      \csname LTb\endcsname%
      \put(2163,2517){\makebox(0,0)[r]{\strut{}$r(out,out)$}}%
      \csname LTb\endcsname%
      \put(3658,2712){\makebox(0,0)[r]{\strut{}$r(in,in)$}}%
      \csname LTb\endcsname%
      \put(3658,2517){\makebox(0,0)[r]{\strut{}$r(out,in)$}}%
    }%
    \gplbacktext
    \put(0,0){\includegraphics{follower-graph-sampled-assortativity}}%
    \gplfronttext
  \end{picture}%
\endgroup

%% file: fig/retweet-assortativity.pdf_tex
% GNUPLOT: LaTeX picture with Postscript
\begingroup
  \makeatletter
  \providecommand\color[2][]{%
    \GenericError{(gnuplot) \space\space\space\@spaces}{%
      Package color not loaded in conjunction with
      terminal option `colourtext'%
    }{See the gnuplot documentation for explanation.%
    }{Either use 'blacktext' in gnuplot or load the package
      color.sty in LaTeX.}%
    \renewcommand\color[2][]{}%
  }%
  \providecommand\includegraphics[2][]{%
    \GenericError{(gnuplot) \space\space\space\@spaces}{%
      Package graphicx or graphics not loaded%
    }{See the gnuplot documentation for explanation.%
    }{The gnuplot epslatex terminal needs graphicx.sty or graphics.sty.}%
    \renewcommand\includegraphics[2][]{}%
  }%
  \providecommand\rotatebox[2]{#2}%
  \@ifundefined{ifGPcolor}{%
    \newif\ifGPcolor
    \GPcolortrue
  }{}%
  \@ifundefined{ifGPblacktext}{%
    \newif\ifGPblacktext
    \GPblacktextfalse
  }{}%
  % define a \g@addto@macro without @ in the name:
  \let\gplgaddtomacro\g@addto@macro
  % define empty templates for all commands taking text:
  \gdef\gplbacktext{}%
  \gdef\gplfronttext{}%
  \makeatother
  \ifGPblacktext
    % no textcolor at all
    \def\colorrgb#1{}%
    \def\colorgray#1{}%
  \else
    % gray or color?
    \ifGPcolor
      \def\colorrgb#1{\color[rgb]{#1}}%
      \def\colorgray#1{\color[gray]{#1}}%
      \expandafter\def\csname LTw\endcsname{\color{white}}%
      \expandafter\def\csname LTb\endcsname{\color{black}}%
      \expandafter\def\csname LTa\endcsname{\color{black}}%
      \expandafter\def\csname LT0\endcsname{\color[rgb]{1,0,0}}%
      \expandafter\def\csname LT1\endcsname{\color[rgb]{0,1,0}}%
      \expandafter\def\csname LT2\endcsname{\color[rgb]{0,0,1}}%
      \expandafter\def\csname LT3\endcsname{\color[rgb]{1,0,1}}%
      \expandafter\def\csname LT4\endcsname{\color[rgb]{0,1,1}}%
      \expandafter\def\csname LT5\endcsname{\color[rgb]{1,1,0}}%
      \expandafter\def\csname LT6\endcsname{\color[rgb]{0,0,0}}%
      \expandafter\def\csname LT7\endcsname{\color[rgb]{1,0.3,0}}%
      \expandafter\def\csname LT8\endcsname{\color[rgb]{0.5,0.5,0.5}}%
    \else
      % gray
      \def\colorrgb#1{\color{black}}%
      \def\colorgray#1{\color[gray]{#1}}%
      \expandafter\def\csname LTw\endcsname{\color{white}}%
      \expandafter\def\csname LTb\endcsname{\color{black}}%
      \expandafter\def\csname LTa\endcsname{\color{black}}%
      \expandafter\def\csname LT0\endcsname{\color{black}}%
      \expandafter\def\csname LT1\endcsname{\color{black}}%
      \expandafter\def\csname LT2\endcsname{\color{black}}%
      \expandafter\def\csname LT3\endcsname{\color{black}}%
      \expandafter\def\csname LT4\endcsname{\color{black}}%
      \expandafter\def\csname LT5\endcsname{\color{black}}%
      \expandafter\def\csname LT6\endcsname{\color{black}}%
      \expandafter\def\csname LT7\endcsname{\color{black}}%
      \expandafter\def\csname LT8\endcsname{\color{black}}%
    \fi
  \fi
    \setlength{\unitlength}{0.0500bp}%
    \ifx\gptboxheight\undefined%
      \newlength{\gptboxheight}%
      \newlength{\gptboxwidth}%
      \newsavebox{\gptboxtext}%
    \fi%
    \setlength{\fboxrule}{0.5pt}%
    \setlength{\fboxsep}{1pt}%
\begin{picture}(4960.00,3060.00)%
    \gplgaddtomacro\gplbacktext{%
      \colorrgb{0.50,0.50,0.50}%
      \put(756,356){\makebox(0,0)[r]{\strut{}$-0.06$}}%
      \colorrgb{0.50,0.50,0.50}%
      \put(756,777){\makebox(0,0)[r]{\strut{}$-0.04$}}%
      \colorrgb{0.50,0.50,0.50}%
      \put(756,1198){\makebox(0,0)[r]{\strut{}$-0.02$}}%
      \colorrgb{0.50,0.50,0.50}%
      \put(756,1619){\makebox(0,0)[r]{\strut{}$0$}}%
      \colorrgb{0.50,0.50,0.50}%
      \put(756,2039){\makebox(0,0)[r]{\strut{}$0.02$}}%
      \colorrgb{0.50,0.50,0.50}%
      \put(756,2460){\makebox(0,0)[r]{\strut{}$0.04$}}%
      \colorrgb{0.50,0.50,0.50}%
      \put(756,2881){\makebox(0,0)[r]{\strut{}$0.06$}}%
      \colorrgb{0.50,0.50,0.50}%
      \put(1326,178){\makebox(0,0){\strut{}$r(in,in)$}}%
      \colorrgb{0.50,0.50,0.50}%
      \put(2288,178){\makebox(0,0){\strut{}$r(in,out)$}}%
      \colorrgb{0.50,0.50,0.50}%
      \put(3249,178){\makebox(0,0){\strut{}$r(out,in)$}}%
      \colorrgb{0.50,0.50,0.50}%
      \put(4211,178){\makebox(0,0){\strut{}$r(out,out)$}}%
    }%
    \gplgaddtomacro\gplfronttext{%
      \csname LTb\endcsname%
      \put(133,1618){\rotatebox{-270}{\makebox(0,0){\strut{}$r$ (Assortativity)}}}%
      \csname LTb\endcsname%
      \put(1539,2712){\makebox(0,0)[l]{\strut{}Retweet Graph}}%
      \colorrgb{0.77,0.09,0.10}%
      \put(1699,1466){\makebox(0,0){\strut{}\fns{-0.006}}}%
      \colorrgb{0.77,0.09,0.10}%
      \put(2661,1754){\makebox(0,0){\strut{}\fns{0.0077}}}%
      \colorrgb{0.77,0.09,0.10}%
      \put(3622,778){\makebox(0,0){\strut{}\fns{-0.039}}}%
      \colorrgb{0.77,0.09,0.10}%
      \put(4584,2501){\makebox(0,0){\strut{}\fns{0.043}}}%
      \csname LTb\endcsname%
      \put(1539,2517){\makebox(0,0)[l]{\strut{}Following Graph}}%
      \colorrgb{0.18,0.56,0.80}%
      \put(1699,958){\makebox(0,0){\strut{}\fns{-0.03}}}%
      \colorrgb{0.18,0.56,0.80}%
      \put(2661,1406){\makebox(0,0){\strut{}\fns{-0.0089}}}%
      \colorrgb{0.18,0.56,0.80}%
      \put(3622,527){\makebox(0,0){\strut{}\fns{-0.051}}}%
      \colorrgb{0.18,0.56,0.80}%
      \put(4584,1337){\makebox(0,0){\strut{}\fns{-0.012}}}%
    }%
    \gplbacktext
    \put(0,0){\includegraphics{retweet-assortativity}}%
    \gplfronttext
  \end{picture}%
\endgroup

%% file: fig/cluster-triangle-types.pdf_tex
%% Creator: Inkscape inkscape 0.48.4, www.inkscape.org
%% Accompanies image file 'cluster-triangle-types.pdf' (pdf, eps, ps)
\begingroup%
  \makeatletter%
  \providecommand\color[2][]{%
    \errmessage{(Inkscape) Color is used for the text in Inkscape, but the package 'color.sty' is not loaded}%
    \renewcommand\color[2][]{}%
  }%
  \providecommand\transparent[1]{%
    \errmessage{(Inkscape) Transparency is used (non-zero) for the text in Inkscape, but the package 'transparent.sty' is not loaded}%
    \renewcommand\transparent[1]{}%
  }%
  \providecommand\rotatebox[2]{#2}%
  \ifx\svgwidth\undefined%
    \setlength{\unitlength}{251.28000488bp}%
    \ifx\svgscale\undefined%
      \relax%
    \else%
      \setlength{\unitlength}{\unitlength * \real{\svgscale}}%
    \fi%
  \else%
    \setlength{\unitlength}{\svgwidth}%
  \fi%
  \global\let\svgwidth\undefined%
  \global\let\svgscale\undefined%
  \makeatother%
  \begin{picture}(1,0.2722063)%
    \put(0,0){\includegraphics[width=\unitlength]{cluster-triangle-types.pdf}}%
    \put(0.12655121,0.21087488){\color[rgb]{0,0,0}\makebox(0,0)[b]{\smash{$i$}}}%
    \put(0.37590643,0.21087488){\color[rgb]{0,0,0}\makebox(0,0)[b]{\smash{$i$}}}%
    \put(0.62526166,0.21087488){\color[rgb]{0,0,0}\makebox(0,0)[b]{\smash{$i$}}}%
    \put(0.87461689,0.21087488){\color[rgb]{0,0,0}\makebox(0,0)[b]{\smash{$i$}}}%
    \put(0.12584279,0.0263355){\color[rgb]{0,0,0}\makebox(0,0)[b]{\smash{Cycle}}}%
    \put(0.37524466,0.0263355){\color[rgb]{0,0,0}\makebox(0,0)[b]{\smash{Middleman}}}%
    \put(0.62429675,0.0263355){\color[rgb]{0,0,0}\makebox(0,0)[b]{\smash{In}}}%
    \put(0.87337217,0.0263355){\color[rgb]{0,0,0}\makebox(0,0)[b]{\smash{Out}}}%
  \end{picture}%
\endgroup%

%% file: fig/follower-graph-sampled-clustering-coefficient.pdf_tex
% GNUPLOT: LaTeX picture with Postscript
\begingroup
  \makeatletter
  \providecommand\color[2][]{%
    \GenericError{(gnuplot) \space\space\space\@spaces}{%
      Package color not loaded in conjunction with
      terminal option `colourtext'%
    }{See the gnuplot documentation for explanation.%
    }{Either use 'blacktext' in gnuplot or load the package
      color.sty in LaTeX.}%
    \renewcommand\color[2][]{}%
  }%
  \providecommand\includegraphics[2][]{%
    \GenericError{(gnuplot) \space\space\space\@spaces}{%
      Package graphicx or graphics not loaded%
    }{See the gnuplot documentation for explanation.%
    }{The gnuplot epslatex terminal needs graphicx.sty or graphics.sty.}%
    \renewcommand\includegraphics[2][]{}%
  }%
  \providecommand\rotatebox[2]{#2}%
  \@ifundefined{ifGPcolor}{%
    \newif\ifGPcolor
    \GPcolortrue
  }{}%
  \@ifundefined{ifGPblacktext}{%
    \newif\ifGPblacktext
    \GPblacktextfalse
  }{}%
  % define a \g@addto@macro without @ in the name:
  \let\gplgaddtomacro\g@addto@macro
  % define empty templates for all commands taking text:
  \gdef\gplbacktext{}%
  \gdef\gplfronttext{}%
  \makeatother
  \ifGPblacktext
    % no textcolor at all
    \def\colorrgb#1{}%
    \def\colorgray#1{}%
  \else
    % gray or color?
    \ifGPcolor
      \def\colorrgb#1{\color[rgb]{#1}}%
      \def\colorgray#1{\color[gray]{#1}}%
      \expandafter\def\csname LTw\endcsname{\color{white}}%
      \expandafter\def\csname LTb\endcsname{\color{black}}%
      \expandafter\def\csname LTa\endcsname{\color{black}}%
      \expandafter\def\csname LT0\endcsname{\color[rgb]{1,0,0}}%
      \expandafter\def\csname LT1\endcsname{\color[rgb]{0,1,0}}%
      \expandafter\def\csname LT2\endcsname{\color[rgb]{0,0,1}}%
      \expandafter\def\csname LT3\endcsname{\color[rgb]{1,0,1}}%
      \expandafter\def\csname LT4\endcsname{\color[rgb]{0,1,1}}%
      \expandafter\def\csname LT5\endcsname{\color[rgb]{1,1,0}}%
      \expandafter\def\csname LT6\endcsname{\color[rgb]{0,0,0}}%
      \expandafter\def\csname LT7\endcsname{\color[rgb]{1,0.3,0}}%
      \expandafter\def\csname LT8\endcsname{\color[rgb]{0.5,0.5,0.5}}%
    \else
      % gray
      \def\colorrgb#1{\color{black}}%
      \def\colorgray#1{\color[gray]{#1}}%
      \expandafter\def\csname LTw\endcsname{\color{white}}%
      \expandafter\def\csname LTb\endcsname{\color{black}}%
      \expandafter\def\csname LTa\endcsname{\color{black}}%
      \expandafter\def\csname LT0\endcsname{\color{black}}%
      \expandafter\def\csname LT1\endcsname{\color{black}}%
      \expandafter\def\csname LT2\endcsname{\color{black}}%
      \expandafter\def\csname LT3\endcsname{\color{black}}%
      \expandafter\def\csname LT4\endcsname{\color{black}}%
      \expandafter\def\csname LT5\endcsname{\color{black}}%
      \expandafter\def\csname LT6\endcsname{\color{black}}%
      \expandafter\def\csname LT7\endcsname{\color{black}}%
      \expandafter\def\csname LT8\endcsname{\color{black}}%
    \fi
  \fi
    \setlength{\unitlength}{0.0500bp}%
    \ifx\gptboxheight\undefined%
      \newlength{\gptboxheight}%
      \newlength{\gptboxwidth}%
      \newsavebox{\gptboxtext}%
    \fi%
    \setlength{\fboxrule}{0.5pt}%
    \setlength{\fboxsep}{1pt}%
\begin{picture}(4960.00,3060.00)%
    \gplgaddtomacro\gplbacktext{%
      \colorrgb{0.50,0.50,0.50}%
      \put(667,569){\makebox(0,0)[r]{\strut{}$0$}}%
      \colorrgb{0.50,0.50,0.50}%
      \put(667,1147){\makebox(0,0)[r]{\strut{}$0.01$}}%
      \colorrgb{0.50,0.50,0.50}%
      \put(667,1725){\makebox(0,0)[r]{\strut{}$0.02$}}%
      \colorrgb{0.50,0.50,0.50}%
      \put(667,2303){\makebox(0,0)[r]{\strut{}$0.03$}}%
      \colorrgb{0.50,0.50,0.50}%
      \put(667,2881){\makebox(0,0)[r]{\strut{}$0.04$}}%
      \colorrgb{0.50,0.50,0.50}%
      \put(756,391){\makebox(0,0){\strut{}$0$}}%
      \colorrgb{0.50,0.50,0.50}%
      \put(1131,391){\makebox(0,0){\strut{}$0.1$}}%
      \colorrgb{0.50,0.50,0.50}%
      \put(1506,391){\makebox(0,0){\strut{}$0.2$}}%
      \colorrgb{0.50,0.50,0.50}%
      \put(1881,391){\makebox(0,0){\strut{}$0.3$}}%
      \colorrgb{0.50,0.50,0.50}%
      \put(2255,391){\makebox(0,0){\strut{}$0.4$}}%
      \colorrgb{0.50,0.50,0.50}%
      \put(2630,391){\makebox(0,0){\strut{}$0.5$}}%
      \colorrgb{0.50,0.50,0.50}%
      \put(3005,391){\makebox(0,0){\strut{}$0.6$}}%
      \colorrgb{0.50,0.50,0.50}%
      \put(3380,391){\makebox(0,0){\strut{}$0.7$}}%
      \colorrgb{0.50,0.50,0.50}%
      \put(3755,391){\makebox(0,0){\strut{}$0.8$}}%
      \colorrgb{0.50,0.50,0.50}%
      \put(4130,391){\makebox(0,0){\strut{}$0.9$}}%
      \colorrgb{0.50,0.50,0.50}%
      \put(4505,391){\makebox(0,0){\strut{}$1$}}%
    }%
    \gplgaddtomacro\gplfronttext{%
      \csname LTb\endcsname%
      \put(133,1725){\rotatebox{-270}{\makebox(0,0){\strut{}$\frac{1}{\alpha}\overline{C}$}}}%
      \csname LTb\endcsname%
      \put(2724,124){\makebox(0,0){\strut{}$\alpha$ (Link Sampling Rate)}}%
      \csname LTb\endcsname%
      \put(2119,2712){\makebox(0,0)[r]{\strut{}Out}}%
      \csname LTb\endcsname%
      \put(2119,2517){\makebox(0,0)[r]{\strut{}Middleman}}%
      \csname LTb\endcsname%
      \put(3525,2712){\makebox(0,0)[r]{\strut{}Cycle}}%
      \csname LTb\endcsname%
      \put(3525,2517){\makebox(0,0)[r]{\strut{}In}}%
    }%
    \gplbacktext
    \put(0,0){\includegraphics{follower-graph-sampled-clustering-coefficient}}%
    \gplfronttext
  \end{picture}%
\endgroup

%% file: fig/clustering-coefficient.pdf_tex
% GNUPLOT: LaTeX picture with Postscript
\begingroup
  \makeatletter
  \providecommand\color[2][]{%
    \GenericError{(gnuplot) \space\space\space\@spaces}{%
      Package color not loaded in conjunction with
      terminal option `colourtext'%
    }{See the gnuplot documentation for explanation.%
    }{Either use 'blacktext' in gnuplot or load the package
      color.sty in LaTeX.}%
    \renewcommand\color[2][]{}%
  }%
  \providecommand\includegraphics[2][]{%
    \GenericError{(gnuplot) \space\space\space\@spaces}{%
      Package graphicx or graphics not loaded%
    }{See the gnuplot documentation for explanation.%
    }{The gnuplot epslatex terminal needs graphicx.sty or graphics.sty.}%
    \renewcommand\includegraphics[2][]{}%
  }%
  \providecommand\rotatebox[2]{#2}%
  \@ifundefined{ifGPcolor}{%
    \newif\ifGPcolor
    \GPcolortrue
  }{}%
  \@ifundefined{ifGPblacktext}{%
    \newif\ifGPblacktext
    \GPblacktextfalse
  }{}%
  % define a \g@addto@macro without @ in the name:
  \let\gplgaddtomacro\g@addto@macro
  % define empty templates for all commands taking text:
  \gdef\gplbacktext{}%
  \gdef\gplfronttext{}%
  \makeatother
  \ifGPblacktext
    % no textcolor at all
    \def\colorrgb#1{}%
    \def\colorgray#1{}%
  \else
    % gray or color?
    \ifGPcolor
      \def\colorrgb#1{\color[rgb]{#1}}%
      \def\colorgray#1{\color[gray]{#1}}%
      \expandafter\def\csname LTw\endcsname{\color{white}}%
      \expandafter\def\csname LTb\endcsname{\color{black}}%
      \expandafter\def\csname LTa\endcsname{\color{black}}%
      \expandafter\def\csname LT0\endcsname{\color[rgb]{1,0,0}}%
      \expandafter\def\csname LT1\endcsname{\color[rgb]{0,1,0}}%
      \expandafter\def\csname LT2\endcsname{\color[rgb]{0,0,1}}%
      \expandafter\def\csname LT3\endcsname{\color[rgb]{1,0,1}}%
      \expandafter\def\csname LT4\endcsname{\color[rgb]{0,1,1}}%
      \expandafter\def\csname LT5\endcsname{\color[rgb]{1,1,0}}%
      \expandafter\def\csname LT6\endcsname{\color[rgb]{0,0,0}}%
      \expandafter\def\csname LT7\endcsname{\color[rgb]{1,0.3,0}}%
      \expandafter\def\csname LT8\endcsname{\color[rgb]{0.5,0.5,0.5}}%
    \else
      % gray
      \def\colorrgb#1{\color{black}}%
      \def\colorgray#1{\color[gray]{#1}}%
      \expandafter\def\csname LTw\endcsname{\color{white}}%
      \expandafter\def\csname LTb\endcsname{\color{black}}%
      \expandafter\def\csname LTa\endcsname{\color{black}}%
      \expandafter\def\csname LT0\endcsname{\color{black}}%
      \expandafter\def\csname LT1\endcsname{\color{black}}%
      \expandafter\def\csname LT2\endcsname{\color{black}}%
      \expandafter\def\csname LT3\endcsname{\color{black}}%
      \expandafter\def\csname LT4\endcsname{\color{black}}%
      \expandafter\def\csname LT5\endcsname{\color{black}}%
      \expandafter\def\csname LT6\endcsname{\color{black}}%
      \expandafter\def\csname LT7\endcsname{\color{black}}%
      \expandafter\def\csname LT8\endcsname{\color{black}}%
    \fi
  \fi
    \setlength{\unitlength}{0.0500bp}%
    \ifx\gptboxheight\undefined%
      \newlength{\gptboxheight}%
      \newlength{\gptboxwidth}%
      \newsavebox{\gptboxtext}%
    \fi%
    \setlength{\fboxrule}{0.5pt}%
    \setlength{\fboxsep}{1pt}%
\begin{picture}(4960.00,3060.00)%
    \gplgaddtomacro\gplbacktext{%
      \colorrgb{0.50,0.50,0.50}%
      \put(578,356){\makebox(0,0)[r]{\strut{}$0$}}%
      \colorrgb{0.50,0.50,0.50}%
      \put(578,917){\makebox(0,0)[r]{\strut{}$0.1$}}%
      \colorrgb{0.50,0.50,0.50}%
      \put(578,1478){\makebox(0,0)[r]{\strut{}$0.2$}}%
      \colorrgb{0.50,0.50,0.50}%
      \put(578,2039){\makebox(0,0)[r]{\strut{}$0.3$}}%
      \colorrgb{0.50,0.50,0.50}%
      \put(578,2600){\makebox(0,0)[r]{\strut{}$0.4$}}%
      \colorrgb{0.50,0.50,0.50}%
      \put(1170,178){\makebox(0,0){\strut{}Cycle}}%
      \colorrgb{0.50,0.50,0.50}%
      \put(2176,178){\makebox(0,0){\strut{}Middleman}}%
      \colorrgb{0.50,0.50,0.50}%
      \put(3183,178){\makebox(0,0){\strut{}In}}%
      \colorrgb{0.50,0.50,0.50}%
      \put(4189,178){\makebox(0,0){\strut{}Out}}%
    }%
    \gplgaddtomacro\gplfronttext{%
      \csname LTb\endcsname%
      \put(133,1618){\rotatebox{-270}{\makebox(0,0){\strut{}$C$ (Clustering Coefficient)}}}%
      \csname LTb\endcsname%
      \put(1361,2712){\makebox(0,0)[l]{\strut{}Retweet Graph}}%
      \csname LTb\endcsname%
      \put(1361,2517){\makebox(0,0)[l]{\strut{}Following Graph}}%
      \colorrgb{0.77,0.09,0.10}%
      \put(1348,1000){\makebox(0,0)[l]{\strut{}\fns{0.12}}}%
      \colorrgb{0.77,0.09,0.10}%
      \put(2354,2110){\makebox(0,0)[l]{\strut{}\fns{0.32}}}%
      \colorrgb{0.77,0.09,0.10}%
      \put(4367,1955){\makebox(0,0)[l]{\strut{}\fns{0.29}}}%
      \colorrgb{0.77,0.09,0.10}%
      \put(3361,451){\makebox(0,0)[l]{\strut{}\fns{0.0011}}}%
      \colorrgb{0.18,0.56,0.80}%
      \put(1348,417){\makebox(0,0)[l]{\strut{}\fns{0.015}}}%
      \colorrgb{0.18,0.56,0.80}%
      \put(2354,423){\makebox(0,0)[l]{\strut{}\fns{0.017}}}%
      \colorrgb{0.18,0.56,0.80}%
      \put(4367,481){\makebox(0,0)[l]{\strut{}\fns{0.027}}}%
      \colorrgb{0.18,0.56,0.80}%
      \put(3361,245){\makebox(0,0)[l]{\strut{}\fns{0.00077}}}%
    }%
    \gplbacktext
    \put(0,0){\includegraphics{clustering-coefficient}}%
    \gplfronttext
  \end{picture}%
\endgroup

%% file: implications.tex
\section{Implications for the Design of Decentralized Microblogging
  Architectures}
\label{sec:implications}

The preceding sections characterized tweet behavior---total quantity,
average rate, and interevent time---and the retweet graph
structure. Although interesting in their own right, in this section we
discuss a particular application---the implications for the design of
performance-constrained, decentralized microblogging platforms like
Shout~\cite{shout}.  In such systems, bandwidth and energy---scarce
resources---must be carefully allocated to achieve some notion of
fairness.  We discuss implications for such allocation strategies.

\noindent
\textbf{Power Law Participation Momentum:} Most users quit after a few
contributions, so greedy allocation of resources to new users is
wasteful.  For example, a routing scheme prioritizing messages from
users with more contributions would implicitly direct bandwidth away
from temporary users.\footnote{We do not consider how malicious users
  might manipulate such schemes, but resistance to such attacks would
  be important for any practical protocol.}  The known power law form
of the momentum function suggests enables the design of optimal
allocation strategies. For example, consider storing old messages by
distributing them across participating nodes. Nodes with more
contributions are more reliable (less likely to leave the network) and
thus require a lower storage replication factor. These failure
probabilities can be easily modeled.

\noindent
\textbf{Heavy-Tailed Rate Distribution:} The two-phase tweet rate
distribution has implications for short-term message delivery and
long-term message storage.  The message generation rate may be modeled
as lognormal---messages are naturally better-distributed in the
network than a power law would suggest, reducing points of congestion
and better balancing bandwidth use.  In the long term, however, the
average tweet rates follow the asymptotic power law with its much
heavier tail, posing issues for archiving and retrieval of tweets.
For example, sharding messages across nodes by author will result in a
few nodes storing and serving the majority of the archived content.
The archiving system must be designed to handle the power law
distribution.

\noindent
\textbf{Heavy-tailed Interevent Distribution:} Simulations and other
performance analysis must use heavy-tail distributions for the
interevent times. Standard Poisson distributions will grossly
underpredict these times, increasing simulated congestion and
resulting in over-provisioned designs.

\noindent
\textbf{Small-World, Assortative, Clustered Retweet Graph:} In a
centralized platform, a single entity can moderate bad behavior,
reject spammers, and ensure fair division of resources. Participants
in a decentralized platform must perform these same tasks themselves
without implicit trust in others.  The implicit retweet graph seems to
encode some information about the real-world relationships of users
that could be inferred for such purposes.  The higher assortativity is
more indicative of a real world network than a social network and the
high clustering implies that users have some commonalities around
which they gather.  We explore this direction in the next section,
using spammer detection via connectivity in the retweet graph as an example.

%% file: spam.tex
\section{Leveraging the Retweet Graph for Spammer Detection}
\label{sec:spam}

Spam is a problem for many communication
platforms~\cite{thomas11nov,yang12apr}, but is particularly concerning
for decentralized, censorship-resistant microblogging platforms.
Twitter, as a centralized service, can decree what constitutes spam,
use its full knowledge of user behavior to detect
violators~\cite{benevenuto10jul,chen11apr,mccord11sep,song11sep,thomas12apr,wang10jul,yang11sep},
and limit the creation of new accounts. However, at its root such
filtering is a form of censorship.  In a censorship-less and
decentralized network, filtering must be applied individually and
locally.\footnote{Distributed filtering mechanisms can prevent spam
  from propagating through the network, but a local mechanism is
  needed for at least the first hop.}  We develop a detection approach
based on the structure of the retweet graph.

\subsection{Approach Overview and Background}
\label{sec:spam-overview}

Detection approaches can focus either on individual messages,
\emph{spam detection}, or on the sender, \emph{spammer detection}. The
latter is most applicable to microblogs, because the short message
lengths---less than 250 characters---make content analysis
difficult~\cite{benevenuto10jul}.  Spammer detection takes two forms
differentiated by the default presumption.  \emph{Blacklisting}
assumes that users are not spammers until proven otherwise, while
\emph{whitelisting} presumes the opposite.  The former is a
non-starter in registrar-less, decentralized networks~\cite{shout}
because blacklisted accounts are easily and cheaply replaced. Some
form of whitelisting is required.

Whitelisting presents its own issue.  Manual whitelisting, akin to
\emph{following} someone on Twitter, is time-consuming and prevents
previously-unacquainted users from connecting.  We develop a method
for automatically whitelisting users based on the intuition that
non-spammers will rarely retweet spammers. This approach is easily
bootstrapped by whitelisting just a few friends and, in decentralized
networks, requires only local information---retweets overhead from
neighbors. Blacklisting is used to block previously-good accounts that
have started sending spam. Becoming whitelisted requires some
effort---an account must generate content that others find worthy of
sharing---and thus such accounts are not quickly or easily replaced.

Many researchers have considered spam detection in
Twitter~\cite{benevenuto10jul,chen11apr,mccord11sep,song11sep,thomas12apr,wang10jul,yang11sep}. We
survey the two most relevant works here.

Benevenuto~\cite{benevenuto10jul} studied the classification
performance of 60 tweet and tweeter attributes, ranging from hashtags
per tweet to the ratio of followers to friends.  Aside from the
obvious inclusion of URLs and account age\footnote{Twitter actively
  removes spammer accounts, biasing the collected data.}, the most
sensitive attributes were related to social behavior---ratio of
followers to friends, number of replies to messages, etc.  Noting that
spammers can easily alter the content of tweets, they suggest focusing
on these harder-to-manipulate attributes for detection.  Their
proposed classifier has a 70\% true positive rate (TPR) and a 4\%
false positive rate (FPR).

Song, Lee, and Kim~\cite{song11sep} developed an approach based on the
followers graph that is similar to our proposal for the retweet graph.
In particular, they consider two metrics in the graph:
\emph{distance}---measured as the shortest path between two
nodes---and \emph{connectivity}---measured via max-flow and random
walk\footnote{The random walk metric is easily manipulated, because it
  requires treating the unidirectional edges as bidirectional,
  removing the asymmetry between spammers and non-spammers. We do not
  consider it further.}.  A classifier over these attributes had 95\%
TPR and 4\% FPR, while the inclusion of attributes like URLs per tweet
improved the performance to 99\% TPR and 1\% FPR.

\begin{figure}
  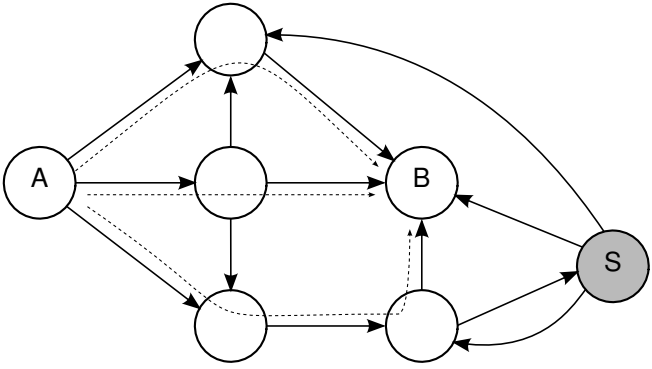
  \caption{Portion of a retweet graph showing how spammers are less
    connected. Non-spammer $B$ is connected to non-spammer $A$ by
    three independent paths, the shortest of which has length
    two. Spammer $S$ is connected by only a single length-three path.}
  \label{fig:retweet-graph-example}
\end{figure}

Decentralized systems may not include explicit social relationships
(e.g, Shout~\cite{shout}), so the followers graph cannot be used.
Instead, we consider the implicit retweet graph.  Intuitively, content
from spammers will not be heavily retweeted, and thus they will be
less connected to non-spammers in the graph, as illustrated in
\autoref{fig:retweet-graph-example}.  Node $A$ is connected to
non-spammer $B$ by three edge independent paths, the shortest of which
has length two.  Spammer $S$, on the other hand, is only connected via
a single path of length three.  $A$ separates spammers by classifying
nodes based on their distance from and edge-independent connectivities
(i.e., max-flow with unit-weighted edges) to itself.

This scheme can be incorporated as follows.  Each participant in the
system maintains his own partial\footnote{Only some messages sent by
  other will be heard.  E.g., in Shout~\cite{shout} only those message
  broadcast in the vicinity of the node will be heard and included.}
list of past messages sent by himself and others.  A partial retweet
graph is constructed from this dataset, with one vertex per sender and
directional edges linking each retweeter to the corresponding
retweetees. Denoting the participant's own vertex as the
root\footnote{Trusting one's own vertex as non-spammer breaks the
  otherwise problematic symmetry between the non-spammer and spammer
  portions of the graph.}, the remaining participants are classified
by two attributes, their distance from the root and the maximum flow
from the root to them.  Users that are classified as non-spammers are
whitelisted.

This approach presents two bootstrapping problems.  How does a new
user with no recorded history construct a retweet graph and do the
messages from a new user that has never been retweeted ever get seen?
For the first question, a user can copy the tweet history from a
trusted friend or bootstrap by explicitly whitelisting his
friends. For the second, the user can ask his friends to whitelist
him, so they can then see and retweet his messages, linking him to the
graph. We also anticipate that some (particularly bored) users will
choose to view all incoming tweets, retweeting some that are not spam.

The following sections analyze the performance of this classification
procedure on our 10\% sample of the retweet graph and synthetic graphs
for parameter sweeps.

\subsection{Performance on the Twitter Retweet Graph}

\begin{figure}
  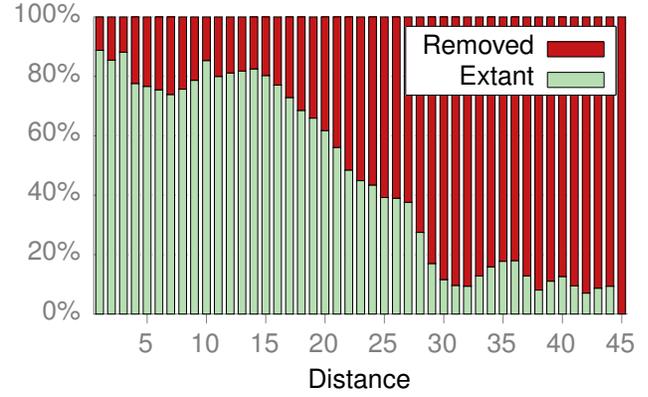
  \caption{Percentage of removed and extant Twitter users as a
    function of distance from benign users in the retweet graph. Most
    removed users are spammers, so this graph shows that distance is
    highly correlated with spammer behavior.}
  \label{fig:spam-classification-distance}
\end{figure}

We first consider the performance on our 10\% sample of the retweet
graph.  This sample is problematic because most of the paths between
non-spammers are not included (90\% of edges are missing) but is
sufficient to show that the hypothesized differences exist.

We randomly chose 100 source--destination pairs of users who distances
in the retweet graph ranged from 1 to 45, for 4500 pairs in total.  We
obtained ground truth classification for these 9000 users by querying
the Twitter API to determine if the account had been removed in the 18
months following the initial collection.  Twitter actively seeks out
and bans spammers, so the majority of the spammers will have been
removed.  Some non-spammers will have also deleted their own accounts,
so we refer to these categories as \emph{removed} and
\emph{extant}. We believe that most removed users were
spammers~\cite{thomas11nov}.

We consider only the pairs whose source node is
extant. \autoref{fig:spam-classification-distance} shows the
percentage of destination nodes in each category by the distance from
their sources.  Clearly, distance in the retweet graph is correlated
with spammer tendencies.  A classifier over this attribute alone
achieves a TPR of 75\% with an FPR of 25\%.

The second attribute, connectivity, shows no correlation in the 10\%
sample graph because the majority of edges are missing. Most pairs
with between one and ten independent paths in the original graph
contains only zero or one paths in the sampled graph, making it
impossible to distinguish a non-spam node linked by ten paths from a
spam node linked by one.  Instead, we turn to synthetic retweet graphs
to study the performance of the combined classifier.

\subsection{Performance on Synthetic Retweet Graphs}
\label{sec:spam-synthetic}

\begin{figure}
  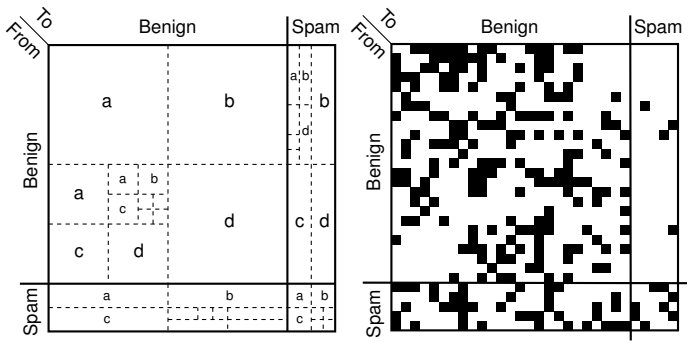
  \caption{Illustration of the modified R-MAT algorithm for generating
    synthetic retweet graphs and a resulting adjacency matrix.  Fewer
    edges are placed in the benign--spam quadrant to model the lower
    likelihood of such retweets. Within each quadrant, edges are
    cascaded in proportion to probabilities $a$, $b$, $c$, and $d$ to
    generate a scale-free, small-world structure.}
  \label{fig:spam-adjacency-matrix}
\end{figure}

The analysis in \autoref{sec:retweet-graph} showed that the retweet
graph is scale-free and small-world, enabling the generation of
synthetic retweet graphs using the R-MAT (\emph{R}ecursive {Mat}rix),
an algorithm designed to generate a variety of such
networks~\cite{chakrabarti04apr}. Although metrics like assortativity
and clustering are not directly controllable---R-MAT cannot capture
the differences between the followers and retweet
graphs~\footnote{Unfortunately, this prevents us from comparing
  retweet-based with follower-based spam detection. A full sample of
  the retweet graph would be needed.}---it is sufficient for our
purposes as we depend only on the connectivity implied by the
small-world structure and limited number of incoming edges to spammer
nodes.

R-MAT produces scale-free, small-world graphs by treating edge
assignment in the adjacency matrix as a two-dimensional binomial
cascade.  We modify the procedure to generate relatively fewer edges
from benign to spammer nodes (B--S) than the other possibilities
(B--B, S--S, S--B), modeling the notion that non-spammers rarely
retweet spammers.

The modified R-MAT process is illustrated in
\autoref{fig:spam-adjacency-matrix}.  We desire a graph with some
number of benign and spammer nodes, some number of non-B--S edges, and
a relatively smaller number of B--S edges.  The adjacency matrix is
divided into four quadrants and the edges split among the B--B, S--S,
and S--B quadrants in proportion to their areas.  Within each
quadrant, the R-MAT algorithm is used to place the edges.  For each
edge, the sub-quadrant in which to place the edge is chosen according
to probabilities $a$, $b$, $c$, and $d$ ($a+b+c+d=1$). The process
recurses until a single cell is selected for the edge. The result of
the process for a small graph is shown in
\autoref{fig:spam-adjacency-matrix}.

The parameters $a$, $b$, $c$, and $d$ are obtained via
AutoMAT-fast~\cite{chakrabarti04apr}, i.e., fitting the degree
distribution of the retweet graph to that of the model.  The R-MAT
process is essentially a two-dimensional binomial cascade, with the
out-edges assigned to the upper and lower halves with probabilities $p
\triangleq a + b$ and $1-p$ and the in-edges assigned to the left and
right halves with probabilities $q \triangleq a + c$ and $1-q$.
Letting $N=2^n$ be the number of nodes and $E$ the number of edges to
assign, then the expected number of nodes $c_k$ with out-degree $k$ is
\begin{equation}
  c_k = \sum_{i=0}^n \binom{n}{i} B\left(k; E, p^{n-i}(1-p)^i\right)
\end{equation}
where $B(k;a,b)$ is the mass function of the binomial distribution
$B(a,b)$. The in-edge distribution is computed similarly. Fitting to
the retweet graph, we obtain $a=0.52$, $b=0.18$, $c=0.17$, and
$d=0.13$.

We fix the fraction of spam nodes to 10\% and assume that spammer
retweet behavior mimics that of benign nodes retweeting each
other. Differences are not in the attackers' interest, as they would
enable additional classification methods.

\begin{figure}
  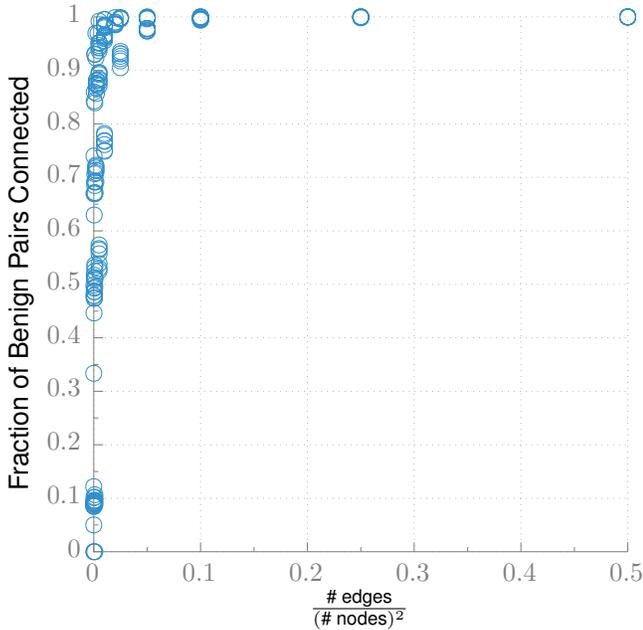
  \caption{Connectivity of benign pairs as a function of the benign
    edge density. Above 5\%, almost all pairs are connected. We expect
    that density does not grow with network size, so this limits the
    network size for which the false positive rate is acceptable.  For
    large networks, the technique will only work within clusters.}
  \label{fig:vary-connected}
\end{figure}

The performance of the classifier is primarily affected by two
metrics---the fraction of possible B--B edges that are present and the
number of B--S edges per spammer vertex---so we conduct parameter
sweeps of these values. 

If the B--B edge density is too low, many benign pairs will not be
connected and the false positive rate will be
high. \autoref{fig:vary-connected} plots this density against the
fraction of benign pairs that are connected for a variety of network
sizes.  Above 5\%, most pairs are connected and above 10\%,
essentially all pairs are connected. We expect the number of edges in
a retweet graph to (above some point) grow linearly in the number of
users, so this relationship places a limit on the network size for
which the technique is usable.  For larger networks (e.g., the world
population), the technique will only work within clusters for which
the edge density is high enough---users outside of one's own cluster
will be identified as spammers.  For example, the average out-degree
of Twitter, 75, would support 25000 participants. However, social
relationships are clustered, so this limitation should rarely be an
issue in practice.\footnote{This limitation does prevent the discovery
  of content from outside of one's own group, possible with
  centralized Twitter today. Content can still traverse two groups if
  seen and retweeted by a member of both.} In a network like
Shout~\cite{shout}, the effective community size is already limited by
geography.

\begin{figure}
  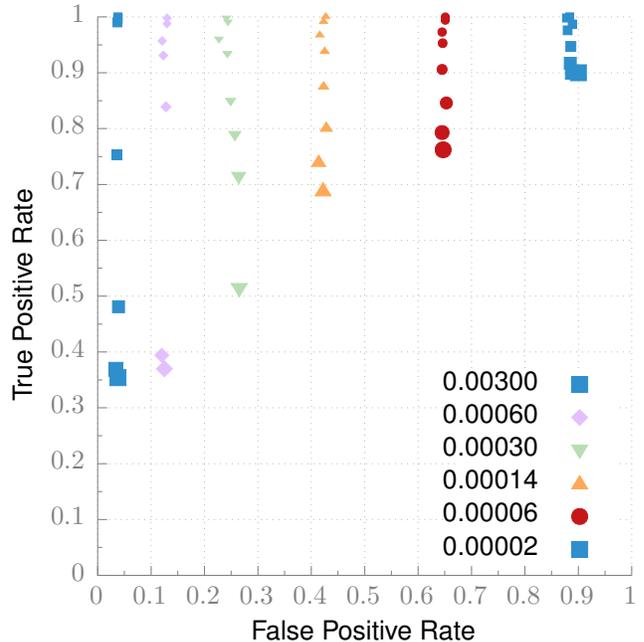
  \caption{Performance of J48 classifier over distance and
    connectivity attributes in the synthetic graphs. The benign edge
    density (marker symbol and color) range from 0.00002 to 0.003 and
    the number of B--S edges per spammer node (marker size) ranges
    from 0.01 to 1.  Each marker is a single point on the resulting
    ROC curve.}
  \label{fig:spam-tpr-fpr-by-edge-rates}
\end{figure}

\autoref{fig:spam-tpr-fpr-by-edge-rates} shows the classification
performance. We use the J48 decision tree classifier over both the
distance and connectivity (max-flow) attributes, using 10-fold cross
validation.  We sweep both the benign edge density (marker symbol and
color) from 0.0002 to 0.003 and the number of B--S edges per spammer
(marker size) from 0.01 to 1.  To reduce clutter, a \emph{single}
point\footnote{The selected points are generally near the knees of the
  curves, but within a class are intentionally chosen to have similar
  FPRs.} from each resulting ROC curve is plotted.  Two trends are
immediately clear.  Decreasing the benign edge density increases the
FPR, but an FPR below 5\% requires just a 0.3\% edge density.
Increasing the B--S rate (number of B--S edges per spammer node)
decreases the true positive rate.  If less than one-tenth of spammers
are retweeted by benign nodes, the TPR is universally above 98\%. The
sensitivity to B--S rate increases with edge density because the
spammer nodes are more interconnected (we hold the S--B and S--S
densities equal to the B--B density).

In summary, \autoref{fig:spam-classification-distance} shows that the
inter-node distance in the retweet graph is highly correlated with
being a spammer, enabling detection.  Simulations on the synthetic
graphs show that inter-node distance and inter-node max flow can
identify spammers with greater than 98\% TPR and less than 5\% FPR
when fewer than one-tenth of spammers are retweeted and at least 0.3\%
of possible edges between benign nodes are present.  For a
community-sized network of 25000 participants, this implies an average
node degree of 75, i.e., that of the Twitter retweet graph.  For
larger networks, the classification works best within smaller
sub-clusters where the edge density is higher.

%% file: fig/retweet-graph-example.pdf_tex
%% Creator: Inkscape inkscape 0.48.4, www.inkscape.org
%% Accompanies image file 'retweet-graph-example.pdf' (pdf, eps, ps)
\begingroup%
  \makeatletter%
  \providecommand\color[2][]{%
    \errmessage{(Inkscape) Color is used for the text in Inkscape, but the package 'color.sty' is not loaded}%
    \renewcommand\color[2][]{}%
  }%
  \providecommand\transparent[1]{%
    \errmessage{(Inkscape) Transparency is used (non-zero) for the text in Inkscape, but the package 'transparent.sty' is not loaded}%
    \renewcommand\transparent[1]{}%
  }%
  \providecommand\rotatebox[2]{#2}%
  \ifx\svgwidth\undefined%
    \setlength{\unitlength}{251.28000488bp}%
    \ifx\svgscale\undefined%
      \relax%
    \else%
      \setlength{\unitlength}{\unitlength * \real{\svgscale}}%
    \fi%
  \else%
    \setlength{\unitlength}{\svgwidth}%
  \fi%
  \global\let\svgwidth\undefined%
  \global\let\svgscale\undefined%
  \makeatother%
  \begin{picture}(1,0.57306589)%
    \put(0,0){\includegraphics[width=\unitlength]{retweet-graph-example.pdf}}%
    \put(0.07164256,0.27969617){\color[rgb]{0,0,0}\makebox(0,0)[b]{\smash{A}}}%
    \put(0.64451725,0.27969617){\color[rgb]{0,0,0}\makebox(0,0)[b]{\smash{B}}}%
    \put(0.93131136,0.154338){\color[rgb]{0,0,0}\makebox(0,0)[b]{\smash{S}}}%
  \end{picture}%
\endgroup%

%% file: fig/spam-classification-distance.pdf_tex
% GNUPLOT: LaTeX picture with Postscript
\begingroup
  \makeatletter
  \providecommand\color[2][]{%
    \GenericError{(gnuplot) \space\space\space\@spaces}{%
      Package color not loaded in conjunction with
      terminal option `colourtext'%
    }{See the gnuplot documentation for explanation.%
    }{Either use 'blacktext' in gnuplot or load the package
      color.sty in LaTeX.}%
    \renewcommand\color[2][]{}%
  }%
  \providecommand\includegraphics[2][]{%
    \GenericError{(gnuplot) \space\space\space\@spaces}{%
      Package graphicx or graphics not loaded%
    }{See the gnuplot documentation for explanation.%
    }{The gnuplot epslatex terminal needs graphicx.sty or graphics.sty.}%
    \renewcommand\includegraphics[2][]{}%
  }%
  \providecommand\rotatebox[2]{#2}%
  \@ifundefined{ifGPcolor}{%
    \newif\ifGPcolor
    \GPcolortrue
  }{}%
  \@ifundefined{ifGPblacktext}{%
    \newif\ifGPblacktext
    \GPblacktextfalse
  }{}%
  % define a \g@addto@macro without @ in the name:
  \let\gplgaddtomacro\g@addto@macro
  % define empty templates for all commands taking text:
  \gdef\gplbacktext{}%
  \gdef\gplfronttext{}%
  \makeatother
  \ifGPblacktext
    % no textcolor at all
    \def\colorrgb#1{}%
    \def\colorgray#1{}%
  \else
    % gray or color?
    \ifGPcolor
      \def\colorrgb#1{\color[rgb]{#1}}%
      \def\colorgray#1{\color[gray]{#1}}%
      \expandafter\def\csname LTw\endcsname{\color{white}}%
      \expandafter\def\csname LTb\endcsname{\color{black}}%
      \expandafter\def\csname LTa\endcsname{\color{black}}%
      \expandafter\def\csname LT0\endcsname{\color[rgb]{1,0,0}}%
      \expandafter\def\csname LT1\endcsname{\color[rgb]{0,1,0}}%
      \expandafter\def\csname LT2\endcsname{\color[rgb]{0,0,1}}%
      \expandafter\def\csname LT3\endcsname{\color[rgb]{1,0,1}}%
      \expandafter\def\csname LT4\endcsname{\color[rgb]{0,1,1}}%
      \expandafter\def\csname LT5\endcsname{\color[rgb]{1,1,0}}%
      \expandafter\def\csname LT6\endcsname{\color[rgb]{0,0,0}}%
      \expandafter\def\csname LT7\endcsname{\color[rgb]{1,0.3,0}}%
      \expandafter\def\csname LT8\endcsname{\color[rgb]{0.5,0.5,0.5}}%
    \else
      % gray
      \def\colorrgb#1{\color{black}}%
      \def\colorgray#1{\color[gray]{#1}}%
      \expandafter\def\csname LTw\endcsname{\color{white}}%
      \expandafter\def\csname LTb\endcsname{\color{black}}%
      \expandafter\def\csname LTa\endcsname{\color{black}}%
      \expandafter\def\csname LT0\endcsname{\color{black}}%
      \expandafter\def\csname LT1\endcsname{\color{black}}%
      \expandafter\def\csname LT2\endcsname{\color{black}}%
      \expandafter\def\csname LT3\endcsname{\color{black}}%
      \expandafter\def\csname LT4\endcsname{\color{black}}%
      \expandafter\def\csname LT5\endcsname{\color{black}}%
      \expandafter\def\csname LT6\endcsname{\color{black}}%
      \expandafter\def\csname LT7\endcsname{\color{black}}%
      \expandafter\def\csname LT8\endcsname{\color{black}}%
    \fi
  \fi
    \setlength{\unitlength}{0.0500bp}%
    \ifx\gptboxheight\undefined%
      \newlength{\gptboxheight}%
      \newlength{\gptboxwidth}%
      \newsavebox{\gptboxtext}%
    \fi%
    \setlength{\fboxrule}{0.5pt}%
    \setlength{\fboxsep}{1pt}%
\begin{picture}(4960.00,3060.00)%
    \gplgaddtomacro\gplbacktext{%
      \colorrgb{0.50,0.50,0.50}%
      \put(578,640){\makebox(0,0)[r]{\strut{}0\%}}%
      \colorrgb{0.50,0.50,0.50}%
      \put(578,1088){\makebox(0,0)[r]{\strut{}20\%}}%
      \colorrgb{0.50,0.50,0.50}%
      \put(578,1536){\makebox(0,0)[r]{\strut{}40\%}}%
      \colorrgb{0.50,0.50,0.50}%
      \put(578,1985){\makebox(0,0)[r]{\strut{}60\%}}%
      \colorrgb{0.50,0.50,0.50}%
      \put(578,2433){\makebox(0,0)[r]{\strut{}80\%}}%
      \colorrgb{0.50,0.50,0.50}%
      \put(578,2881){\makebox(0,0)[r]{\strut{}100\%}}%
      \colorrgb{0.50,0.50,0.50}%
      \put(1070,391){\makebox(0,0){\strut{}5}}%
      \colorrgb{0.50,0.50,0.50}%
      \put(1517,391){\makebox(0,0){\strut{}10}}%
      \colorrgb{0.50,0.50,0.50}%
      \put(1964,391){\makebox(0,0){\strut{}15}}%
      \colorrgb{0.50,0.50,0.50}%
      \put(2411,391){\makebox(0,0){\strut{}20}}%
      \colorrgb{0.50,0.50,0.50}%
      \put(2858,391){\makebox(0,0){\strut{}25}}%
      \colorrgb{0.50,0.50,0.50}%
      \put(3306,391){\makebox(0,0){\strut{}30}}%
      \colorrgb{0.50,0.50,0.50}%
      \put(3753,391){\makebox(0,0){\strut{}35}}%
      \colorrgb{0.50,0.50,0.50}%
      \put(4200,391){\makebox(0,0){\strut{}40}}%
      \colorrgb{0.50,0.50,0.50}%
      \put(4647,391){\makebox(0,0){\strut{}45}}%
    }%
    \gplgaddtomacro\gplfronttext{%
      \csname LTb\endcsname%
      \put(2679,124){\makebox(0,0){\strut{}Distance}}%
      \csname LTb\endcsname%
      \put(3998,2406){\makebox(0,0)[r]{\strut{}Extant}}%
      \csname LTb\endcsname%
      \put(3998,2637){\makebox(0,0)[r]{\strut{}Removed}}%
    }%
    \gplbacktext
    \put(0,0){\includegraphics{spam-classification-distance}}%
    \gplfronttext
  \end{picture}%
\endgroup

%% file: fig/spam-adjacency-matrix.pdf_tex
%% Creator: Inkscape inkscape 0.48.4, www.inkscape.org
%% Accompanies image file 'spam-adjacency-matrix.pdf' (pdf, eps, ps)
\begingroup%
  \makeatletter%
  \providecommand\color[2][]{%
    \errmessage{(Inkscape) Color is used for the text in Inkscape, but the package 'color.sty' is not loaded}%
    \renewcommand\color[2][]{}%
  }%
  \providecommand\transparent[1]{%
    \errmessage{(Inkscape) Transparency is used (non-zero) for the text in Inkscape, but the package 'transparent.sty' is not loaded}%
    \renewcommand\transparent[1]{}%
  }%
  \providecommand\rotatebox[2]{#2}%
  \ifx\svgwidth\undefined%
    \setlength{\unitlength}{251.28000488bp}%
    \ifx\svgscale\undefined%
      \relax%
    \else%
      \setlength{\unitlength}{\unitlength * \real{\svgscale}}%
    \fi%
  \else%
    \setlength{\unitlength}{\svgwidth}%
  \fi%
  \global\let\svgwidth\undefined%
  \global\let\svgscale\undefined%
  \makeatother%
  \begin{picture}(1,0.50143266)%
    \put(0,0){\includegraphics[width=\unitlength]{spam-adjacency-matrix.pdf}}%
    \put(0.5173885,0.44364843){\color[rgb]{0,0,0}\rotatebox{-45}{\makebox(0,0)[b]{\smash{\sss{From}}}}}%
    \put(0.54640931,0.47642392){\color[rgb]{0,0,0}\rotatebox{-45}{\makebox(0,0)[b]{\smash{\sss{To}}}}}%
    \put(0.73463864,0.46243214){\color[rgb]{0,0,0}\makebox(0,0)[b]{\smash{\sss{Benign}}}}%
    \put(0.95749759,0.46243214){\color[rgb]{0,0,0}\makebox(0,0)[b]{\smash{\sss{Spam}}}}%
    \put(0.54033037,0.26866486){\color[rgb]{0,0,0}\rotatebox{90}{\makebox(0,0)[b]{\smash{\sss{Benign}}}}}%
    \put(0.54033037,0.04103093){\color[rgb]{0,0,0}\rotatebox{90}{\makebox(0,0)[b]{\smash{\sss{Spam}}}}}%
    \put(0.03047426,0.47724916){\color[rgb]{0,0,0}\rotatebox{-45}{\makebox(0,0)[b]{\smash{\sss{To}}}}}%
    \put(0,0.44412604){\color[rgb]{0,0,0}\rotatebox{-45}{\makebox(0,0)[b]{\smash{\sss{From}}}}}%
    \put(0.22143665,0.46243214){\color[rgb]{0,0,0}\makebox(0,0)[b]{\smash{\sss{Benign}}}}%
    \put(0.44429561,0.46243214){\color[rgb]{0,0,0}\makebox(0,0)[b]{\smash{\sss{Spam}}}}%
    \put(0.02584208,0.27151608){\color[rgb]{0,0,0}\rotatebox{90}{\makebox(0,0)[b]{\smash{\sss{Benign}}}}}%
    \put(0.02584208,0.04388196){\color[rgb]{0,0,0}\rotatebox{90}{\makebox(0,0)[b]{\smash{\sss{Spam}}}}}%
    \put(0.13056537,0.35061927){\color[rgb]{0,0,0}\makebox(0,0)[b]{\smash{\sss{a}}}}%
    \put(0.3146772,0.35061927){\color[rgb]{0,0,0}\makebox(0,0)[b]{\smash{\sss{b}}}}%
    \put(0.08875512,0.12562185){\color[rgb]{0,0,0}\makebox(0,0)[b]{\smash{\sss{c}}}}%
    \put(0.3146772,0.17090853){\color[rgb]{0,0,0}\makebox(0,0)[b]{\smash{\sss{d}}}}%
    \put(0.08875512,0.21364364){\color[rgb]{0,0,0}\makebox(0,0)[b]{\smash{\sss{a}}}}%
    \put(0.17851421,0.12562185){\color[rgb]{0,0,0}\makebox(0,0)[b]{\smash{\sss{d}}}}%
    \put(0.15430828,0.23638229){\color[rgb]{0,0,0}\makebox(0,0)[b]{\smash{\tiny{a}}}}%
    \put(0.20198665,0.23638229){\color[rgb]{0,0,0}\makebox(0,0)[b]{\smash{\tiny{b}}}}%
    \put(0.15430828,0.19163795){\color[rgb]{0,0,0}\makebox(0,0)[b]{\smash{\tiny{c}}}}%
    \put(0.45724926,0.35061927){\color[rgb]{0,0,0}\makebox(0,0)[b]{\smash{\sss{b}}}}%
    \put(0.45724926,0.17090853){\color[rgb]{0,0,0}\makebox(0,0)[b]{\smash{\sss{d}}}}%
    \put(0.41963671,0.17090853){\color[rgb]{0,0,0}\makebox(0,0)[b]{\smash{\sss{c}}}}%
    \put(0.41103791,0.39115373){\color[rgb]{0,0,0}\makebox(0,0)[b]{\smash{\tiny{a}}}}%
    \put(0.42864221,0.39115373){\color[rgb]{0,0,0}\makebox(0,0)[b]{\smash{\tiny{b}}}}%
    \put(0.42864221,0.30981374){\color[rgb]{0,0,0}\makebox(0,0)[b]{\smash{\tiny{d}}}}%
    \put(0.41963671,0.0618062){\color[rgb]{0,0,0}\makebox(0,0)[b]{\smash{\tiny{a}}}}%
    \put(0.45724926,0.0618062){\color[rgb]{0,0,0}\makebox(0,0)[b]{\smash{\tiny{b}}}}%
    \put(0.41963671,0.0271003){\color[rgb]{0,0,0}\makebox(0,0)[b]{\smash{\tiny{c}}}}%
    \put(0.13056537,0.0618062){\color[rgb]{0,0,0}\makebox(0,0)[b]{\smash{\tiny{a}}}}%
    \put(0.3146772,0.0618062){\color[rgb]{0,0,0}\makebox(0,0)[b]{\smash{\tiny{b}}}}%
    \put(0.13056537,0.0271003){\color[rgb]{0,0,0}\makebox(0,0)[b]{\smash{\tiny{c}}}}%
  \end{picture}%
\endgroup%

%% file: fig/vary-connected.pdf_tex
% GNUPLOT: LaTeX picture with Postscript
\begingroup
  \makeatletter
  \providecommand\color[2][]{%
    \GenericError{(gnuplot) \space\space\space\@spaces}{%
      Package color not loaded in conjunction with
      terminal option `colourtext'%
    }{See the gnuplot documentation for explanation.%
    }{Either use 'blacktext' in gnuplot or load the package
      color.sty in LaTeX.}%
    \renewcommand\color[2][]{}%
  }%
  \providecommand\includegraphics[2][]{%
    \GenericError{(gnuplot) \space\space\space\@spaces}{%
      Package graphicx or graphics not loaded%
    }{See the gnuplot documentation for explanation.%
    }{The gnuplot epslatex terminal needs graphicx.sty or graphics.sty.}%
    \renewcommand\includegraphics[2][]{}%
  }%
  \providecommand\rotatebox[2]{#2}%
  \@ifundefined{ifGPcolor}{%
    \newif\ifGPcolor
    \GPcolortrue
  }{}%
  \@ifundefined{ifGPblacktext}{%
    \newif\ifGPblacktext
    \GPblacktextfalse
  }{}%
  % define a \g@addto@macro without @ in the name:
  \let\gplgaddtomacro\g@addto@macro
  % define empty templates for all commands taking text:
  \gdef\gplbacktext{}%
  \gdef\gplfronttext{}%
  \makeatother
  \ifGPblacktext
    % no textcolor at all
    \def\colorrgb#1{}%
    \def\colorgray#1{}%
  \else
    % gray or color?
    \ifGPcolor
      \def\colorrgb#1{\color[rgb]{#1}}%
      \def\colorgray#1{\color[gray]{#1}}%
      \expandafter\def\csname LTw\endcsname{\color{white}}%
      \expandafter\def\csname LTb\endcsname{\color{black}}%
      \expandafter\def\csname LTa\endcsname{\color{black}}%
      \expandafter\def\csname LT0\endcsname{\color[rgb]{1,0,0}}%
      \expandafter\def\csname LT1\endcsname{\color[rgb]{0,1,0}}%
      \expandafter\def\csname LT2\endcsname{\color[rgb]{0,0,1}}%
      \expandafter\def\csname LT3\endcsname{\color[rgb]{1,0,1}}%
      \expandafter\def\csname LT4\endcsname{\color[rgb]{0,1,1}}%
      \expandafter\def\csname LT5\endcsname{\color[rgb]{1,1,0}}%
      \expandafter\def\csname LT6\endcsname{\color[rgb]{0,0,0}}%
      \expandafter\def\csname LT7\endcsname{\color[rgb]{1,0.3,0}}%
      \expandafter\def\csname LT8\endcsname{\color[rgb]{0.5,0.5,0.5}}%
    \else
      % gray
      \def\colorrgb#1{\color{black}}%
      \def\colorgray#1{\color[gray]{#1}}%
      \expandafter\def\csname LTw\endcsname{\color{white}}%
      \expandafter\def\csname LTb\endcsname{\color{black}}%
      \expandafter\def\csname LTa\endcsname{\color{black}}%
      \expandafter\def\csname LT0\endcsname{\color{black}}%
      \expandafter\def\csname LT1\endcsname{\color{black}}%
      \expandafter\def\csname LT2\endcsname{\color{black}}%
      \expandafter\def\csname LT3\endcsname{\color{black}}%
      \expandafter\def\csname LT4\endcsname{\color{black}}%
      \expandafter\def\csname LT5\endcsname{\color{black}}%
      \expandafter\def\csname LT6\endcsname{\color{black}}%
      \expandafter\def\csname LT7\endcsname{\color{black}}%
      \expandafter\def\csname LT8\endcsname{\color{black}}%
    \fi
  \fi
    \setlength{\unitlength}{0.0500bp}%
    \ifx\gptboxheight\undefined%
      \newlength{\gptboxheight}%
      \newlength{\gptboxwidth}%
      \newsavebox{\gptboxtext}%
    \fi%
    \setlength{\fboxrule}{0.5pt}%
    \setlength{\fboxsep}{1pt}%
\begin{picture}(4960.00,4960.00)%
    \gplgaddtomacro\gplbacktext{%
      \colorrgb{0.50,0.50,0.50}%
      \put(578,747){\makebox(0,0)[r]{\strut{}$0$}}%
      \colorrgb{0.50,0.50,0.50}%
      \put(578,1150){\makebox(0,0)[r]{\strut{}$0.1$}}%
      \colorrgb{0.50,0.50,0.50}%
      \put(578,1554){\makebox(0,0)[r]{\strut{}$0.2$}}%
      \colorrgb{0.50,0.50,0.50}%
      \put(578,1957){\makebox(0,0)[r]{\strut{}$0.3$}}%
      \colorrgb{0.50,0.50,0.50}%
      \put(578,2361){\makebox(0,0)[r]{\strut{}$0.4$}}%
      \colorrgb{0.50,0.50,0.50}%
      \put(578,2764){\makebox(0,0)[r]{\strut{}$0.5$}}%
      \colorrgb{0.50,0.50,0.50}%
      \put(578,3167){\makebox(0,0)[r]{\strut{}$0.6$}}%
      \colorrgb{0.50,0.50,0.50}%
      \put(578,3571){\makebox(0,0)[r]{\strut{}$0.7$}}%
      \colorrgb{0.50,0.50,0.50}%
      \put(578,3974){\makebox(0,0)[r]{\strut{}$0.8$}}%
      \colorrgb{0.50,0.50,0.50}%
      \put(578,4378){\makebox(0,0)[r]{\strut{}$0.9$}}%
      \colorrgb{0.50,0.50,0.50}%
      \put(578,4781){\makebox(0,0)[r]{\strut{}$1$}}%
      \colorrgb{0.50,0.50,0.50}%
      \put(667,569){\makebox(0,0){\strut{}$0$}}%
      \colorrgb{0.50,0.50,0.50}%
      \put(1472,569){\makebox(0,0){\strut{}$0.1$}}%
      \colorrgb{0.50,0.50,0.50}%
      \put(2277,569){\makebox(0,0){\strut{}$0.2$}}%
      \colorrgb{0.50,0.50,0.50}%
      \put(3082,569){\makebox(0,0){\strut{}$0.3$}}%
      \colorrgb{0.50,0.50,0.50}%
      \put(3887,569){\makebox(0,0){\strut{}$0.4$}}%
      \colorrgb{0.50,0.50,0.50}%
      \put(4692,569){\makebox(0,0){\strut{}$0.5$}}%
    }%
    \gplgaddtomacro\gplfronttext{%
      \csname LTb\endcsname%
      \put(133,2764){\rotatebox{-270}{\makebox(0,0){\strut{}Fraction of Benign Pairs Connected}}}%
      \csname LTb\endcsname%
      \put(2679,302){\makebox(0,0){\strut{}$\frac{\text{\# edges}}{(\text{\# nodes})^2}$}}%
    }%
    \gplbacktext
    \put(0,0){\includegraphics{vary-connected}}%
    \gplfronttext
  \end{picture}%
\endgroup

%% file: fig/spam-tpr-fpr-by-edge-rates.pdf_tex
% GNUPLOT: LaTeX picture with Postscript
\begingroup
  \makeatletter
  \providecommand\color[2][]{%
    \GenericError{(gnuplot) \space\space\space\@spaces}{%
      Package color not loaded in conjunction with
      terminal option `colourtext'%
    }{See the gnuplot documentation for explanation.%
    }{Either use 'blacktext' in gnuplot or load the package
      color.sty in LaTeX.}%
    \renewcommand\color[2][]{}%
  }%
  \providecommand\includegraphics[2][]{%
    \GenericError{(gnuplot) \space\space\space\@spaces}{%
      Package graphicx or graphics not loaded%
    }{See the gnuplot documentation for explanation.%
    }{The gnuplot epslatex terminal needs graphicx.sty or graphics.sty.}%
    \renewcommand\includegraphics[2][]{}%
  }%
  \providecommand\rotatebox[2]{#2}%
  \@ifundefined{ifGPcolor}{%
    \newif\ifGPcolor
    \GPcolortrue
  }{}%
  \@ifundefined{ifGPblacktext}{%
    \newif\ifGPblacktext
    \GPblacktextfalse
  }{}%
  % define a \g@addto@macro without @ in the name:
  \let\gplgaddtomacro\g@addto@macro
  % define empty templates for all commands taking text:
  \gdef\gplbacktext{}%
  \gdef\gplfronttext{}%
  \makeatother
  \ifGPblacktext
    % no textcolor at all
    \def\colorrgb#1{}%
    \def\colorgray#1{}%
  \else
    % gray or color?
    \ifGPcolor
      \def\colorrgb#1{\color[rgb]{#1}}%
      \def\colorgray#1{\color[gray]{#1}}%
      \expandafter\def\csname LTw\endcsname{\color{white}}%
      \expandafter\def\csname LTb\endcsname{\color{black}}%
      \expandafter\def\csname LTa\endcsname{\color{black}}%
      \expandafter\def\csname LT0\endcsname{\color[rgb]{1,0,0}}%
      \expandafter\def\csname LT1\endcsname{\color[rgb]{0,1,0}}%
      \expandafter\def\csname LT2\endcsname{\color[rgb]{0,0,1}}%
      \expandafter\def\csname LT3\endcsname{\color[rgb]{1,0,1}}%
      \expandafter\def\csname LT4\endcsname{\color[rgb]{0,1,1}}%
      \expandafter\def\csname LT5\endcsname{\color[rgb]{1,1,0}}%
      \expandafter\def\csname LT6\endcsname{\color[rgb]{0,0,0}}%
      \expandafter\def\csname LT7\endcsname{\color[rgb]{1,0.3,0}}%
      \expandafter\def\csname LT8\endcsname{\color[rgb]{0.5,0.5,0.5}}%
    \else
      % gray
      \def\colorrgb#1{\color{black}}%
      \def\colorgray#1{\color[gray]{#1}}%
      \expandafter\def\csname LTw\endcsname{\color{white}}%
      \expandafter\def\csname LTb\endcsname{\color{black}}%
      \expandafter\def\csname LTa\endcsname{\color{black}}%
      \expandafter\def\csname LT0\endcsname{\color{black}}%
      \expandafter\def\csname LT1\endcsname{\color{black}}%
      \expandafter\def\csname LT2\endcsname{\color{black}}%
      \expandafter\def\csname LT3\endcsname{\color{black}}%
      \expandafter\def\csname LT4\endcsname{\color{black}}%
      \expandafter\def\csname LT5\endcsname{\color{black}}%
      \expandafter\def\csname LT6\endcsname{\color{black}}%
      \expandafter\def\csname LT7\endcsname{\color{black}}%
      \expandafter\def\csname LT8\endcsname{\color{black}}%
    \fi
  \fi
    \setlength{\unitlength}{0.0500bp}%
    \ifx\gptboxheight\undefined%
      \newlength{\gptboxheight}%
      \newlength{\gptboxwidth}%
      \newsavebox{\gptboxtext}%
    \fi%
    \setlength{\fboxrule}{0.5pt}%
    \setlength{\fboxsep}{1pt}%
\begin{picture}(4960.00,4960.00)%
    \gplgaddtomacro\gplbacktext{%
      \colorrgb{0.50,0.50,0.50}%
      \put(578,569){\makebox(0,0)[r]{\strut{}$0$}}%
      \colorrgb{0.50,0.50,0.50}%
      \put(578,990){\makebox(0,0)[r]{\strut{}$0.1$}}%
      \colorrgb{0.50,0.50,0.50}%
      \put(578,1411){\makebox(0,0)[r]{\strut{}$0.2$}}%
      \colorrgb{0.50,0.50,0.50}%
      \put(578,1833){\makebox(0,0)[r]{\strut{}$0.3$}}%
      \colorrgb{0.50,0.50,0.50}%
      \put(578,2254){\makebox(0,0)[r]{\strut{}$0.4$}}%
      \colorrgb{0.50,0.50,0.50}%
      \put(578,2675){\makebox(0,0)[r]{\strut{}$0.5$}}%
      \colorrgb{0.50,0.50,0.50}%
      \put(578,3096){\makebox(0,0)[r]{\strut{}$0.6$}}%
      \colorrgb{0.50,0.50,0.50}%
      \put(578,3517){\makebox(0,0)[r]{\strut{}$0.7$}}%
      \colorrgb{0.50,0.50,0.50}%
      \put(578,3939){\makebox(0,0)[r]{\strut{}$0.8$}}%
      \colorrgb{0.50,0.50,0.50}%
      \put(578,4360){\makebox(0,0)[r]{\strut{}$0.9$}}%
      \colorrgb{0.50,0.50,0.50}%
      \put(578,4781){\makebox(0,0)[r]{\strut{}$1$}}%
      \colorrgb{0.50,0.50,0.50}%
      \put(667,391){\makebox(0,0){\strut{}$0$}}%
      \colorrgb{0.50,0.50,0.50}%
      \put(1070,391){\makebox(0,0){\strut{}$0.1$}}%
      \colorrgb{0.50,0.50,0.50}%
      \put(1472,391){\makebox(0,0){\strut{}$0.2$}}%
      \colorrgb{0.50,0.50,0.50}%
      \put(1875,391){\makebox(0,0){\strut{}$0.3$}}%
      \colorrgb{0.50,0.50,0.50}%
      \put(2277,391){\makebox(0,0){\strut{}$0.4$}}%
      \colorrgb{0.50,0.50,0.50}%
      \put(2680,391){\makebox(0,0){\strut{}$0.5$}}%
      \colorrgb{0.50,0.50,0.50}%
      \put(3082,391){\makebox(0,0){\strut{}$0.6$}}%
      \colorrgb{0.50,0.50,0.50}%
      \put(3485,391){\makebox(0,0){\strut{}$0.7$}}%
      \colorrgb{0.50,0.50,0.50}%
      \put(3887,391){\makebox(0,0){\strut{}$0.8$}}%
      \colorrgb{0.50,0.50,0.50}%
      \put(4290,391){\makebox(0,0){\strut{}$0.9$}}%
      \colorrgb{0.50,0.50,0.50}%
      \put(4692,391){\makebox(0,0){\strut{}$1$}}%
    }%
    \gplgaddtomacro\gplfronttext{%
      \csname LTb\endcsname%
      \put(133,2675){\rotatebox{-270}{\makebox(0,0){\strut{}True Positive Rate}}}%
      \csname LTb\endcsname%
      \put(2679,124){\makebox(0,0){\strut{}False Positive Rate}}%
      \csname LTb\endcsname%
      \put(3998,764){\makebox(0,0)[r]{\strut{}0.00002}}%
      \csname LTb\endcsname%
      \put(3998,1013){\makebox(0,0)[r]{\strut{}0.00006}}%
      \csname LTb\endcsname%
      \put(3998,1262){\makebox(0,0)[r]{\strut{}0.00014}}%
      \csname LTb\endcsname%
      \put(3998,1511){\makebox(0,0)[r]{\strut{}0.00030}}%
      \csname LTb\endcsname%
      \put(3998,1760){\makebox(0,0)[r]{\strut{}0.00060}}%
      \csname LTb\endcsname%
      \put(3998,2009){\makebox(0,0)[r]{\strut{}0.00300}}%
    }%
    \gplbacktext
    \put(0,0){\includegraphics{spam-tpr-fpr-by-edge-rates}}%
    \gplfronttext
  \end{picture}%
\endgroup

%% file: conc.tex
\section{Conclusion}
\label{sec:conc}

We have presented an initial characterization of aggregate user
behavior, describing the distributions of lifetime contributions,
tweet rates, and inter-tweet durations. These behaviors are thought to
be common across communication platforms, but our results differ from
prior analysis, suggesting future study to determine the true extent
of the similarities.  Our retweet graph analysis revealed structural
differences from the followers graph that are more consistent with
real world social networks. Explaining the underlying causes of the
observed differences---we conjecture that retweets more closely mirror
real-world relationships and trust---is an open problem.  Finally, we
developed a method for detecting spammers via their low connectivity
in the retweet graph.

%% file: appendices.tex
\appendices

\input{retweet-identification}
\input{mle}

%% file: retweet-identification.tex
\section{Procedures for Identifying Retweets}
\label{sec:retweet-identification}

Retweeting was not an official feature in Twitter's early years, but
instead developed organically. A variety of syntaxes appeared (e.g.,
\texttt{RT@username}, \texttt{retweeting username}, and \texttt{via
  username}) and are still used today. We detect these retweets using
the following (Java) regular expression.
\begin{verbatim}
Pattern.compile(
 "(?:^|[\\W])(?:rt|retweet(?:ing)?|via)" + 
 "\\s*:?\\s*@\\s*([a-zA-Z0-9_]{1,20})"   +
 "(?:\$|\\W)"
)
\end{verbatim}
In 2009, Twitter
officially\footnote{https://blog.twitter.com/2009/project-retweet-phase-one}
added support for retweeting to their backend schema and the user
interface. These retweets are identified by the Twitter API.

%% file: mle.tex
\section{Derivation of the EM Method}
\label{sec:em-derivation}

Using the same notation as \autoref{sec:em-summary}, the likelihood to
maximize is
\begin{align}
  \mathcal{L}_C(\phi|f,g) &= \log p(f,g|\phi) \\
                          &\propto \log p(f|\phi) \\
                          &\propto \log \prod_{1\leq j\leq i} \left( \phi_i c_{i,j} \right)^{f_{i,j}} \\
                          &= \sum_{1\leq j\leq i} f_{i,j}\log \left( \phi_i c_{i,j} \right).
\end{align}
The expected likelihood under an estimate $\phi^{k}$ is
\begin{align}
  \mathcal{Q}(\phi,\phi^{(k)}) &\triangleq \E_{f|g,\phi^{(k)}}\left[ \mathcal{L}_C(\phi|f,g) \right] \\
                               &= \sum_{1\leq j\leq i} \E_{\phi^{(k)}} \left[ f_{i,j}|g \right] \log\left( \phi_i c_{i,j} \right)
\end{align}
and the iterative maximization step is
\begin{equation}
  \phi^{(k+1)} \triangleq \argmax_{\phi} \mathcal{Q}(\phi,\phi^{(k)}).
\end{equation}
The maximum is computed under the constraint
$\sum_{1\leq i}\phi_{i} = 1$ using Lagrangian multipliers. Defining the Lagrangian
\begin{equation}
  L(\phi,\lambda) \triangleq \sum_{1\leq j\leq i} \E_{\phi^{(k)}} \left[f_{i,j}|g \right] 
  \log\left( \phi_i c_{i,j} \right) + \lambda(1 - \sum_{1\leq i}\phi_i),
\end{equation}
the associated partial derivatives are
\begin{align}
\frac{\partial L}{\partial \phi_i}   &= \frac{\E_{\phi^{(k)}} \left[ f_{i,j}|g \right]}{\phi_i} - \lambda\text{, and}\\
  \frac{\partial L}{\partial \lambda} &= 1 - \sum_{1\leq i}\phi_i
\end{align}
Solving for
\begin{equation}
\phi_i = \frac{\E_{\phi^{(k)}} \left[ f_i|g \right]}{\sum_{1\leq l} \E_{\phi^{(k)}} \left[ f_l|g \right] }
\end{equation}
and defining 
\begin{equation}
  \gamma \triangleq \sum_{1\leq l} \E_{\phi^{(k)}} \left[ f_l|g \right] = \sum_{1\leq l} g_l 
\end{equation}
yields
\begin{align}
\phi^{(k+1)}_i &= \frac{\E_{\phi^{(k)}} \left[ f_i|g \right]}{\gamma} \\
               &= \frac{\phi_i^{(k)}}{\gamma} \sum_j\frac{c_{i,j}g_j}{\sum_{1\leq l}\phi_l^{(k)}c_{l,j}}.
\end{align}
or in matrix form (for fast implementation on a computer)
\begin{equation}
\phi^{(k+1)} = \frac{1}{\gamma} \times \phi^{(k)} \times C\cdot\frac{g}{C^\top\cdot\phi^{(k)}}.
\end{equation}
The original frequencies can be expressed as
\begin{equation}
  \hat{f}_i = \gamma\phi_i\frac{1}{1-B_{0.1}(i,0)}.
\end{equation}